\pdfoutput=1

\documentclass[PAPER, atlasdraft=false, UKenglish, cernpreprint, texmf, orcidlogo]{atlasdoc}

\usepackage{atlaspackage}
\usepackage{atlasbiblatex}
\usepackage{multirow}

\usepackage{atlasphysics}
\graphicspath{{logos/}{figures/}}

\usepackage{ANA-HIGP-2024-33-PAPER-defs}

\AtlasTitle{
Search for Higgs boson pair production in association with top-quark pairs using 196 fb$^{-1}$ of proton--proton collision data at $\sqrt{s}=$ 13 and 13.6 TeV with the ATLAS detector
}

\AtlasAbstract{%
This paper presents the first search for non-resonant Higgs boson pair production in association with a top-quark pair ($t\Bar{t}HH$) using proton--proton collision data collected with the ATLAS detector at the Large Hadron Collider. The data sample corresponds to  an integrated luminosity of 196 fb$^{-1}$, comprising  140 fb$^{-1}$ at a centre-of-mass energy of $\sqrt{s}=13$ TeV and 56 fb$^{-1}$ at 13.6 TeV. The search targets three distinct final states expected from $t\Bar{t}HH$ decays: (i) one lepton (electron or muon) with at least five $b$-quarks, (ii) at least two $b$-quarks accompanied by two leptons with the same electric charges or multiple leptons, and (iii) at least three $b$-quarks with two photons.
The $t\Bar{t}HH$ production cross-section, relative to its Standard Model prediction, is measured to be $\mu_{t\Bar{t}HH}=-3^{+11}_{-12}$. This result corresponds to a 95\% confidence-level upper limit of 20 times the Standard Model prediction for the $t\Bar{t}HH$ production cross-section, with an expected limit of 21.
The Higgs effective field theory Wilson coefficient $c_{t\Bar{t}HH}$ is also constrained at the same confidence level to the range of $-3.9<c_{t\Bar{t}HH}<3.3$, compared with the expected range of $-4.0<c_{t\Bar{t}HH}<3.5$.
}

\AtlasRefCode{HIGP-2024-33}

\PreprintIdNumber{CERN-EP-2026-045}

\AtlasJournal{JHEP}
\AtlasJournalRef{JHEP 08 (2026) 020}
\AtlasDOI{10.1007/JHEP08(2026)020}

\AtlasCoverEgroupAnalysisTeam{atlas-higp-2024-33-analysis-team@cern.ch}

\hypersetup{pdftitle={ATLAS document},pdfauthor={The ATLAS Collaboration}}

\begin{document}

\maketitle

\tableofcontents

\clearpage
%

\section{Introduction}

The Standard Model of particle physics (SM) has proven to be an immensely successful theory in describing fundamental particles and their interactions. A crucial aspect of this theory is the Brout--Englert--Higgs mechanism~\cite{Englert:1964et,Higgs:1964pj,Guralnik:1964eu} for the electroweak symmetry breaking (EWSB), which predicts the existence of a neutral scalar particle known as the Higgs boson ($H$). Since its discovery in 2012 by the ATLAS and CMS collaborations~\cite{HIGG-2012-27,CMS-HIG-12-028}, substantial progress has been made in understanding the production processes and decay modes of the Higgs boson and its couplings to fermions and gauge bosons~\cite{HIGG-2021-23, CMS-HIG-22-001}. However, the trilinear and quartic self-couplings of the Higgs boson remain only weakly constrained. These self-couplings are linked to the Higgs potential, which drives the EWSB, and can be studied through Higgs boson pair (\HH) production.

Following the discovery of the Higgs boson, the search for \HH\ production has become a central focus of the physics programme at the Large Hadron Collider (LHC). Among the primary \HH\ production processes in proton--proton ($pp$) collisions at the LHC, gluon--gluon fusion (\ggF) and vector-boson fusion (\VBF) processes have garnered the most attention due to their higher cross-sections. Various \HH\ decay channels, such as \bbbb, \bbyy, and \bbtt, have been explored~\cite{HDBS-2021-18,CMS-HIG-20-011}. The production of \HH\ in association with a vector boson, known as the \VHH\ process, has also been investigated~\cite{HDBS-2019-31,CMS-HIG-22-006}. No evidence for \HH\ production has been observed at the LHC\@. The combined 95\% confidence level~(CL) upper limit on the \HH\ production cross-section, expressed relative to the SM prediction, is 2.9  from ATLAS~\cite{HDBS-2021-18}, 3.5 from CMS~\cite{CMS-HIG-20-011} and 2.5 from their combination~\cite{CMS:2026nuu}.%

\begin{figure}[!htb]
\centering
\subfloat[]{\includegraphics[width=0.32\textwidth]{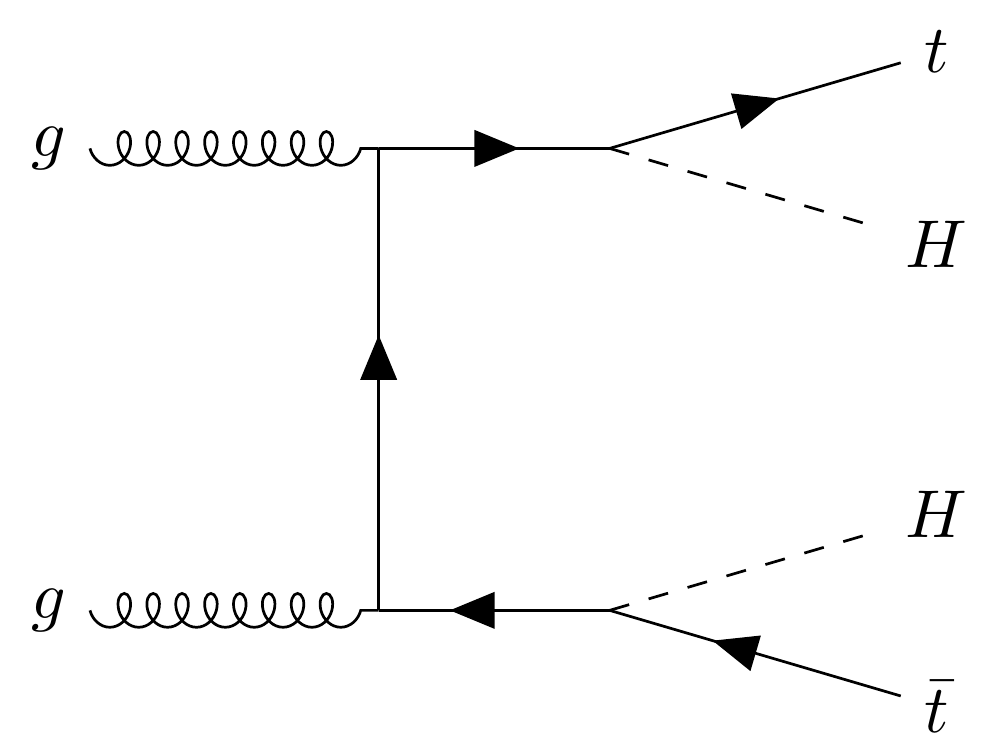}} %
\subfloat[]{\includegraphics[width=0.32\textwidth]{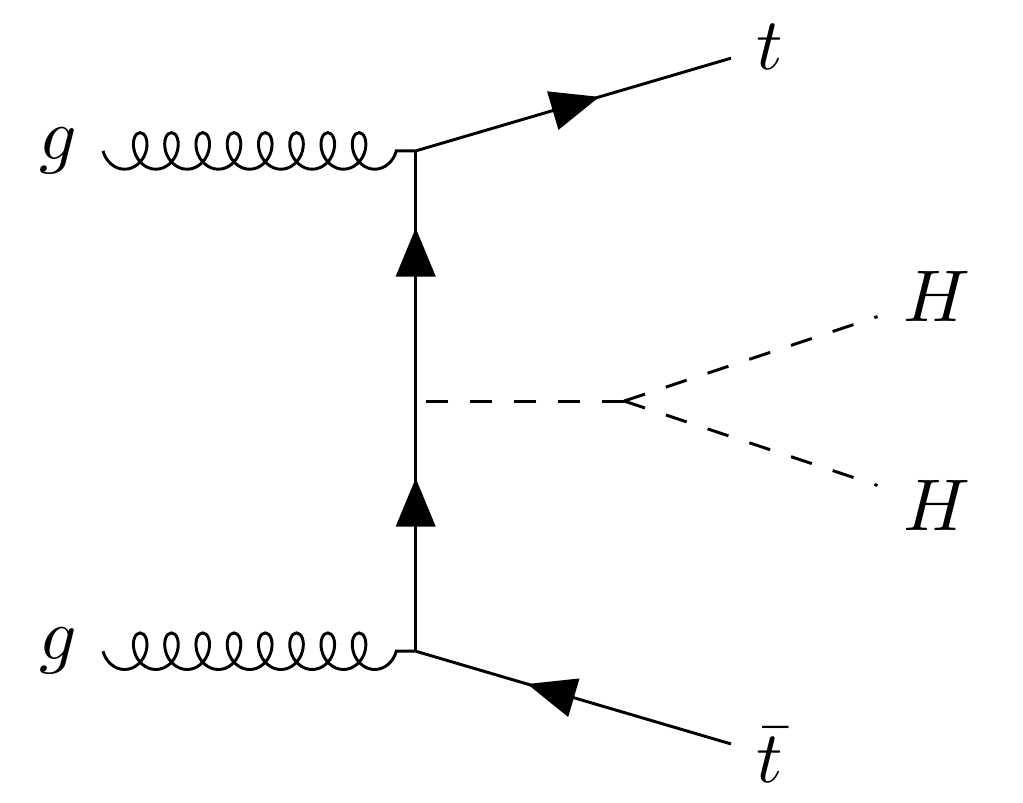}} %
\subfloat[]{\includegraphics[width=0.31\textwidth]{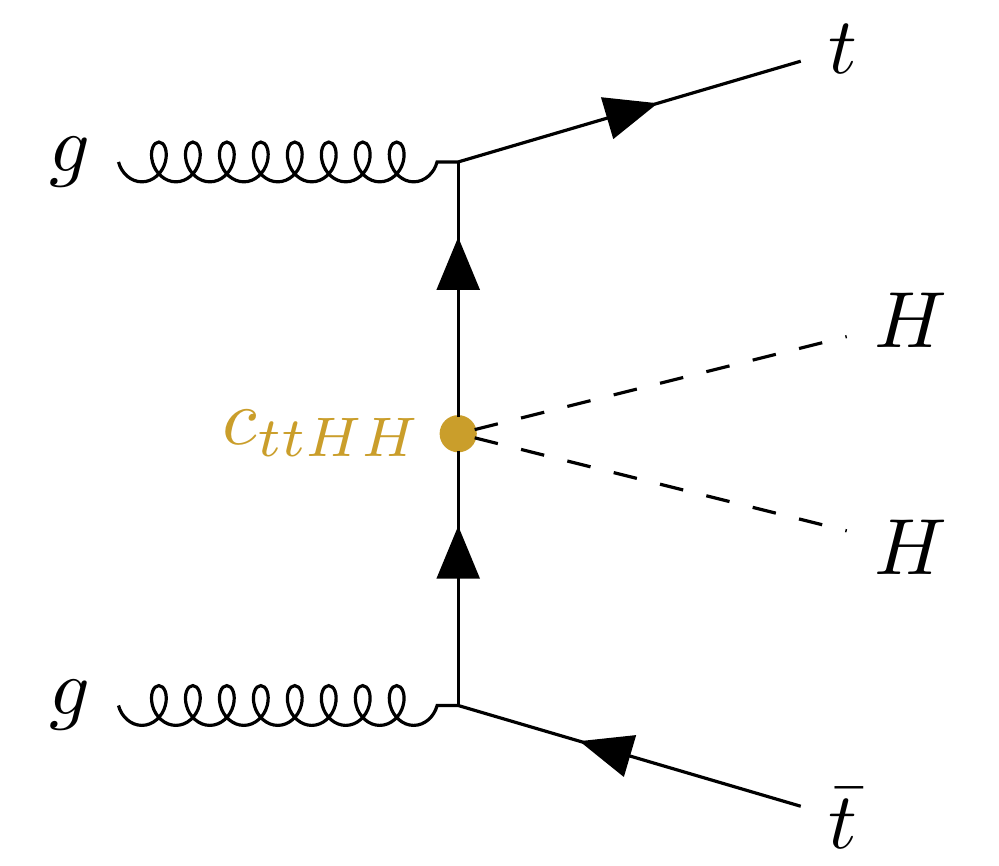}\label{fig:Feynman-Diagram-c}} %

\caption{Examples of leading-order Feynman diagrams for the non-resonant \ttHH\ process. Each diagram depicts a distinct subprocess arising from different Higgs boson couplings: (a) the Higgs--top Yukawa coupling, (b) the Higgs--top Yukawa coupling and Higgs boson trilinear self-coupling,  and (c) the non-SM \ttHH\ quartic coupling (highlighted with a circle), as described by the Higgs effective field theory.}
\label{fig:Feynman-Diagram}
\end{figure}

This paper presents the first search for Higgs boson pair production in association with a top-quark pair (\ttHH), focusing on non-resonant \HH\ production as predicted by the SM\@. The \ttHH\ process has the third largest \HH\ production cross-section in the SM, after the \ggF\ and \VBF\ $HH$ processes. Examples of the leading Feynman diagrams for the process are shown in Figure~\ref{fig:Feynman-Diagram}.

In the SM, the \ttHH\ process has a predicted cross-section of $\sigma_{\ttHH}=0.756^{\;\,+4.3\%}_{-15.0\%}\ (\text{scale})\pm 3.4\%\ (\text{PDF}+\alphas)$~fb for $pp$ collisions at $\sqrt{s}=13$~\TeV and $0.860^{\;\,+4.2\%}_{-14.0\%}\ (\text{scale})\pm 3.3\%\ (\text{PDF}+\alphas)$~fb at 13.6~\TeV, calculated at next-to-leading order (NLO) in quantum chromodynamics (QCD) for a top-quark mass of $m_t=172.5\,\GeV$ and a Higgs boson mass of $m_H=125\,\GeV$~\cite{Frederix:2014hta,deFlorian:2016spz}. The ``PDF+\alphas'' uncertainty accounts for uncertainties in the parton distribution functions and strong coupling constant, the ``scale'' uncertainty is due to the finite order of QCD calculations. Approximately 81\% of the cross-section arises from the Higgs--top Yukawa coupling diagram shown in Figure~\ref{fig:Feynman-Diagram}~(a), while about 8\% originates from the diagram involving the Higgs boson self-coupling shown in Figure~\ref{fig:Feynman-Diagram}~(b). The remaining 11\% results from constructive interference between the two diagrams. Although the \ttHH\ cross-section is sensitive to deviations of the Higgs boson self-coupling from its SM value, its dependence on the coupling modifier is considerably weaker than in the \ggF\ and \VBF\ $HH$ production modes~\cite{Frederix:2014hta}.

The analysis utilises the full LHC Run~2 data sample (2015--2018), corresponding to an integrated luminosity of 140 \ifb collected with the ATLAS detector at a centre-of-mass energy $\sqrt{s}=13\,\TeV$ and a Run~3 data sample (2022--2023) corresponding to 56~\ifb at $\sqrt{s}=13.6\,\TeV$.
The analysis is performed using the framework documented in Ref.~\cite{atlas_collaboration_2025_15927813}, developed within the ATLAS Collaboration for collision data analysis. The statistical interpretation is carried out using \textsc{TRExFitter} framework~\cite{aly_2025_16684099}.

Since the top quark decays almost exclusively into a $W$ boson and a $b$-quark,  \ttHH\ events can result in a variety of final states with two or more $b$-quarks, depending on the decay modes of the $W$ and Higgs bosons.
The search leverages the large branching ratio of the \Hbb\ decay and the distinctive experimental signatures of leptons and photons. Three specific final states are considered. The first, labelled as \OneL, is characterised by a single lepton ($\ell$ = electron or muon) and at least five $b$-quarks, capturing signals primarily from $W\to\ell\nu$ and \HHbbbb\ decays. The second, referred to as \SSML, requires two or more $b$-quarks together with two leptons with the same electric charges (same-sign dilepton pair) or more than two leptons. It is designed to select events in which at least one $W$ boson decays leptonically, with additional leptons potentially arising from Higgs boson decays (e.g. $H\to\tau\tau, WW^*,ZZ^*$ decays). The third, referred to as \bbyy, consists of at least two $b$-quarks together with two photons,  targeting \HHbbyy\ decays.%

The effect of new physics introduced by a new \ttHH\ quartic interaction, shown in Figure \ref{fig:Feynman-Diagram-c}, can be modelled using an effective field theory (EFT) approach.
The results are interpreted in the Higgs effective field theory (HEFT) framework~\cite{Feruglio:1992wf, Buchalla:2012qq, ALONSO2013926, Brivio:2013pma, Buchalla_2014, Brivio_2019}, as it allows the \ttHH\ quartic interaction via the Wilson coefficient \cttHH\ to be directly targeted without affecting other Higgs boson couplings. Unlike the \ggF\ $HH$ channel, which includes additional Higgs--gluon couplings, the \ttHH\ channel is only affected by them at higher loop-order in weakly interacting theories, making the interpretation more robust.

This paper is structured as follows: Section~\ref{sec:detector} provides a brief overview of the ATLAS detector, and Section~\ref{sec:samples} describes the data and Monte Carlo~(MC) samples used in the analysis. The reconstruction and identification of the physics objects, along with the event-selection criteria, are summarised in Section~\ref{sec:objectsAndEvents}. The definitions of the signal and background regions, together with the background estimation methods, are presented in Section~\ref{sec:backgrounds}, while the multivariate techniques employed for signal--background separation are outlined in  Section~\ref{sec:mva}. Theoretical and experimental uncertainties are discussed in Section~\ref{sec:systematics}, and the final results and the HEFT interpretation are given in Sections~\ref{sec:results} and \ref{sec:results_eft}, respectively. Finally, Section \ref{sec:conclusion} provides the conclusions.

%
%
%
%
%
%


\section{ATLAS detector}
\label{sec:detector}

\newcommand{\AtlasCoordFootnote}{%
ATLAS uses a right-handed coordinate system with its origin at the nominal interaction point (IP)
in the centre of the detector and the \(z\)-axis along the beam pipe.
The \(x\)-axis points from the IP to the centre of the LHC ring,
and the \(y\)-axis points upwards.
Polar coordinates \((r,\phi)\) are used in the transverse plane,
\(\phi\) being the azimuthal angle around the \(z\)-axis.
The pseudorapidity is defined in terms of the polar angle \(\theta\) as \(\eta = -\ln \tan(\theta/2)\) and is equal to the rapidity
$ y = \frac{1}{2} \ln \left( \frac{E + p_z}{E - p_z} \right) $ in the relativistic limit.
Angular distance is measured in units of \(\Delta R \equiv \sqrt{(\Delta y)^{2} + (\Delta\phi)^{2}}\).}

The ATLAS experiment~\cite{PERF-2007-01,GENR-2019-02} at the LHC is a multipurpose particle detector
with a forward--backward symmetric cylindrical geometry and a near \(4\pi\) coverage in
solid angle.\footnote{\AtlasCoordFootnote}
It consists of an inner tracking detector surrounded by a thin superconducting solenoid
providing a \qty{2}{\tesla} axial magnetic field, electromagnetic (EM) and hadronic calorimeters, and a muon spectrometer.
The inner detector~(ID) covers the pseudorapidity range \(|\eta| < 2.5\).
It consists of silicon pixel, silicon microstrip, and transition radiation tracking detectors.
Lead/liquid-argon (LAr) sampling calorimeters provide EM energy measurements
with high granularity within the region \(|\eta|< 3.2\).
A steel/scintillator-tile hadronic calorimeter covers the central pseudorapidity range (\(|\eta| < 1.7\)).
The endcap and forward regions are instrumented with LAr calorimeters
for EM and hadronic energy measurements up to \(|\eta| = 4.9\).
The muon spectrometer surrounds the calorimeters and is based on
three large superconducting air-core toroidal magnets with eight coils each.
The field integral of the toroids ranges between \num{2.0} and \qty{6.0}{\tesla\metre}
across most of the detector.
The muon spectrometer includes a system of precision tracking chambers up to \(|\eta| = 2.7\) and fast detectors for triggering up to \(|\eta| = 2.4\).
The luminosity is measured mainly by the LUCID--2 detector that is located close to the beampipe.
A two-level trigger system was used to select events~\cite{TRIG-2016-01,TRIG-2022-01}.
The first-level trigger is implemented in hardware and used a subset of the detector information
to accept events at a rate close to \qty{100}{\kHz}.
This is followed by a software-based trigger that
reduced the accepted rate of complete events to \qty{1.25}{\kHz} and \qty{3}{\kHz} on average in Run~2 and Run~3, respectively,
depending on the data-taking conditions.
A software suite~\cite{SOFT-2022-02} is used in data simulation, in the reconstruction
and analysis of real and simulated data, in detector operations, and in the trigger and data acquisition
systems of the experiment.


\section{Data and Monte Carlo samples}
\label{sec:samples}

The data used in the \OneL\ and \SSML\ channels were collected using single-lepton (electron or muon) triggers with varying transverse momentum (\pT) thresholds and isolation requirements~\cite{TRIG-2018-05,TRIG-2018-01,TRIG-2022-01}.  The lowest \pT threshold ranged from 24 to 26~\GeV for electrons and from 20 to 26~\GeV for muons throughout the data-taking period covered by this search. Events in the \bbyy\ channel were collected with diphoton triggers~\cite{TRIG-2018-05,TRIG-2022-01}, with nominal transverse momentum of 35 (25)~\GeV for the leading (sub-leading) photon. Events are included in the search only if all relevant detector subsystems were confirmed to be in optimal operating condition~\cite{DAPR-2018-01}, resulting in a total integrated luminosity of $140.1\pm 1.2$ \ifb for the Run~2 data sample~\cite{DAPR-2021-01} and $56.0 \pm 1.1$~\ifb for the Run~3 data sample~\cite{ATL-DAPR-PUB-2023-001,ATL-DAPR-PUB-2024-001}.

MC simulations play a central role in signal modelling, efficiency determination, background estimation, and event selection optimisation.
The simulated samples were produced using a variety of event generators, along with different programs for parton showering, hadronisation, and underlying event modelling, collectively referred to as soft-QCD processes below. The decays of bottom and charm hadrons were simulated using the \evtgen\ program~\cite{Lange:2001uf}, except for samples generated by \Sherpa~\cite{Bothmann:2019yzt}. Top quarks were decayed at leading order~(LO) using \madspin~\cite{Frixione:2007zp, Artoisenet:2012st} to preserve spin correlations. All simulations assumed a top-quark mass of 172.5~\GeV, a Higgs boson mass of 125~\GeV, and the corresponding decay branching ratios~\cite{deFlorian:2016spz}.
Simulated processes were normalised to the most accurate theoretical cross-section predictions available.

Table~\ref{tab:MC-samples} summarises the nominal signal and background samples used in this search, and additional details of the signal and major background processes are provided in this section. Alternative samples employing different generators or parton-shower models are also used to evaluate modelling uncertainties.
All samples were generated with statistics far exceeding those of the data, in most cases by more than two orders of magnitude.
The MC samples were processed through an ATLAS detector simulation program~\cite{SOFT-2010-01}, either using  \GEANT~\cite{Agostinelli:2002hh} or a fast simulation with parameterised calorimeter showers~\cite{ATL-PHYS-PUB-2010-013}. Pile-up effects, i.e. multiple interactions occurring within the same or nearby bunch crossings, were modelled by overlaying~\cite{SIMU-2020-01} the hard-scatter events with simulated inelastic $pp$ collisions generated using the EPOS~\cite{Porteboeuf:2010um} and \Pythia~\cite{Sjostrand:2014zea} programs. The distribution of the average number of interactions was matched to that observed in the data for each data-taking period.

For samples generated with \powhegbox~\cite{Frixione:2007nw,Nason:2004rx,Frixione:2007vw,Alioli:2010xd,Re:2010bp,Alioli:2009je}, \mg\,(\mdg)~\cite{Alwall:2011uj}, and \mgamc\,(\mgmc)~\cite{Alwall:2014hca}, soft-QCD processes were modelled using \Pythia\ with the \nnpdftwo~\cite{Ball:2012cx} parton distribution function~(PDF) set and the A14 set of tuned parameters (tune)~\cite{ATL-PHYS-PUB-2014-021}. Samples generated with \sherpa\ used its native implementation of soft-QCD modelling. Events generated with the \herwigseven~\cite{Bahr:2008pv,Bellm:2017jjp,Bellm:2019zci} generator, together with the \mmhtnlo~\cite{Harland-Lang:2014zoa} PDF set and H7-UE tune, are used to assess systematic uncertainties associated with the modelling of soft-QCD processes.

The \ttHH\ signal samples were generated to cover all the final states considered in this analysis, with decay filters applied to ensure orthogonality at generator level. For the \OneL\ channel, separate samples were produced for fully hadronic, single-leptonic, and dileptonic decays of the top-quark pairs, with the Higgs boson pair decaying as \HHbbbb. To target the \SSML\ final state, inclusive samples were generated with all possible top-quark and Higgs boson decays except \HHbbbb, and were filtered to contain either a same-sign dilepton pair or at least three leptons. In both channels, the \ggF\ and \VBF\ $HH$ contributions are found to be negligible and therefore are not considered. Dedicated samples with \HHbbyy\ decays were also produced. The signal samples were modelled with version~3.3.1 of the \mg\,(\mdg) generator~\cite{Alwall:2011uj} at LO using \nnpdftwo, and were interfaced to \pythia\ for parton-shower and underlying-event modelling using the A14 tune. The renormalisation ($\mu_\text{r}$) and factorisation ($\mu_\text{f}$) scales were set to \HTscale$/2$, where \HTscale\ is defined as the sum of the transverse masses $\sqrt{p_\text{T}^2+m^2}$ of all final-state partons. The \ttHH\ samples were normalised using the SM cross-section computed at NLO in QCD~\cite{Frederix:2014hta}.

\begin{table}[!htb]
\centering
\caption{List of MC event generators and their associated matrix-element (ME) orders, parton showers, PDFs, and underlying set of tuned parameters used for simulating signal and background processes.
When different from the Run~2 configuration, Run~3 settings are indicated in parentheses.
}
\vspace*{-2mm}
\footnotesize
\begin{tabular}{llllll}
\toprule
Process        &  Generator         & ME    & Parton          & PDF set           & Tune  \\
&        & order   & Shower         &                  &    \\[1mm] \hline \\[-3mm]
\ttHH          &  \mdg           &  LO     & \pythia        & \nnpdftwo        & A14 \\[1mm] \hline \\[-3mm]

\ttbar (5FS)   &  \powhegbox     &  NLO    & \pythia        & \nnpdfnlo      & A14 \\

\tttt          &  \mgmc         &  NLO    & \pythia        & \nnpdfnlo      & A14 \\
\ttt           &  \mgmc         &  LO     & \pythia        & \nnpdftwo        & A14 \\[1mm] \hline \\[-3mm]

\ttW           & \sherpa         & MEPS@NLO    & \sherpa        & \nnpdfnnlo     & \Sherpa \\
\ttZ           & \mgmc          & NLO     & \pythia        & \nnpdfnnlo       & A14 \\[1mm] \hline \\[-3mm]

\ttWW          & \mgmc (\mdg)          & LO      & \pythia        & \nnpdfnlo        & A14 \\
\ttWZ,\,\ttWH,\,\ttZZ   & \mdg             & LO      & \pythia        & \nnpdftwo      & A14 \\[1mm] \hline \\[-3mm]

$tW$       &  \powhegbox     &  NLO    & \pythia        & \nnpdfnlo        & A14 \\
\tZ            &  \mgmc        &  LO     & \pythia        & \nnpdfnlonf      & A14 \\
$t(q)b$ ($t$-channel)  &  \powhegbox     &  NLO    & \pythia        & \nnpdfnlonf    & A14  \\
$t(q)b$ ($s$-channel)  &  \powhegbox     &  NLO     & \pythia       & \nnpdfnlo       & A14         \\
\tWZ,\,\tHq,\,\tHW  &  \mgmc         &  NLO    & \pythia        & \nnpdfnlo        & A14 \\[1mm] \hline \\[-3mm]

$V/VV/VVV+\text{jets}$   & \sherpa    &  MEPS@NLO$\dag$      & \sherpa        & \nnpdfnlo     & \sherpa \\

$\gamma\gamma+\textrm{jets}$  & \Sherpa     &  NLO    & \Sherpa          & \nnpdfnnlo   & A14 \\
$\gamma\gamma+t\bar{t}$       & \mdg         &  LO       & \pythia          & \nnpdftwo    & A14 \\[1mm] \hline \\[-3mm]

\ggF\ $H$        & \nnlops~\cite{Hamilton:2013fea,Hamilton:2015nsa}  &  NLO      & \pythia        & \pdflhcone   & AZNLO (A14) \\
\VBF\ $H$        & \powhegbox   &  NLO      & \pythia        & \pdflhcone (\pdflhctwo)    & AZNLO (A14) \\
$VH$             & \powhegbox   &  NLO      & \pythia        & \pdflhcone (\pdflhctwo)   & AZNLO (A14) \\
\ttH           &  \powhegbox     &  NLO    & \pythia        & \nnpdfnlo (\pdflhctwo)     & A14 \\[1mm] \hline \\[-3mm]
\ggF\ $HH$       & \powhegbox   &  NLO      & \pythia        & \pdflhctwo     & A14 \\
\VBF\ $HH$       & \mdg          &  NLO      & \pythia        & \nnpdfnlo    & A14 \\

\bottomrule
\multicolumn{6}{l}{$\dag$ For the subdominant $gg\to VV$ contribution, the ME calculation is at leading order.}\\
\end{tabular}
\label{tab:MC-samples}
\end{table}

The three search channels have distinct background compositions due to their differing signal characteristics. In the \OneL\ channel, the dominant background arises from \ttbar\ production accompanied by additional heavy-flavour jets. The \SSML\ channel features a complex mixture, including contributions from top-quark pair production in association with a $W$, $Z$, or Higgs boson, four top-quark production, and events containing charge-misidentified leptons or leptons originating from heavy-flavour decays. The \bbyy\ channel is affected by both continuum and resonant backgrounds: the former primarily from $\gamma\gamma+\textrm{jets}$ processes, and the latter from single-Higgs or non-\ttHH\ di-Higgs production. In all channels, MC simulation plays a central role in estimating background contributions, as detailed in Section~\ref{sec:backgrounds}. A brief overview of the simulation of the major background processes is provided below.

Top-quark pair production (\ttbar) is a dominant background source in both the \OneL\ and \SSML\ channels. The process was modelled using a \ttbar\ sample generated in the five-flavour scheme (5FS) with the \powhegbox\ generator at NLO, in which $b$-quarks are treated as massless, and the \nnpdfnlo~\cite{Ball:2014uwa} PDF set. The $h_\textrm{damp}$ parameter was set to 1.5 times of the top-quark mass ($m_t$) in \powhegbox~\cite{ATL-PHYS-PUB-2016-020}, and the \texttt{pThard} parameter was set zero in \Pythia.\footnote{The $h_\textrm{damp}$ parameter regulates the modelling of low- and high-\pT\ radiation within \powhegbox, while \texttt{pThard} controls how \Pythia\ vetos its own radiation to avoid double-counting emissions already generated by \powhegbox.} The $\mu_\text{r}$ and $\mu_\text{f}$ scales were set to the \powhegbox\ default choice of $\sqrt{m_t^2+\pT^2}$, where \pT is the top-quark transverse momentum. The simulated sample was normalised to the SM prediction at next-to-next-to-leading order (NNLO) in QCD including next-to-next-to-leading logarithmic (NNLL) resummation, as computed with \textsc{Top++2}~\cite{Czakon:2011xx}.%

Top-quark pair production in association with a $W$ boson (\ttW) is one of the dominant background processes in the \SSML\ channel. It was simulated using the \sherpa\ generator~\cite{Bothmann:2019yzt}, which provides matrix-element calculations at NLO accuracy in QCD for up to one additional parton, and at LO accuracy for up to two additional partons~\cite{Cascioli:2011va,Denner:2016kdg,Buccioni:2019sur},  merged with the \Sherpa\ parton shower through the MEPS@NLO prescription~\cite{Hoeche:2012yf}. The $\mu_\text{r}$ and $\mu_\text{f}$ scales were set to be \HTscale$/2$ as for the \ttHH\ signal sample.
Electroweak corrections up to orders $\alpha^2\alphas^2$ and $\alpha^3$~\cite{Kallweit:2015dum,Gutschow:2018tuk} were included, along with real-emission contributions from subleading corrections at order $\alpha^3\alphas$~\cite{Frederix:2017wme}.
The process was normalised to the SM cross-section computed at NNLO in QCD~\cite{Buonocore:2023ljm}.

Top-quark pair production in association with a $Z$ boson (\ttZ) or a Higgs boson (\ttH) constitute two other important background sources in the \SSML\ channel. The \ttZ\ sample was simulated using \mgmc\ at NLO with the \nnpdfnlo\ PDF set, with  $\mu_\text{r}$ and $\mu_\text{f}$ both set to \HTscale$/2$. It was normalised to the SM cross-section calculated at NLO in QCD and electroweak corrections~\cite{deFlorian:2016spz,Frixione:2015zaa}. The \ttH\  sample was simulated using \powhegbox\ at NLO with the PDF set of \nnpdfnlo\ for Run~2 and \pdflhctwo~\cite{Butterworth:2015oua} for Run~3. The $\mu_\text{r}$ and $\mu_\text{f}$ scales were set to the geometric mean of the transverse mass of the three outgoing partons. The process was normalised to the SM cross-section calculated at NLO in QCD and electroweak corrections~\cite{deFlorian:2016spz,Beenakker:2002nc,Dawson:2003zu,Zhang:2014gcy,Frixione:2014qaa}.

The four-top-quark production process (\tttt) also contributes significantly in the \SSML\ channel. It was simulated using  the \mgamc\ generator, which incorporates matrix-element calculations at NLO accuracy in QCD\@. The $\mu_\text{r}$ and $\mu_\text{f}$ scales were set to $H^\text{scale}_T/2$. The simulated sample was normalised to the SM cross-section computed at NLO in QCD, including electroweak corrections~\cite{Frederix:2017wme}. Together with the three-top-quark process (\ttt), modelled at LO using \mgamc\ as described in Ref.~\cite{HIGP-2024-08}, these contributions are collectively referred to as the multi-top background in the following.

In the \OneL\ and \SSML\ channels, additional minor background processes, such as $V/VV/VVV+\text{jets}$ and $t\bar{t}+VV$, are modelled using \Sherpa\ at NLO and \mg\ at LO, respectively, as detailed in Refs.~\cite{HIGP-2024-08,HIGG-2020-24,TOPQ-2019-30}.

For the \bbyy\ channel, the non-resonant diphoton background arises from processes with two prompt photons and events in which one or both photon candidates originate from misidentified jets. The normalisation of this background is determined entirely from data, while \sherpa\ MC samples with two prompt photons were used to model its kinematics. The resonant background contributions from single Higgs boson production (primarily \ggF\ $H$ and \VBF\ $H$)  and non-\ttHH\ Higgs boson pair production (mainly \ggF\ $HH$ and \VBF\ $HH$) were modelled using MC samples, most of which were generated with the \powhegbox\ generator as described in Ref.~\cite{HIGP-2025-10}.


\section{Analysis strategy, object identification, and event selection}
\label{sec:objectsAndEvents}

Although all three channels target \ttHH\ production, their distinct signatures and background compositions require dedicated analysis strategies. The \OneL\ and \SSML\ final states feature relatively high signal and background rates, enabling analyses to be performed both on the combined Run~2 and Run~3 data samples and separately for each data-taking period. In contrast, the \bbyy\ final state offers a cleaner signature but suffers from limited numbers of events, motivating an analysis based solely on the combined data sample.

The \OneL\ and \SSML\ channels adopt similar methodologies, relying on both simulation and data for background estimation and employing transformer-based machine learning techniques~\cite{Vaswani:2017lxt} to enhance signal-to-background discrimination. Potential \ttHH\ contributions are extracted through simultaneous fits to the transformer outputs in the signal regions and to selected fit variables in the background control regions.
In the \bbyy\ channel, potential signals are identified using a strategy analogous to Ref.~\cite{HIGP-2025-10}, based on fits to the diphoton invariant-mass spectra. This method leverages the narrow Higgs boson resonance and the excellent diphoton mass resolution of the ATLAS detector, enabling a fully data-driven determination of the continuum background from the resonance sidebands.

The physics objects used in this analysis are summarised in Section~\ref{sec:objects}, with many shared across the three channels. The event selection criteria specific to each channel are then described in Sections~\ref{sec:1Lselection}, \ref{sec:MLselection}, and \ref{sec:bbyyselection}. To be considered, an event must have a primary vertex (PV), defined as the vertex reconstructed with at least two ID tracks with $\pT>500\,\MeV$. If multiple PVs are present, the one with the largest sum of $p_\text{T}^2$ of its associated tracks is chosen~\cite{ATL-PHYS-PUB-2015-026}.

\subsection{Object definition}
\label{sec:objects}

Electrons are reconstructed by matching topological energy clusters in the EM calorimeter with tracks in the ID and are identified using likelihood-based criteria that incorporate calorimeter shower shape, track-to-cluster matching, and associated track quality requirements~\cite{EGAM-2018-01,EGAM-2021-01,EGAM-2021-02}. Electron candidates must have $\pT>15\,\GeV$ and $|\eta_\text{cluster}|<2.47$, excluding the transition region between the barrel and endcap calorimeters ($1.37<|\eta_\text{cluster}|<1.52$). In the \OneL\ channel, they are required to satisfy the \textit{loose} identification likelihood and \textit{loose} isolation criteria~\cite{EGAM-2018-01}, whereas the \SSML\ channel requires the \textit{tight} identification likelihood and  \textit{tight} isolation criteria. Unless otherwise stated, the isolation criteria are based on combined tracking and calorimeter information.

Muons are reconstructed by combining tracks from the ID with those from the muon spectrometer~\cite{MUON-2018-03,MUON-2022-01}. Muon candidates must have $\pT>15\,\GeV$ and $|\eta|<2.5$. In the \OneL\ channel, they must satisfy the \textit{loose} identification quality and \textit{loose} track-based isolation requirements, while in the \SSML\ channel they are required to meet the \textit{medium} identification quality and \textit{tight} isolation criteria~\cite{PERF-2015-10}.

To select leptons originating from the primary vertex, the transverse impact parameter, normalised by its estimated uncertainty ($|d_0|/\sigma(d_0)$), must be less than five for electrons and three for muons, while the longitudinal impact parameter must satisfy $|z_0\sin\theta|<0.5$~mm for both lepton flavours. %

Photons are reconstructed from  the topological energy clusters in the EM calorimeter~\cite{EGAM-2018-01,EGAM-2021-01,EGAM-2021-02}. Photon candidates are required to have $|\eta_\text{cluster}|<1.37$ or $1.52<|\eta_\text{cluster}|<2.37$. Photons are required to satisfy \textit{tight} identification likelihood and \textit{loose} track-based isolation criteria.

Jets are reconstructed using a particle-flow algorithm~\cite{PERF-2015-09} that combines measurements from the ID and the calorimeters. The anti-$k_t$ algorithm~\cite{Cacciari:2008gp} with a radius parameter of $R=0.4$ is employed. Jets are required to have $\pT>20\,\GeV$ and $|\eta|<2.5$ and a calibration procedure is performed as described in Refs.~\cite{JETM-2018-05,JETM-2022-01}. To mitigate pile-up effects, the jet-vertex tagger (JVT)~\cite{PERF-2014-03} is applied to jets with $\pT<60\,\GeV$ and $|\eta|<2.4$.

Jets originating from the hadronisation of $b$-quarks are identified (\textit{$b$-tagged}) using the GN2 algorithm~\cite{FTAG-2023-05}, a transformer-based neural network designed to predict the flavour of reconstructed jets. The network takes as input per-track quantities and their associated uncertainties, along with the jet \pT and $\eta$. Its output scores are combined into a $b$-tagging discriminant. Five thresholds, referred to as working points (WP), are defined with average efficiencies of 65\%, 70\%, 77\%, 85\%, and 90\% for tagging $b$-quark jets in simulated $\ttbar$ events. A pseudo-continuous $b$-tagging~(PCBT) score, $w_\text{PCBT}$, is assigned to each jet and represents the number of WPs it satisfies. The score starts at one for jets failing to satisfy any WP and increases by one for each successively tighter WP satisfied. It reaches a maximum value of $w_\text{PCBT}=6$ for jets satisfying the tightest 65\% WP\@. In the following, $N_b^{90\%\, (85\%,77\%)}$ denotes the number of $b$-tagged jets satisfying the 90\% (85\%, 77\%) WP\@. Another relevant variable is $\sum_{\text{jet}} w_{\text{PCBT}}$, defined as the sum of the $w_{\text{PCBT}}$ scores of all jets in the event for which a valid $w_\text{PCBT}$ is assigned.

The missing transverse momentum, with its magnitude denoted by \met, is defined as the negative vectorial sum of the \pT of all reconstructed and calibrated objects in the event. This sum also includes  momenta from ID tracks matched to the primary vertex but not associated with any other reconstructed objects (soft term)~\cite{JETM-2020-03}.

A sequential overlap-removal procedure is applied independently in each channel to ensure that the same calorimeter energy deposit or track is not assigned to multiple reconstructed objects. The procedure uses the lepton selection described above for the \OneL\ channel, whereas for the \SSML\ channel the isolation requirements are removed from the lepton selection. For the \bbyy\ channel, the leptons considered in the removal procedure are selected using the \textit{medium} identification requirement together with \textit{loose} isolation criteria. The impact of the overlap-removal procedure on the search is negligible.

Channel-specific event selections are then applied to maximise the sensitivity of each channel.

\subsection{\OneL\ selections}
\label{sec:1Lselection}
In the \OneL\ channel, candidate events are required to contain exactly one lepton (electron or muon) with transverse momentum $\pT > 27\,\GeV$, satisfying \textit{tight} identification and isolation requirements. The selected lepton must be matched to a corresponding trigger-level object. \(\met\) is required to exceed \(20~\GeV\), consistent with the presence of a neutrino from the leptonic top-quark decay.

Events are required to have at least six jets and  at least four $b$-tagged jets identified with the GN2 algorithm using the 85\% WP, $N_{b}^{85\%}\geq 4$. The nominal misidentification rates for this WP are approximately 20\% for $c$-jets and 1\% for light-flavour jets.

To mitigate the parton-shower limitations in the modelling of high jet multiplicities and low jet transverse momenta, an upper limit of eleven jets is imposed. Since signal \ttHH\ events are expected to exhibit substantial hadronic activity, the scalar sum of jet transverse momenta, \HTjets, is required to be greater than 400~\GeV.

\subsection{\SSML\ selections}
\label{sec:MLselection}
The event selection for the \SSML\ channel follows the strategy developed for studies of the \tttt\ production process, using the procedures detailed in Refs.~\cite{EXOT-2019-26,TOPQ-2021-08}. Candidate events are required to contain at least two leptons with $\pT>15\,\GeV$, with the leading lepton required to satisfy $\pT>27\,\GeV$ and to be matched to a corresponding trigger-level object. Events are categorised into two sub-channels: same-sign dilepton (\SSLL) and trilepton (\ThreeL). The \SSLL\ sub-channel includes events with exactly two leptons of the same electric charge, while the \ThreeL\ sub-channel includes events with three or more leptons.

To reduce background from charge mis-identification in $Z\to e^+e^-$ processes, \SSLL\ events with same-sign electrons must satisfy $|m_{ee}-m_{Z}|>10\,\GeV$. In the \ThreeL\ sub-channel, all opposite-sign same-flavour lepton pairs are required to meet  $|m_{\ell\ell}-m_{Z}|>10\,\GeV$ to suppress contamination from $Z$ boson decays.

\subsection{\texorpdfstring{\boldmath{$b\bar{b}\gamma\gamma$}}{bbγγ} selections}
\label{sec:bbyyselection}

The \bbyy\ channel targets the $HH\!\to\!b\bar{b}\gamma\gamma$ decays of  \ttHH\ events. The selection of the two photons follows the long-established ATLAS criteria~\cite{HIGG-2012-27,HIGP-2025-10}. Events must contain exactly two photons with an invariant mass in the range of $105 < m_{\gamma\gamma}< 160$~\GeV. The leading (subleading) photon is required to satisfy the condition on the relative transverse momentum of $p_\text{T}/m_{\gamma\gamma}>35\%(25\%)$, ensuring a uniform acceptance across the $m_{\gamma\gamma}$ spectrum. This selection guarantees that both photons carry sufficient transverse momentum to be compatible with a Higgs boson decay, while suppressing backgrounds containing photons originating from QCD processes. In addition, events must contain at least two $b$-tagged jets selected with the 90\% working point, $N^{90\%}_{b}\geq 2$. %
%
%
%


%

\section{Signal regions, control regions, and background estimation}
\label{sec:backgrounds}

Background contributions across the three channels are estimated by using one of three approaches: fully simulation-based, hybrid simulation-data, or fully data-driven methods. Processes that are well modelled or negligible are taken directly from MC simulation, while backgrounds that are difficult to simulate reliably use MC to describe their kinematic properties and data control regions (CR) to determine their normalisation. In the \bbyy\ search, where a narrow $H\to\gamma\gamma$ resonance is targeted, the continuum background is extracted from the diphoton mass sidebands in data defined as $105<m_{\gamma\gamma}<120\ \text{\GeV} \cup 130<m_{\gamma\gamma}<160\ \text{\GeV}$. The following sections outline the specific strategies adopted for each channel.

\subsection{The \OneL\ channel}

In the \OneL\ channel, two signal-enriched regions (SR), $6b$ and $5b_{\text{Hi}}$, and two background-enriched CRs with negligible signal contamination, $5b_{\text{Lo}}$ and $4b_{\text{Hi}}$, are defined based on the number of $b$-tagged jets satisfying different WP criteria. These regions are summarised in Table \ref{tab:1L-Regions}.
The SRs capture $\sim1.1\%$ of all the produced  inclusive \ttHH\ signal events.

\begin{table}[!htb]
\centering
\caption{Definition of regions in the \OneL\ channel, along with the variables used in the fit. The quantities $N_b^{85\%}$  and $N_b^{77\%}$ represent the number of $b$-tagged jets identified using the 85\% and 77\% efficiency working points, respectively. }
\begin{tabular}{llccl}
\toprule
Region              & Type    & $N_b^{85\%}$  & $N_b^{77\%}$  & Fit variable  \\[1mm]
\hline
$4b_{\text{Hi}}$    & CR      & 4            & 4         & $\sum_\text{jet} w_{\text{PCBT}}$    \\[1mm]
$5b_{\text{Lo}}$  & CR      & 5            & $\leq 4$     & $\sum_\text{jet} w_{\text{PCBT}}$   \\[1mm]
$5b_{\text{Hi}}$  & SR      & 5            & 5           & Transformer output        \\
$6b$              & SR      & $\geq 6$      & --          & Transformer output       \\
\bottomrule
\end{tabular}
\label{tab:1L-Regions}
\end{table}

The dominant background in the \OneL\ channel arises from \ttbar\ events with additional heavy-flavour~(HF) jets, originating from $b$- or $c$-quarks produced primarily via gluon emissions from higher-order QCD processes. Previous ATLAS and CMS analyses~\cite{TOPQ-2017-12,TOPQ-2019-03,TOPQ-2021-26,CMS-TOP-22-009} have shown that MC simulations systematically underestimate the production rate of these events. To correct for this discrepancy, the normalisations of the \ttjets\ background components are adjusted using CRs, with corrections derived according to the flavour composition of the additional jets. Their kinematic distributions are modelled using MC simulation.
The \ttjets\ events are categorised according to the generator-level flavour of the extra jets, identified via a hadron--jet matching algorithm described in Ref.~\cite{HIGG-2020-24}.
Five categories are defined: \ttcl(light), \ttOneb, \ttTwob, \ttThreeb, and \ttFourb.

The normalisation factors (NF) for each \ttjets\ category are determined simultaneously with the signal strength (see Section~\ref{sec:results}) using a profile likelihood fit that incorporates both the signal and CRs defined in Table~\ref{tab:1L-Regions}, along with the corresponding fitting variables in each region.
Pre-fit normalisations are taken directly from the generator cross-section predictions. Each category is treated as an independent process and assigned its own modelling uncertainties. The post-fit normalisations of the dominant components, \ttOneb\ and \ttTwob, are extracted through unconstrained parameters in the fit, allowing them to adjust freely to the observed data. Due to the limited constraining power of the data, the \ttcl, \ttThreeb, and \ttFourb\ components are normalised to their MC predictions, with associated systematic uncertainties discussed in Section~\ref{sec:systematics}. Other minor background contributions ($V/VV/VVV$, \ttV, \ttH, multi-top and $V(V,VV)$+jets) are also estimated by using MC simulation.

\subsection{The \SSML\ channel}
\label{sec:SSML-backgrounds}
In the \SSML\ channel, the SRs are defined as follows. Events in the \SSLL\ category are required to have at least six jets, while those in the \ThreeL\ category must contain at least five jets. In both cases, at least two jets must be $b$-tagged using the 85\% WP, i.e. $N_b^{85\%} \geq 2$. Additionally, the scalar sum of the transverse momentum of jets and leptons, \HTall, must exceed 400\,\GeV. In this channel, the SRs capture approximately $0.75\%$ of all produced \ttHH\ signal events.

Background contributions in the \SSML\ channel are classified into two main types: physics backgrounds and instrumental backgrounds.
Physics backgrounds originate from processes that produce final states similar to those targeted by the \SSML\ selection, including \VV, \VVV, \ttV, \ttH, \ttVV, and multi-top. In these processes, leptons arise from the prompt decays of heavy bosons ($W$, $Z$, and $H$). Among these background sources, \ttV, \ttH, and multi-top constitute the dominant contributions limiting the search sensitivity.

All physics backgrounds, except for \ttW, are estimated by using MC simulations.
For \ttW, its kinematic properties are modelled using MC, while a dedicated CR is employed to constrain its normalisation using data. This is achieved through a template fit to the jet multiplicity distribution ($N_\text{j}$) in the \ttW\ CR\@. This approach accounts for a potentially enhanced \ttW\ production cross-section relative to the SM prediction, motivated by recent measurements~\cite{TOPQ-2019-30,CMS-TOP-21-011}. The selection criteria for this CR are given in Table~\ref{tab:SSML-Regions}. The requirements for this and other CRs described below are designed to enhance contributions from the targeted background sources while remaining orthogonal to the SR.

Instrumental backgrounds originate from leptons with misidentified charge (QmisID) and non-prompt leptons from photon conversions, heavy-flavour quark decays, and jets mis-identified as leptons (Fake/non-prompt). The main process contributing to these backgrounds is \ttbar\ production.
These backgrounds are challenging to simulate accurately and are therefore evaluated using a combination of MC and data-driven techniques, following the methodology described in Ref.~\cite{TOPQ-2021-08}.

QmisID primarily affects the \SSLL\ $ee$ and $e\mu$ channels, as electron identification is particularly susceptible to charge misidentification due to bremsstrahlung radiation followed by photon conversion. In Run~2 data, a boosted decision tree (BDT) based electron charge identification (ECID) tool is employed to significantly reduce the QmisID contribution~\cite{EGAM-2018-01}. The ECID tool combines tracking information with the magnitude and spatial pattern of bremsstrahlung radiation to reject electrons whose charges are likely mismeasured. However, this tool is not yet available for the Run~3 data sample, resulting an  increased QmisID background.

The charge misidentification rate is derived from $Z\to e^{+}e^{-}$ events in data, selected with $|m_{ee}-m_Z|<10\,\GeV$ and no charge requirement. The rate is determined from the fraction of same-sign electron pairs and parameterised as a function of the electron \pT\ and $\eta$. It is then applied to the $ee$ and $e\mu$  data events passing the \SSLL\ selection, excluding the charge requirement, to estimate the QmisID background contribution. In Run~2, the rate ranges from 0.004\% for electrons with $\pT<60\,\GeV$ in the central region to 4\% for those with $\pT>130\,\GeV$ in the forward region. In Run~3, the rate increases by a factor of 2--5 depending on the electron \pT\ and $|\eta|$.

\begin{table}[!htb]
\caption{Definitions of signal and control regions in the \SSML\ channel. In the \ThreeL\ sub-channel, leptons are labelled as $\ell_1$, $\ell_2$, and $\ell_3$ in decreasing order of \pT.  $N_\text{j}$, and $N_b^{85\%}$ denote the numbers of jets and $b$-tagged jets identified using the 85\% working point. }
\small
\begin{tabular}{cccclc}\toprule
Region    &    Channel       &   $N_\text{j}$  &  $N_b^{85\%}$   & Other selection   &  Fit variable \\ \\[-4mm]\midrule
\multirow{2}{*}{SR}             & \SSLL            & $\ge 6$   & \multirow{2}{*}{$\ge 2$} & \multirow{2}{*}{\HTall $>400\,\GeV$} & Transformer \\[0.5mm]
& \ThreeL          & $\ge 5$   &         & & output   \\ \midrule\\[-3mm]

\ttW             & \SSLL\ \em,\, \mm   & $\ge 4$   & $\ge 2$ & $p_\text{T}^{\ell_2}>25\,\GeV$, $|\eta(e)|<1.5$ & $N_\text{j}$ \\[0.5mm]
&                &           &         & If $N_b^{85\%}=2$: \HTall $<400\,\GeV$ or $N_\text{j}<6$ &    \\[0.5mm]
&                &           &         & If $N_b^{85\%}\ge 3$: \HTall $<400\,\GeV$   &   \\[3mm]

Int. Conv. & \SSLL\ $ee,\, \em$  & $4\le N_\text{j}<6$  & $\ge 1$  & Only one $e$ from $\gamma^*\to ee$  &   $N_\text{events}$ \\
&                  &               &          & No $e$ from $\gamma\to ee$           &               \\[1mm]
Mat. Conv.         & \SSLL\ $ee,\, \em$  & $4\le N_\text{j}<6$  & $\ge 1$  & Only one $e$ from $\gamma\to ee$   &   $N_\text{events}$ \\
&                    &             &          & No $e$ from $\gamma^*\to ee$          &  \\[3mm]

HF $\mu$          & \ThreeL\ \emm,\, \mmm   & $\ge 1$   & $=1$  & $\etmiss>35\,\GeV$   & $p_\text{T}^{\ell_3}$ \\
&                &           &       & $100<$ \HTall $<275\,\GeV$  &             \\
&                &           &       & Total lepton charge $=\pm 1$ &             \\[1mm]
HF $e$            & \ThreeL\ \eee,\, \eem  & $\ge 1$   & $=1$  & $\etmiss>35\,\GeV$  &  $p_\text{T}^{\ell_3}$ \\
&                &           &       & $100<$ \HTall $<275\,\GeV$   &              \\
&                &           &       & Total lepton charge $=\pm 1$ &              \\

\bottomrule
\end{tabular}
\label{tab:SSML-Regions}
\end{table}

Backgrounds from photon conversions, $\gamma^{(*)}\to e^+e^-$, occurring either at the primary vertex or within the detector material, can lead to only one of the electrons being reconstructed. A conversion electron is identified through its small invariant mass with a nearby reconstructed track. It is classified as an internal conversion (Int. Conv.), $\gamma^*\to e^+e^-$, if the associated track points to the primary vertex,  or as a material conversion (Mat. Conv.), $\gamma\to e^+e^-$, if the track instead points to a conversion vertex in the detector material.
Leptons from heavy-flavour ($b$- and $c$-quark) decays  generally fail to satisfy the identification criteria designed for prompt leptons from vector boson decays. However, a small fraction satisfy the selection and contribute to the non-prompt lepton background.

Similar to the \ttW\ background, CRs are used to improve the estimates of these fake/non-prompt lepton background contributions. MC predictions are scaled by NFs derived from a template fit to data in all the CRs, defined in Table~\ref{tab:SSML-Regions}. Each CR targets a specific background source: `Int. Conv.' for virtual photon conversions at the primary  vertex, `Mat. Conv.' for real photon conversions in the detector material, `HF $e$' for electrons from heavy flavour decays, and `HF $\mu$' for muons from heavy flavour decays. The selection criteria leverage the characteristic features of these background processes to enhance their contributions while ensuring orthogonality among all CRs and the SR\@. The distributions of event variables used in the fits are listed in the table.

For the Run~3 data, photon conversion CRs (Int. Conv. and Mat. Conv.) are not used due to the larger QmisID background contribution stemming from the absence of the ECID tool. Instead, the NFs obtained from the Run~2 data are used to estimate the conversion background contributions in Run~3 data.

Fake/non-prompt lepton background contributions from light hadron decays or mis-identified jets are expected to be minimal and are estimated by using simulation without additional corrections.

\subsection{The \texorpdfstring{\boldmath{$b\bar{b}\gamma\gamma$}}{bbγγ} channel}

Three SRs are defined for the \bbyy\ channel using BDTs, as described in Section~\ref{sec:mva}.
The background model in this channel comprises a dominant non-resonant (continuum) component, along with several resonant contributions from Higgs boson production in the diphoton invariant-mass spectrum.

The continuum background primarily originates from $\gamma\gamma+$jets events, with smaller contributions from $\gamma+$jet or dijet events where one or two jets are misidentified as photons. Due to the tight event selection, this background is dominated by processes with associated heavy-flavour production. Therefore, two dedicated samples, $\gamma\gamma b\bar{b}$ and \ttyy\, are used to model the continuum background composition and kinematics. MC simulation is used solely for optimising the event selection and for training multivariate classifiers.

The continuum background is modelled with an exponential function chosen using the  spurious-signal procedure~\cite{ATL-PHYS-PUB-2020-028}, which evaluates a broad range of candidate parameterisations. The exponential form is retained because it satisfies statistical criteria ensuring that it does not artificially absorb or generate a signal. The exponential shape parameter is determined from data via an unbinned maximum-likelihood fit to the diphoton invariant mass distribution, $m_{\gamma\gamma}$, performed in the sidebands of all the \bbyy\ SRs described in Section~\ref{sec:mva}. The fit uses independent normalisations for each region while enforcing a common exponential shape correlated across them. Allowing the shape parameter to vary independently in each SR changes the overall sensitivity by less than 1\%, and the resulting parameter values are consistent across regions within statistical uncertainties.

Resonant backgrounds from single Higgs boson production and non-\ttHH\ Higgs boson pair production processes are modelled using simulated samples and are normalised to their corresponding theoretical cross-sections. Their shapes are described by double-sided Crystal Ball (DSCB) functions~\cite{HIGG-2014-04} derived from simulation.

The complete background model is thus constructed as the sum of the analytic continuum and resonant MC templates. A combined likelihood fit simultaneously determines the normalisations and the correlated continuum shape across the signal and sideband regions, while the resonant contributions remain constrained to their theoretical predictions.


\section{Multivariate discriminants}
\label{sec:mva}

Multivariate analysis techniques are employed to exploit the kinematic differences between signal and background events that satisfy the SR selection in the \OneL\ and \SSML\ channels. A transformer architecture is used, incorporating the scaled dot-product attention mechanism described in Ref.~\cite{Vaswani:2017lxt}.

MC samples are used to train the transformer models, which are designed to distinguish among various event classes, including both signals and backgrounds. The training process aims to minimise a loss function that quantifies the fraction of misclassified events within the training data samples.

Signal and background samples used to develop the transformer models are partitioned into three statistically independent subsamples: one for training (training set), one for hyper-parameter optimisation (validation set), and one for performance evaluation (test set). This strategy facilitates the study and mitigation of over-training effects. The test set is also used to predict the expected signal and background distributions of the transformer output scores for the final statistical analyses. A $k$-fold method, with $k=3$ for \OneL\ and $k=4$ for \SSML, is performed by rotating the three sets in order to evaluate independently all data samples and maximise the number of MC events used in the final statistical analyses. The training is performed separately for the Run~2 and Run~3 data samples.

The input features to the transformer consist of kinematic variables, complemented by $b$-tagging. These include the momentum vectors $(\pT, \eta,\phi)$ of leptons, jets and \met, along with the jet $w_{\text{PCBT}}$  and the charge information for leptons.

For the \OneL\ channel, a binary transformer classifier is trained to distinguish  the \ttHH\ signal from all background processes. In contrast, the \SSML\ channel employs a three-class training scheme to separately identify the \ttHH\ signal, the multi-top background, and all other backgrounds. The multi-top contribution  is treated separately due to its comparatively large cross-section uncertainty.
The transformer provides class-likelihood scores that are between zero and one for each event: $p_{\ttHH}$ and $p_\textrm{bkgd}$ in \OneL\ and $p_{\ttHH}$, $p_\text{multi-top}$, and $p_\textrm{others}$ in \SSML. These scores are used to construct discriminants, $D_{\OneL}$ for \OneL\ and $D_{\SSML}$ for \SSML, for the final signal-background separation:
\begin{equation*}
D_{\OneL}=p_{\ttHH},\ \ \ \ \ \ D_{\SSML}=\ln\left(\frac{p_{\ttHH}}{0.2\cdot p_\textrm{multi-top}+0.8\cdot p_\textrm{others}}\right).
\end{equation*}

The weights assigned to the multi-top and other background components in the \SSML\ discriminant are optimised to maximise sensitivity to the \ttHH\ signal. These weights closely reflect the relative contributions of the corresponding background processes in the highest $D_{\SSML}$ regime.

The \bbyy\ channel employs a set of BDTs to define SRs of varying purities, rather than serving as the final discriminant. Signal-to-background discrimination is achieved through a multi-BDT strategy, implemented using the \textsc{XGBoost} algorithm~\cite{Chen:2016btl} and the \textsc{Optuna} framework~\cite{Akiba:2019lwq} for hyperparameter optimisation.

Four independent BDTs, shown in Figure~\ref{fig:bbyy-bdts}, are trained on the combined Run~2 and Run~3 data samples to distinguish the \ttHH\ signal from the four major backgrounds: $\gamma\gamma b\bar{b}$, \ttyy, \ttH, and other Higgs processes. Key input features include the reconstructed di-Higgs boson invariant mass and the kinematics of photons and $b$-tagged jets. Training is validated with independent test samples, and overtraining is assessed by comparing training and test BDT-score distributions. Three mutually exclusive SRs (SR1, SR2, and SR3), ordered from the lowest to highest sensitivity, are defined for events with $N_b^{90\%}\ge 3$ using three thresholds ($c_i^1<c_i^2<c_i^3$) on each of the four BDTs ($i=1,2,3,4$). SR3 contains events satisfying all highest thresholds ($c_i^3$), SR2 consists of events satisfying the intermediate thresholds ($c_i^2$) but fail to satisfy the SR3 requirement. Finally SR1 includes events satisfying the baseline thresholds ($c_i^1$), excluding those already assigned to SR2 or SR3\@. The threshold values, $c_i^k$, are optimised to maximise the expected signal significance within the diphoton mass window $120<m_{\gamma\gamma}<130\,\GeV$, while ensuring at least three sideband events in each SR\@. In the \bbyy\ channel, the SRs retain approximately 0.05\% of all \ttHH\ signal.

\begin{figure}[!htb]
\centering
\subfloat[]{\includegraphics[width=0.43\textwidth]{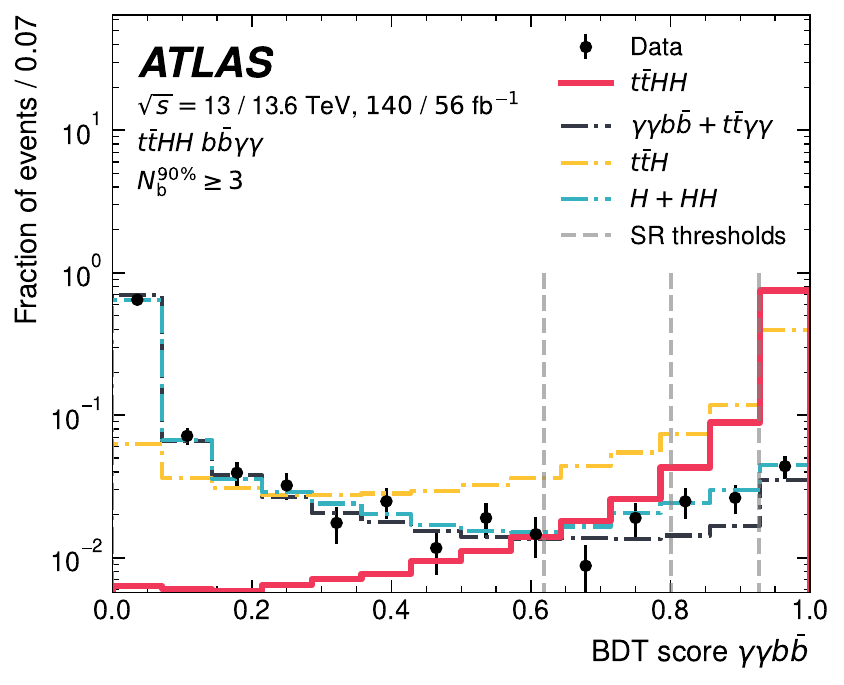}}
\subfloat[]{\includegraphics[width=0.43\textwidth]{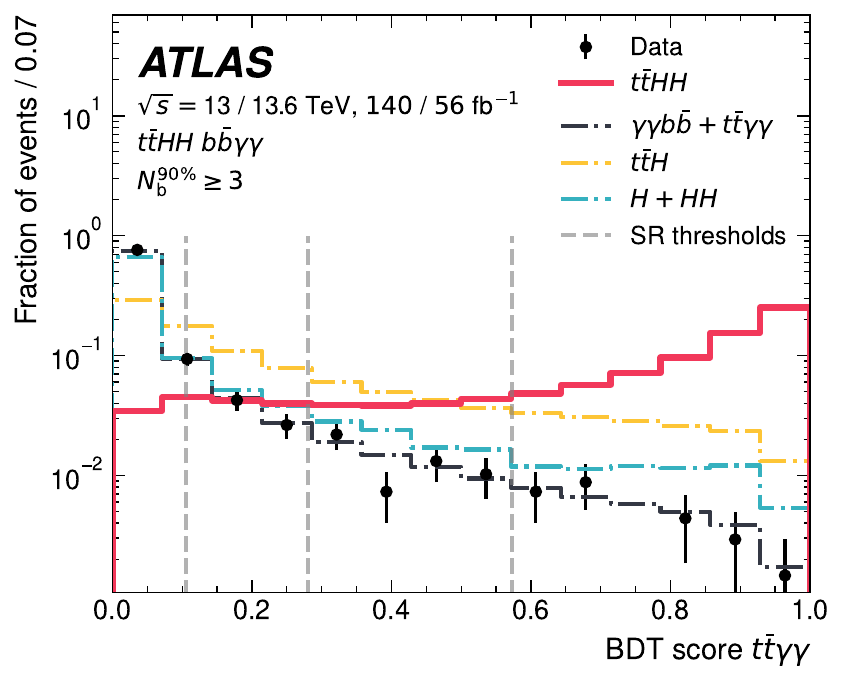}}\\
\subfloat[]{\includegraphics[width=0.43\textwidth]{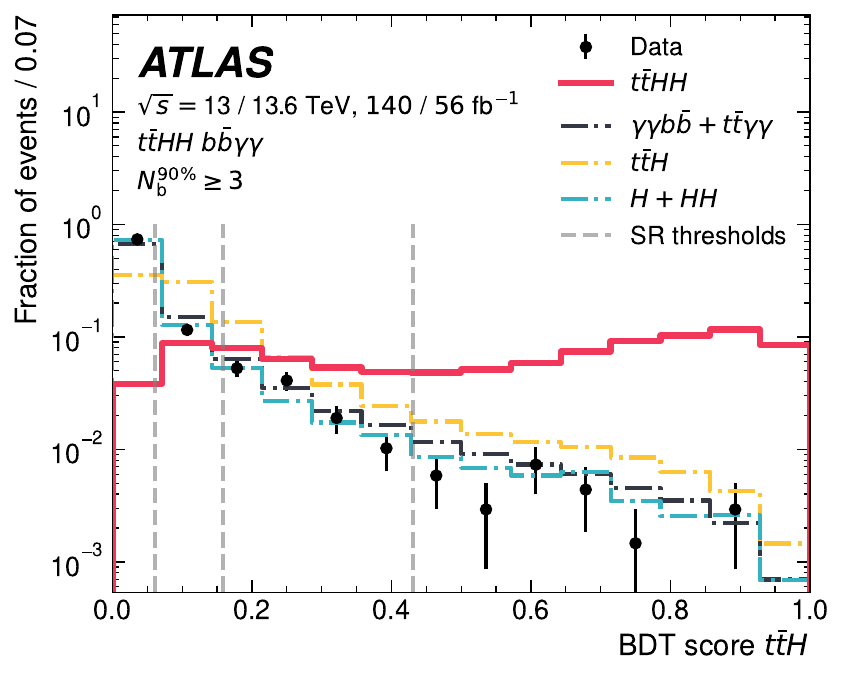}}
\subfloat[]{\includegraphics[width=0.43\textwidth]{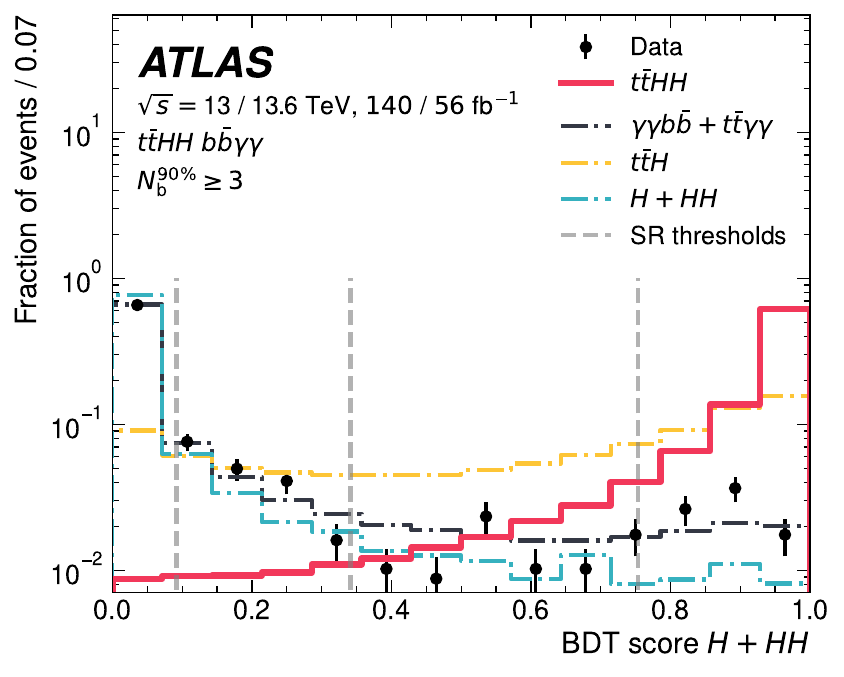}}
\caption{BDT score distributions obtained from the combined Run~2 and Run~3 data samples, used to discriminate the \ttHH\ signal from the main background processes: (a) $\gamma\gamma b\bar{b}$, (b) \ttyy, (c) \ttH, and (d) single Higgs boson and di-Higgs boson production. All histograms are normalised to unity. Vertical dashed lines indicate the BDT score thresholds used to define the signal regions, whose boundaries are determined simultaneously across the four BDT scores. %
}
\label{fig:bbyy-bdts}
\end{figure}


\section{Systematic uncertainties}
\label{sec:systematics}

Systematic uncertainties in the \ttHH\ search are categorised into three main groups: experimental uncertainties, which affect reconstructed objects such as leptons, photons, and jets; theoretical uncertainties, related to the modelling of signal and background processes; and data-driven uncertainties, associated with background estimates derived from data. These uncertainties impact both the shape of kinematic distributions and the normalisation of signal and background contributions.
While the specific objects and processes may vary across channels, the methodologies for evaluating experimental and theoretical uncertainties are similar. In contrast, the estimation of uncertainties for data-driven background contributions is channel-specific.
In the \OneL\ and \SSML\ channels, systematic variations are propagated through the transformer-output templates used in the binned fit. For the \bbyy\ channel, the uncertainties are incorporated via the $m_{\gamma\gamma}$ model (signal DSCB parameters and background-shape parameters) entering the unbinned fit in each category.

All three channels remain statistically limited, especially the \bbyy\ channel. The dominant systematic uncertainties stem from the modelling of \ttbar\ production with at least three $b$-jets in the \OneL\ channel, the theoretical uncertainty on the four-top production cross section in the \SSML\ channel, and the modelling of \ttH\ production with extra heavy-flavour jets in the \bbyy\ channel.

The subsections below first describe systematic uncertainties common to all channels, followed by those specific to each channel  and their corresponding impacts.

\subsection{Experimental uncertainties}
Jet-related uncertainties  arise from several sources, including the jet energy scale (JES), jet energy resolution (JER), flavour tagging, and JVT used for pile-up rejection. JES and JER uncertainties are evaluated using control samples where jets recoil against well-calibrated reference objects -- such as photons, $Z$ bosons, or other calibrated jets~\cite{JETM-2018-05}. Flavour tagging uncertainties originate from uncertainties in tagging efficiencies and misidentification rates. These are assessed using \ttbar\ events for $b$- and $c$-jets, and $Z$+jets events for light-quark and gluon jets~\cite{FTAG-2019-02,FTAG-2019-07,FTAG-2020-08}. Additional uncertainties are considered for the JVT efficiency~\cite{PERF-2014-03}.
Furthermore, a separate uncertainty is applied in simulation to account for differences in detector response between $b$-jets, light-quark jets and gluon jets.

Uncertainties in the reconstruction, identification, isolation and trigger efficiencies of electrons, photons, and muons are also considered, along with uncertainties in their energy scale and resolution~\cite{PERF-2015-10,EGAM-2018-01}. The latter uncertainties for jets and leptons are propagated to the calculation of \met, which also incorporates contributions from the modelling of the underlying event and the reconstruction efficiency of tracks used to compute the soft term~\cite{PERF-2016-07}.

The uncertainty in the integrated luminosity is 0.83\%~\cite{DAPR-2021-01} for the Run~2 data sample, and 2.0\%~\cite{ATL-DAPR-PUB-2023-001,ATL-DAPR-PUB-2024-001} for the Run~3 data sample. These values are derived using the LUCID-2 detector~\cite{LUCID2}, supplemented by measurements from the inner detector and calorimeters. An additional uncertainty is considered for corrections applied to the pile-up profile in simulated samples.

\subsection{Theoretical uncertainties}
Uncertainties in the modelling of signal and background processes arise from several sources:  the PDFs, $\alphas$, missing higher-order QCD  corrections in the partonic MEs, and initial- and final-state radiation (ISR and FSR). PDF uncertainties are evaluated following the PDF4LHC recommendations\,\cite{Butterworth:2015oua}, while $\alphas$ uncertainties are estimated by varying its value within the current measurement range. Uncertainties due to missing high-order corrections are assessed by varying the renormalisation and factorisation scales relative to their central values. Generator-related uncertainties are determined by comparing nominal predictions with those obtained using alternative generators or parton shower models, as specified in Table~\ref{tab:MC-samples}. ISR and FSR uncertainties are evaluated by varying the renormalisation scales in the parton shower, as well as the renormalisation and factorisation scales in the MEs.

\subsection{\OneL\ specific uncertainties}

Modelling uncertainties for $\ttbar$+jets are evaluated by comparing the nominal \POWPY[8] (5FS) sample with four alternative samples: \POWPY[8] configurations with \texttt{pThard=1} and with $h_\text{damp}=3m_t$, a \POWHER[7] sample, and a \powhegboxRes+\herwigseven\ (4FS) sample. Each alternative generator and each $\ttbar$+jets component ($t\bar{t}+c/l$, $t\bar{t}+1b$, $t\bar{t}+2b$, $t\bar{t}+3b$, and $t\bar{t}+\geq4b$) is treated independently.
The normalisations of the $t\bar{t}+1b$ and $t\bar{t}+2b$ components are constrained in data through two independent free NFs. For these two, the alternative samples are normalised to the nominal phase space, so that the uncertainties account for shape and acceptance effects, separated from overall normalisation variations.
For the $t\bar{t}+3b$, $t\bar{t}+\geq4b$, and the subdominant $t\bar{t}+c/l$ components, additional normalisation uncertainties of 40\%, 50\%, and 6\%, respectively, are assigned. These values are derived, for each component, from the sum in quadrature of the relative differences between the nominal prediction and all alternative samples. In this case, the alternative samples are not normalised to the nominal phase space. The resulting uncertainty includes their intrinsic normalisation differences and therefore includes normalisation, shape, and acceptance effects.

\subsection{\SSML\ specific uncertainties}

For the \SSML\ channel, a conservative uncertainty is assigned to the \tttt\ production. The ATLAS Collaboration has previously measured the \tttt\ cross-section to be about two standard deviations above the SM prediction~\cite{TOPQ-2021-08}. To account for both the 15\% theoretical uncertainty~\cite{Frederix:2017wme} and the discrepancy with the ATLAS measurement, normalisation uncertainties of $^{+70\%}_{-15\%}$ are applied, similar to the approach taken in Ref.~\cite{HIGP-2024-08}. These uncertainties are also applied to the minor \tttt\ background in the \OneL\ channel.

Uncertainties in the fake/non-prompt lepton and QmisID background contributions are evaluated following the methodology outlined in Ref.~\cite{TOPQ-2021-08}.
For the fake/non-prompt lepton background, only shape uncertainties in the distributions used in the template fit are considered, as their normalisations are free parameters in the fit. These uncertainties are estimated by comparing data with the background predictions in regions kinematically similar to the control regions.
For the heavy-flavour non-prompt background, a loose lepton selection is employed, while for the conversion background, comparisons are made using $\mu\mu\gamma^{(*)}$ events in the \ThreeL\ sub-channel.

Additional systematic uncertainties associated with the ECID are also included.
As discussed in Section~\ref{sec:SSML-backgrounds}, the photon-conversion backgrounds for Run~3 are estimated by using the NFs derived from Run~2 data. Extra systematic uncertainties are applied to account for differences in event yields between the Run~3 and the Run~2 extrapolations without the ECID tool.

\subsection{\texorpdfstring{\boldmath{$b\bar{b}\gamma\gamma$}}{bbγγ} specific uncertainties}

The dominant uncertainty in the continuum $\gamma\gamma+$jets background arises from the choice of analytic function for its modelling. This background modelling uncertainty is evaluated using the \textit{spurious signal} method that quantifies the fitted signal yield obtained when a signal-plus-background fit is applied to background-only templates~\cite{ATL-PHYS-PUB-2020-028}. In all categories, the resulting bias is treated as a systematic uncertainty in the signal yield.

For resonant processes, a conservative 100\% uncertainty is assigned to single and di-Higgs boson processes to account for modelling limitations in heavy-flavour--jet-associated production. An exception is made for \ttH\ production with additional heavy-flavour jets, for which a 50\% normalisation uncertainty is assigned, derived from the measurement in Ref.~\cite{TOPQ-2017-12,TOPQ-2019-03}.


\section{Statistical analysis and results}
\label{sec:results}

To assess the compatibility of the observed data with background expectations and to search for the presence of a \ttHH\ signal, simultaneous maximum-likelihood fits are performed across the SRs and CRs.
In the \OneL\ and \SSML\ channels, binned likelihood fits to data are performed using the transformer discriminant in the SRs and the fit variables listed in \cref{tab:1L-Regions,tab:SSML-Regions} for the CRs. For the \bbyy\ channel, an unbinned maximum-likelihood fit is performed to the diphoton invariant mass $m_{\gamma\gamma}$ in each of the three BDT-defined SRs.

The parameter-of-interest (POI) in the fit is the signal strength, $\mu_{\ttHH}$, defined as the ratio of the observed \ttHH\ signal cross-section to the SM expectation. The fit also includes normalisation parameters,  which constrain background contributions using data from CRs in the \OneL\ and \SSML\ channels, and background parameters for the \bbyy\ channel. Systematic uncertainties are incorporated via nuisance parameters (NPs).

Likelihood functions are constructed separately for each SR and CR\@. In the \OneL\ and \SSML\ channels, the likelihood is built from binned Poisson probabilities of the corresponding fitting variable. In the \bbyy\ channel, the likelihood is unbinned and defined as the product of per-event probability densities evaluated at the observed $m_{\gamma\gamma}$. Gaussian or log-normal constraints are applied to the NPs in both cases. These individual likelihoods are then aggregated across all regions to form the final likelihood function, which is subsequently maximised in the fit.

For the \OneL\ and \SSML\ channels, the final likelihood function is:
\begin{equation*}{\cal L}(\mu_{\ttHH},\vec{\kappa},\vec{\theta})=\prod_r\left[\prod_{i}^{N^r_\textrm{bins}}\textrm{Pois}(n^r_i|\nu^r_i(\mu_{\ttHH},\vec{\kappa},\vec{\theta}))\right]\cdot G(\vec{\theta}).
\end{equation*}
where $\prod_r$ denotes the product of all regions, $\textrm{Pois}(n^r_i|\nu^r_i(\mu_{\ttHH},\vec{\kappa},\vec{\theta}))$ represents the Poisson probability in bin $i$ from region $r$, with $n^r_i$ and $\nu^r_i(\mu_{\ttHH},\vec{\kappa},\vec{\theta})$ representing the observed and expected event counts, respectively. The variables $\vec{\kappa}$ and $\vec{\theta}$ correspond to NFs and NPs.
The expected count in each bin is given by
\begin{equation}
\nu^r_i(\mu_{\ttHH}, \vec{\kappa},\vec{\theta})=\mu_{\ttHH} S^r_i(\vec{\theta}) + B^r_i(\vec{\kappa},\vec{\theta}).
\label{eq:nu}
\end{equation}
where $S^r_i$ is the expected signal yield from the SM, and $B^r_i$ is the estimated background contribution, in bin $i$ from region $r$. The function $G(\vec{\theta})$ encodes the prior constraints on the NPs.

For \bbyy, the final likelihood function has the form:
\begin{equation*}{\cal L} (\mu_{\ttHH},\vec{\kappa},\vec{\theta}) =\prod_r\left[\textrm{Pois}(n_r|\nu_r(\mu_{\ttHH},\vec{\kappa},\vec{\theta}))\times\prod_{i}^{n_r}f_r(m^i_{\gamma\gamma},\vec{\kappa},\vec{\theta})\right]\cdot G(\vec{\theta}).
\end{equation*}
Here $n_r$ and $\nu_r$ are the observed and expected event counts in region $r$, respectively. The function $f_r(m^i_{\gamma\gamma},\vec{\kappa},\vec{\theta})$ represents the probability density for the diphoton mass of event $i$ in the region $r$, summing over both signal and background contributions, analogous to Eq.~\ref{eq:nu}:
\begin{equation}
f_r(m^i_{\gamma\gamma},\vec{\kappa},\vec{\theta})=\mu_{\ttHH} S_r(m^i_{\gamma\gamma})+B_r(m^i_{\gamma\gamma},\vec{\kappa},\vec{\theta})
\label{eq:fr}
\end{equation}
where $S_r$ and $B_r$ denote the probability densities for the SM signal and background contributions in region $r$, respectively.
The vector $\vec{\kappa}$ parameterises the background model and $G(\vec{\theta})$ represents the NP constraints as described above.

For the \OneL\ and \SSML\ channels, the Run~2 and Run~3 data samples are first fitted independently and then combined. In contrast, the \bbyy\ channel employs a single fit to the two data samples, motivated by the limited numbers of events and the similarity of their kinematic distributions.  Experimental systematic uncertainties are assumed to be correlated across channels within the same run period, but uncorrelated between run periods except for most of the jet-related uncertainties, which are treated as correlated. Theoretical uncertainties in the process cross-sections are considered correlated across both run periods and all channels, while other modelling uncertainties are uncorrelated across run periods to account for differences in the MC generator configurations used in the two runs. The NFs for the \ttOneb, \ttTwob, and \ttW\ processes, introduced to account for differences between data and theoretical predictions, are treated as correlated between the run periods, while all other NFs, related to instrumental backgrounds in the \SSML\ channel, are kept uncorrelated.

\begin{figure}[!htbp]
\centering
\subfloat[]{\includegraphics[width=0.43\textwidth]{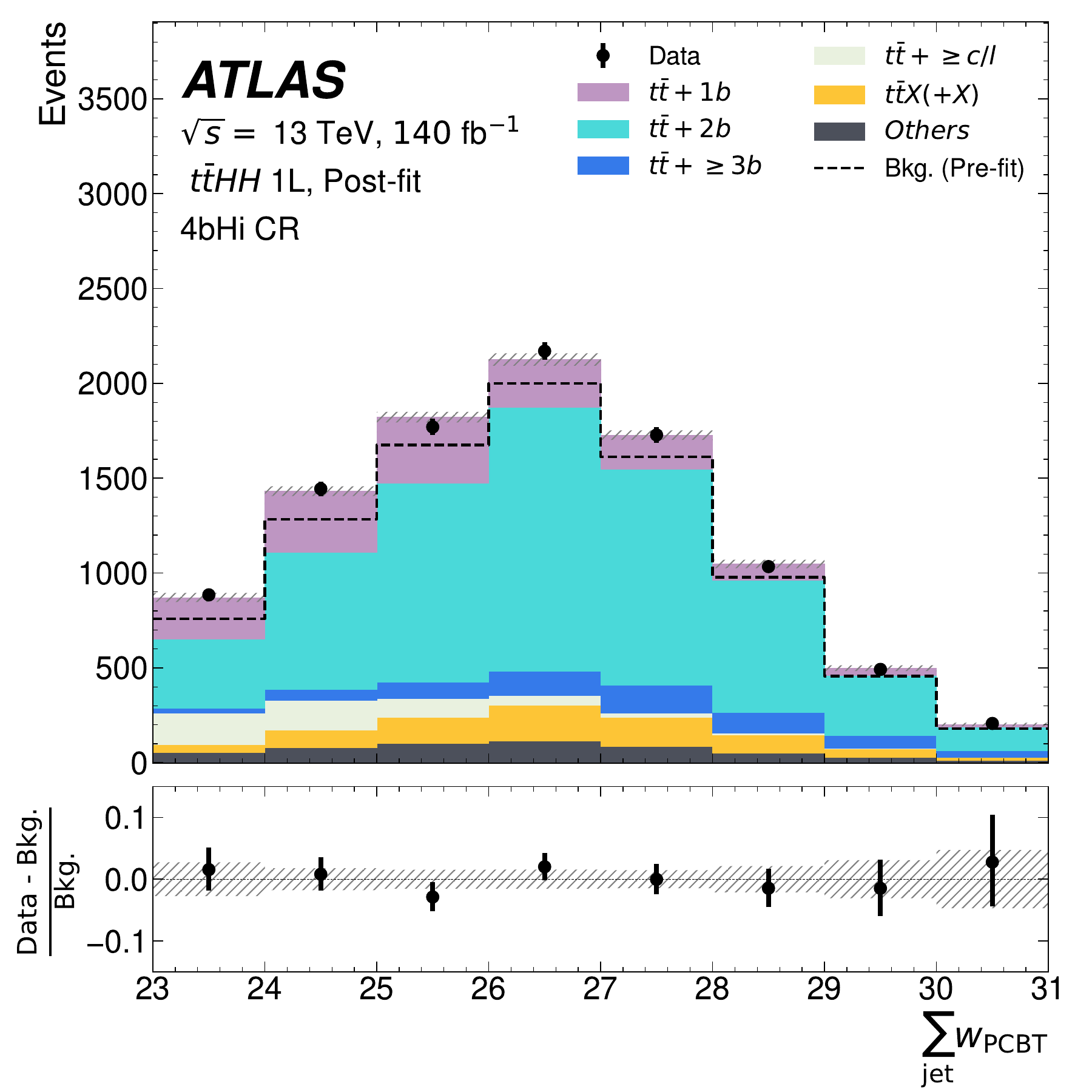}}
\subfloat[]{\includegraphics[width=0.43\textwidth]{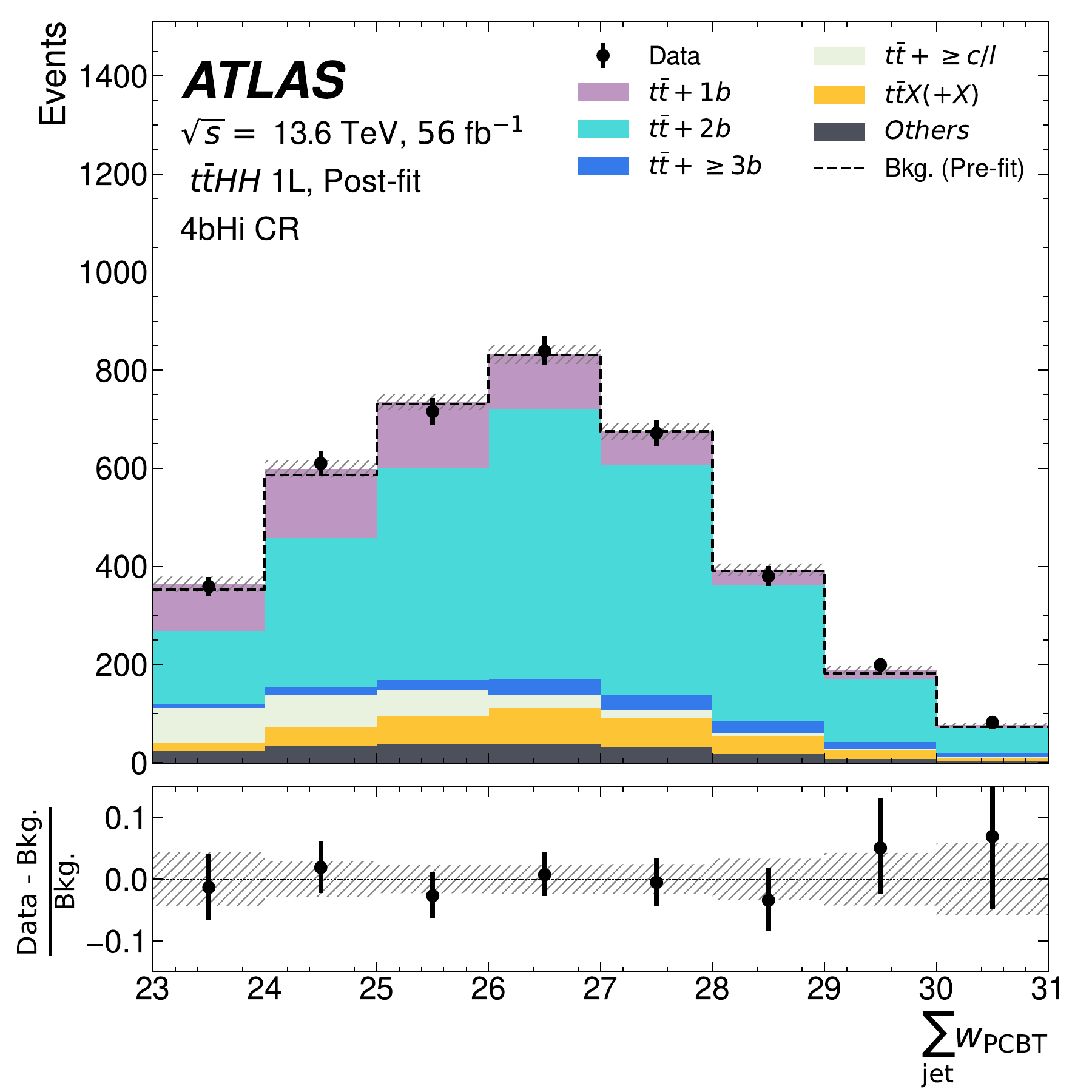}}\\
\subfloat[]{\includegraphics[width=0.43\textwidth]{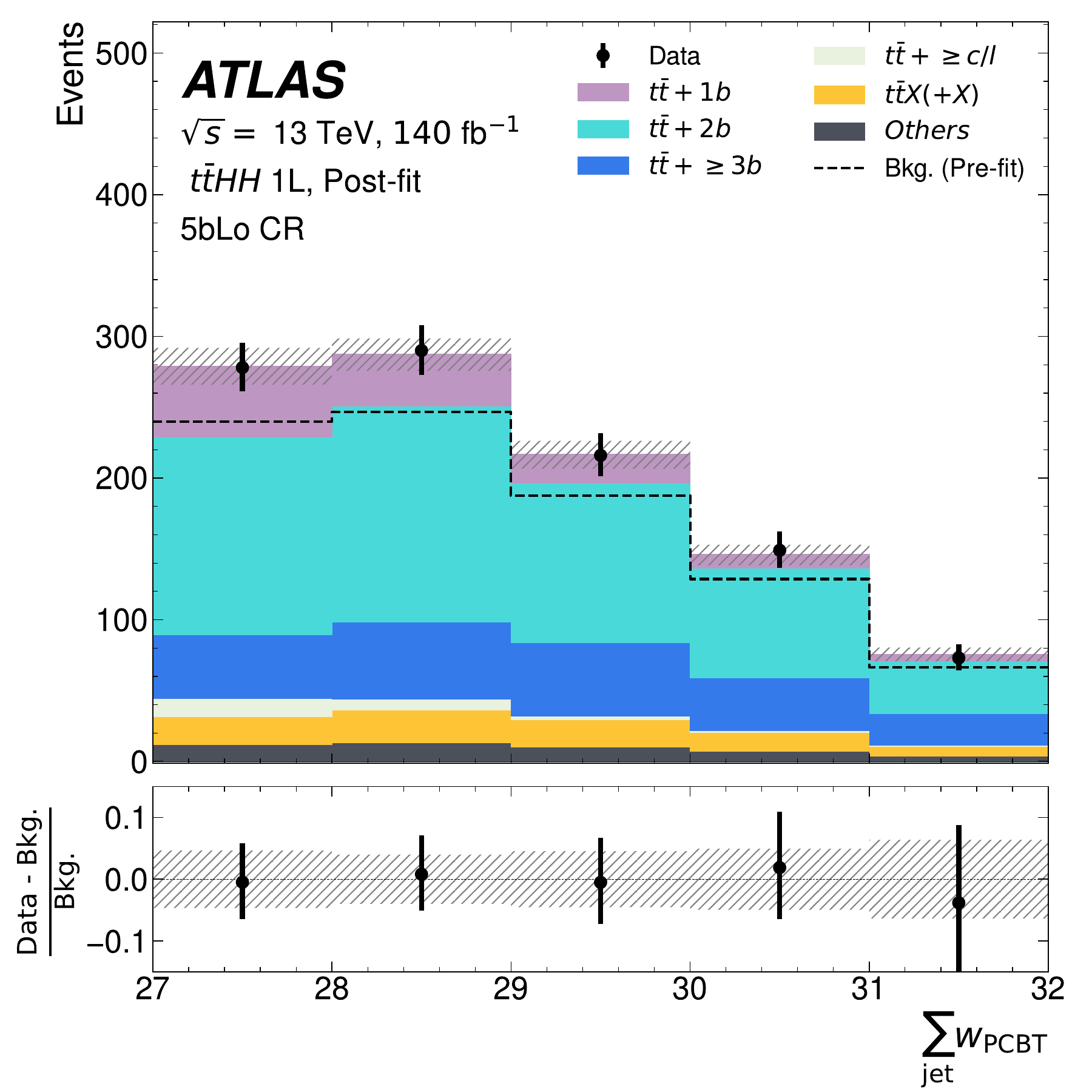}}
\subfloat[]{\includegraphics[width=0.43\textwidth]{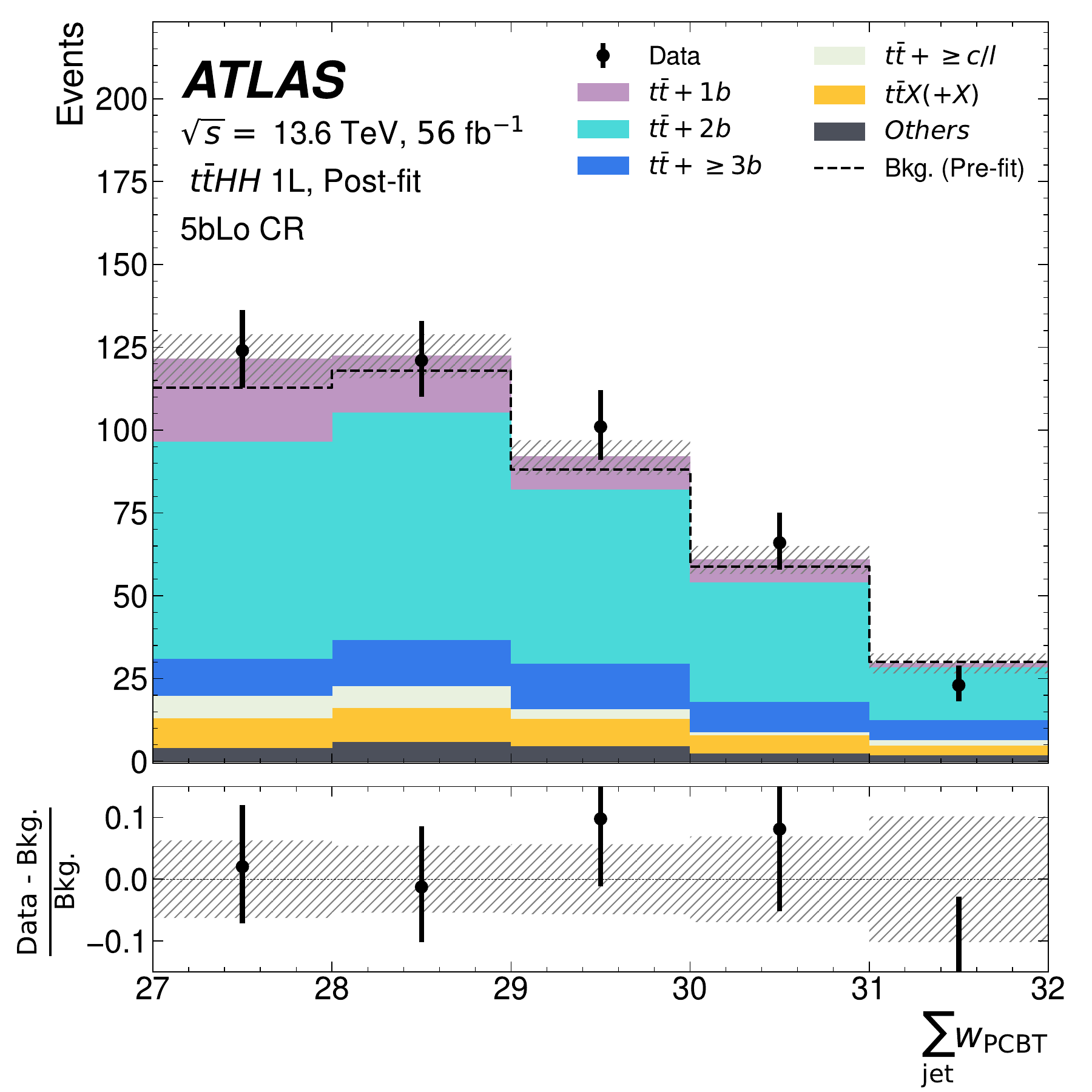}}
\caption{Distributions of the sum of jet pseudo-continuous $b$-tagging scores in the $4b_\text{Hi}$ and $5b_\text{Lo}$ control regions of the \OneL\ channel for the (a, c) Run~2 and (b, d) Run~3 data samples, comparing the data with the post-fit stacked background predictions.
The lower panels display the relative differences between the data and the background predictions.
The hatched bands indicate the combined statistical and systematic uncertainties in the background predictions.
For reference, the pre-fit background predictions are overlaid as dashed histograms.
}
\label{fig:1L-backgrounds}
\end{figure}

The fits are performed under the signal-plus-background hypothesis.
Good agreement is observed between the data and background expectations across all fit variables, for all channels and data-taking periods. Figures~\ref{fig:1L-backgrounds} and~\ref{fig:SSML-backgrounds} show the post-fit background composition in the \OneL\ and \SSML\ CRs, respectively, for the combined Run~2 and Run~3 data samples. In the \OneL\ channel, the post-fit NFs for the \ttOneb\ and \ttTwob\ processes are $1.2 \pm 0.4$ and $0.91 \pm 0.11$, respectively, while in the \SSML\ channel the \ttW\ NF is $1.04 \pm 0.20$. These NFs are consistent within uncertainties between Run~2 and Run~3.

\begin{figure}[!htbp]
\centering
\subfloat[]{\includegraphics[width=0.45\textwidth]{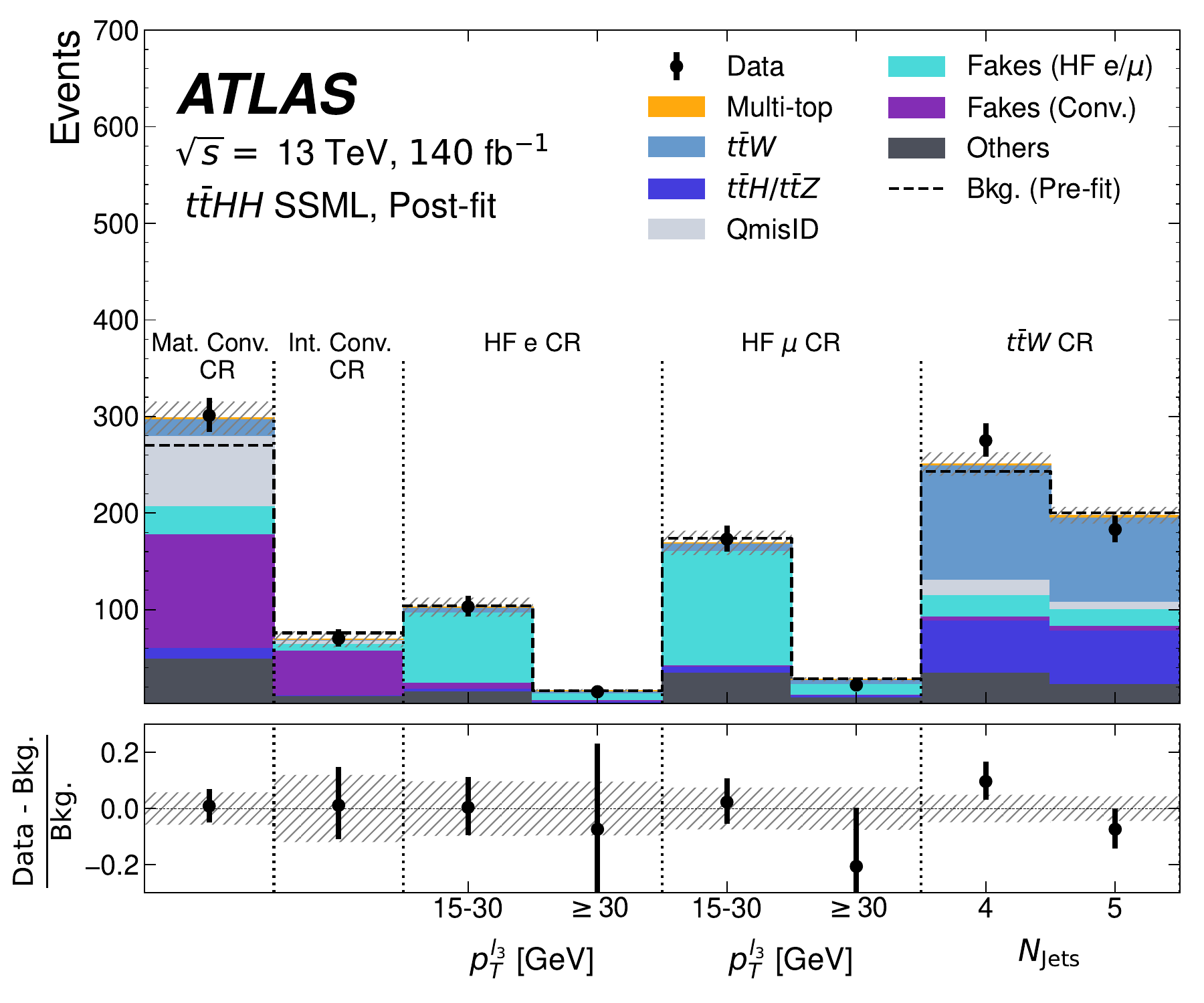}} \hspace*{0.5cm}
\subfloat[]{\includegraphics[width=0.45\textwidth]{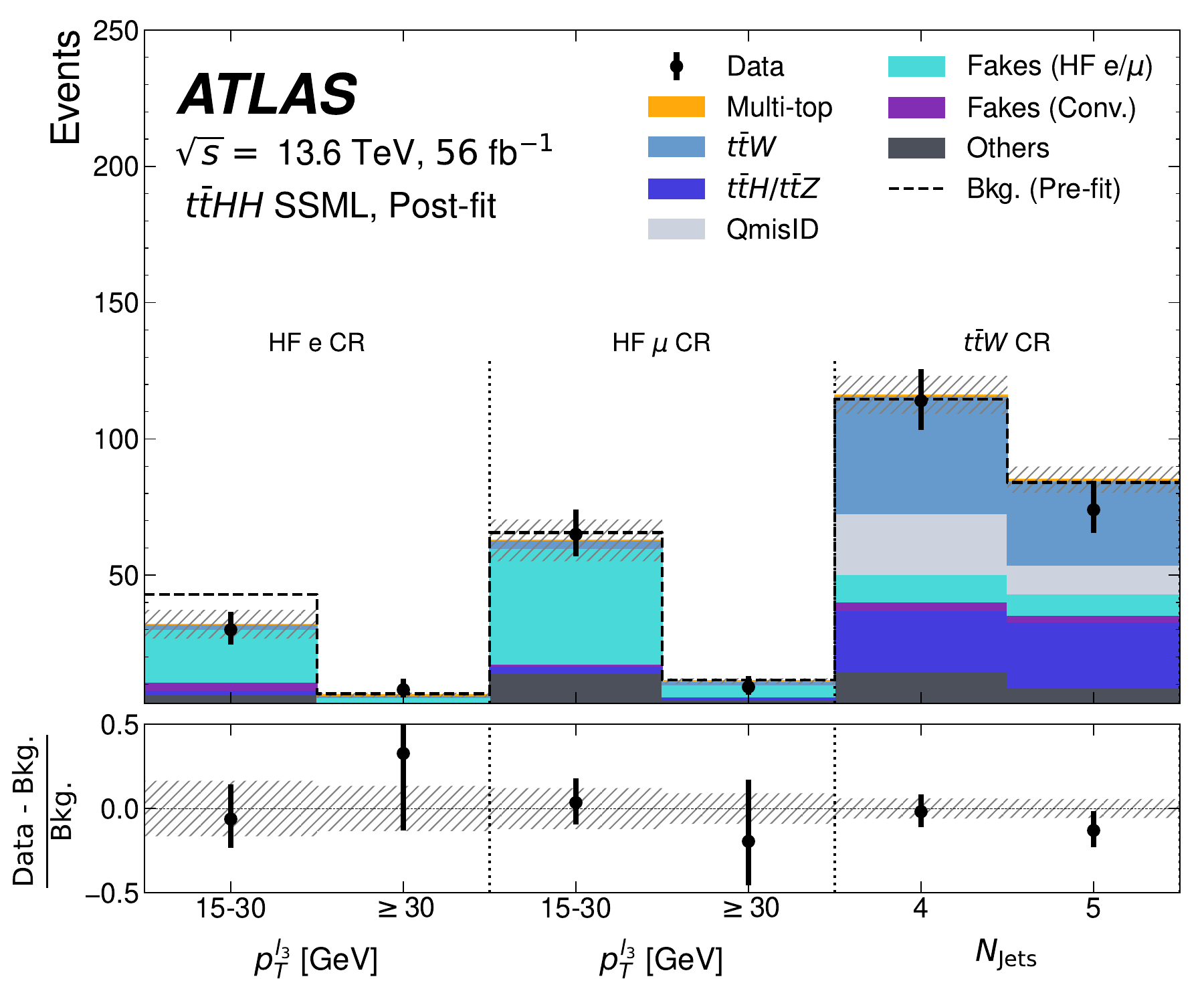}}
\caption{Distributions of the fit variables of the \SSML\ control regions, using the same binning as in the fit, for the (a) Run~2 and (b) Run~3 data samples, comparing the data with the post-fit stacked background predictions.
The lower panels display the relative differences between the data and the background predictions.
The hatched bands indicate the combined statistical and systematic uncertainties in the background predictions.
For reference, the pre-fit background predictions are overlaid as dashed histograms. }
\label{fig:SSML-backgrounds}
\end{figure}

The corresponding post-fit discriminant distributions are shown in Figure \ref{fig:1L-discriminant} for the \OneL\ channel and in Figure \ref{fig:SSML-discriminant} for the \SSML\ channel, presented separately for the Run 2 and Run 3 data samples. Figure \ref{fig:bbyy-discriminant} shows the post-fit diphoton invariant mass distribution for the \bbyy\ channel, combining the Run~2 and Run~3 data samples. The measured signal-strength values are all compatible with the SM prediction of unity,  $\mu_{\ttHH}^{\OneL}=-5^{+15}_{-15}$ from \OneL, $\mu_{\ttHH}^{\SSML}=5^{+17}_{-21}$ from \SSML, and $\mu_{\ttHH}^{b\bar{b}\gamma\gamma}=-13^{+35}_{-6}$ from \bbyy. The asymmetry in the measured signal-strength uncertainty in the \bbyy\ channel arises from the physical constraint that the total expected event yield must remain non-negative.
Combining three channels yields a best-fit signal-strength value of $\mu_{\ttHH}=-3^{+11}_{-12}$.
Distributions of representative kinematic variables in the Run~2 signal regions are shown in Figure~\ref{fig:1L-SR-distribution} for \OneL\ and  Figure~\ref{fig:SSML-SR-distribution} for \SSML, demonstrating good agreement between the observed data and the post-fit background predictions. Comparable agreement is observed in Run~3.

\begin{figure}[!htbp]
\centering
\subfloat[]{\includegraphics[width=0.4\textwidth]{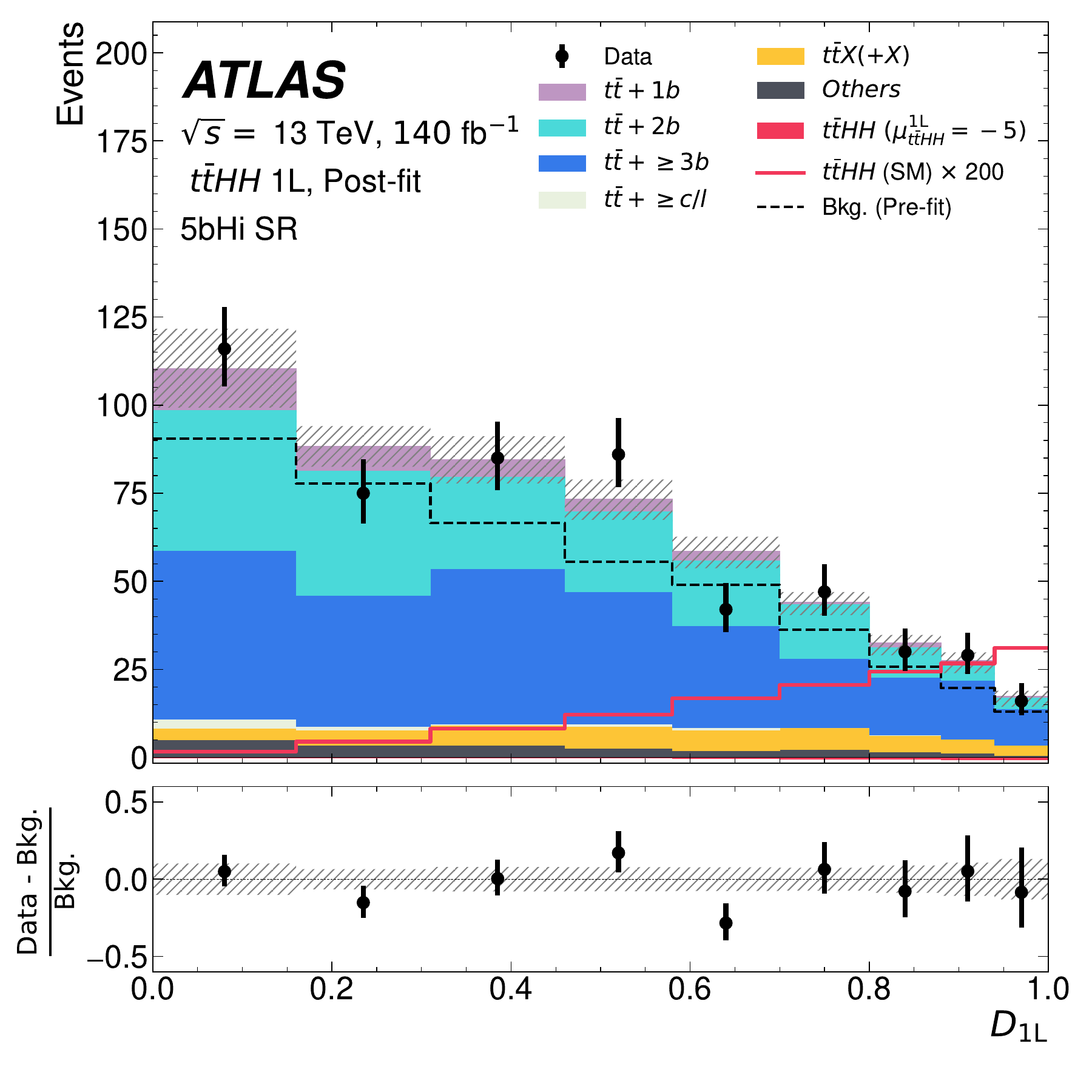}} \hspace*{0.5cm}
\subfloat[]{\includegraphics[width=0.4\textwidth]{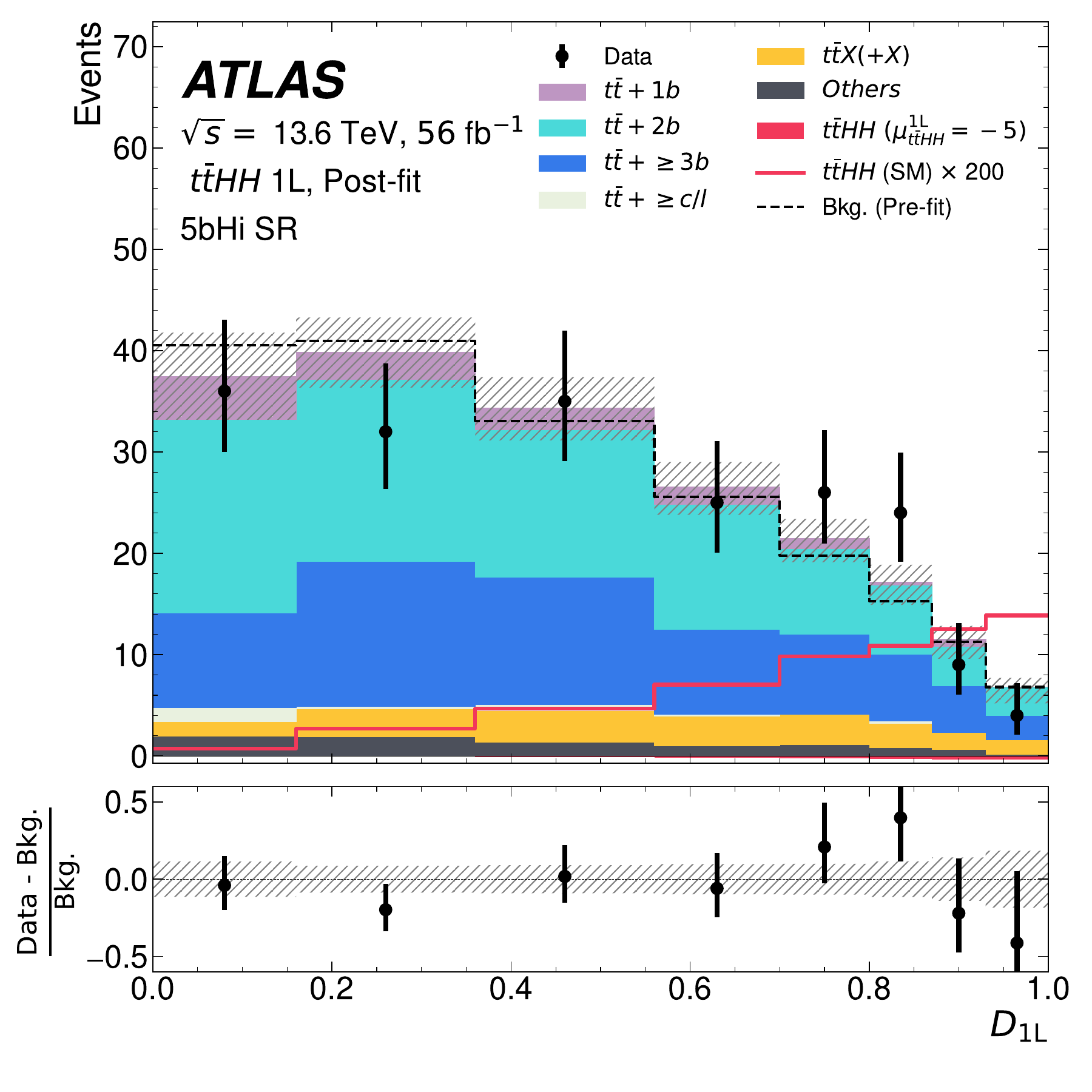}}\\
\subfloat[]{\includegraphics[width=0.4\textwidth]{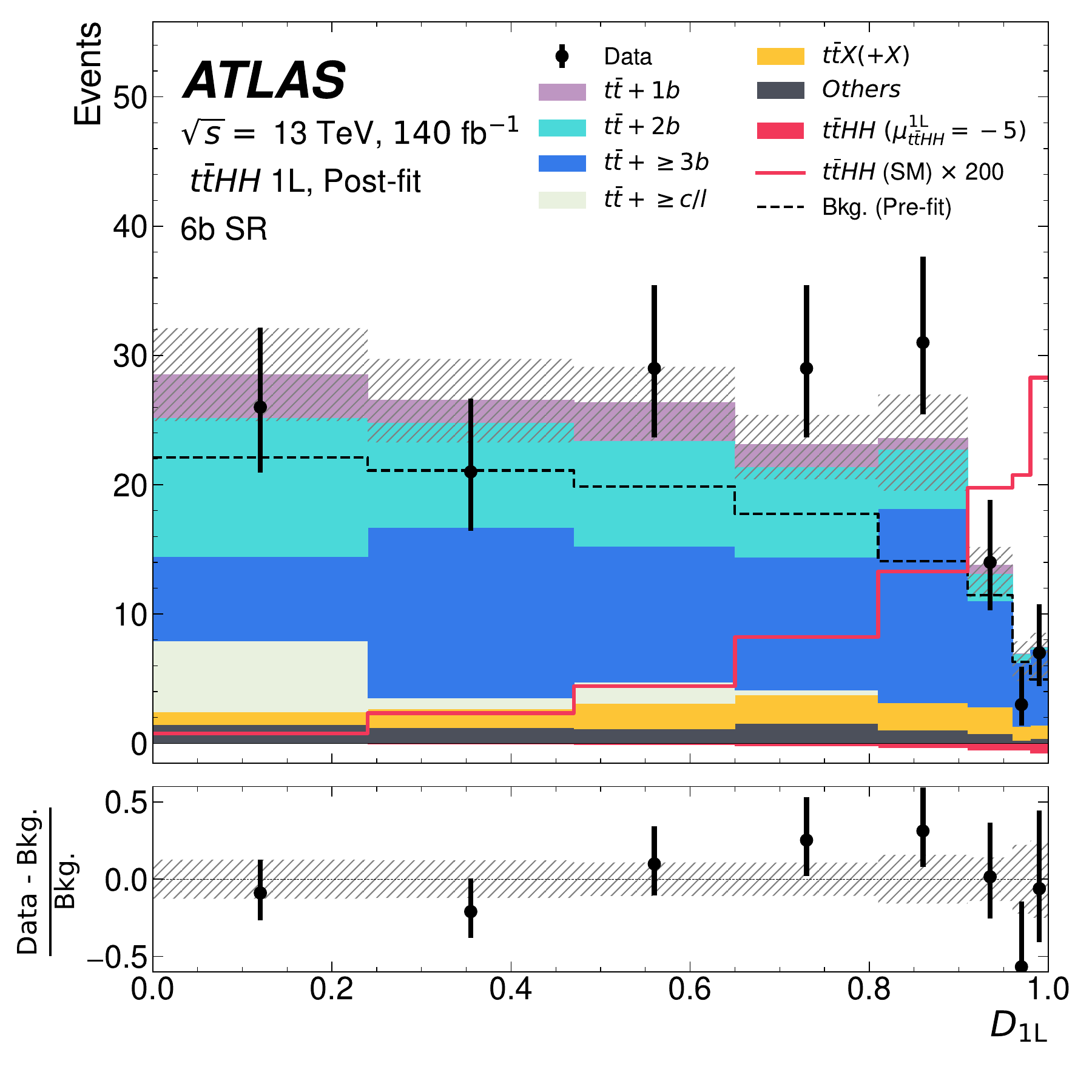}} \hspace*{0.5cm}
\subfloat[]{\includegraphics[width=0.4\textwidth]{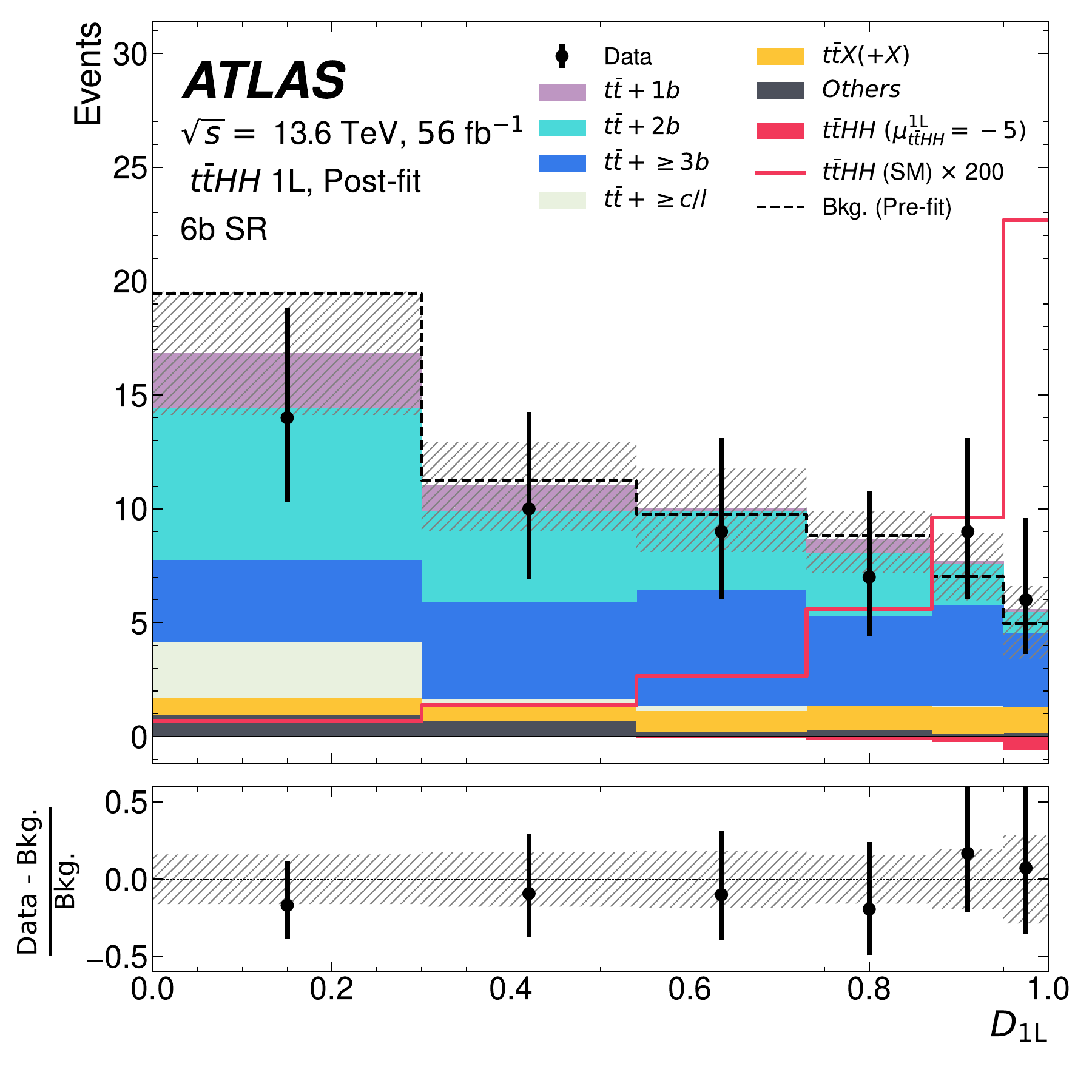}}
\caption{Distributions of the $D_{\OneL}$ discriminant in the  $5b_{\text{Hi}}$ and $6b$ signal regions of the \OneL\ channel for the (a, c) Run~2 and (b, d) Run~3 data samples, comparing the data with the post-fit stacked background predictions.
The observed \ttHH\ signal in this channel, corresponding to a measured signal strength of $\mu^{\text{1L}}_{\ttHH}=-5^{+15}_{-15}$, is overlaid as a negative contribution.
The lower panels display the relative differences between the data and the background predictions.
The hatched bands indicate the combined statistical and systematic uncertainties in the background predictions.
For reference, the pre-fit background predictions are overlaid as dashed lines, and the SM \ttHH\ signal distribution -- scaled up by a factor of 200 -- are shown as solid lines.}
\label{fig:1L-discriminant}
\end{figure}

\begin{figure}[!htbp]
\centering
\subfloat[]{\includegraphics[width=0.40\textwidth]{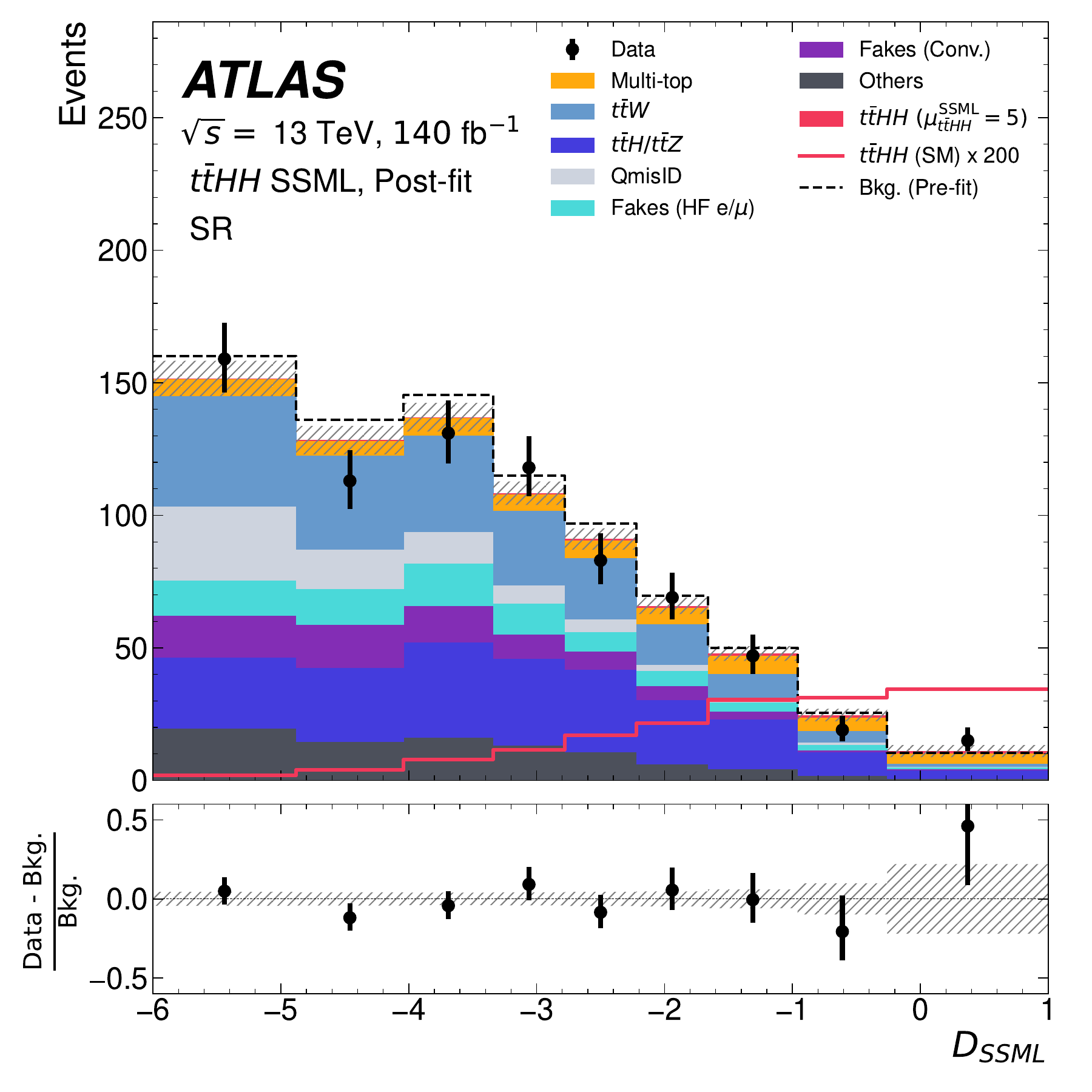}} \hspace*{0.5cm}
\subfloat[]{\includegraphics[width=0.40\textwidth]{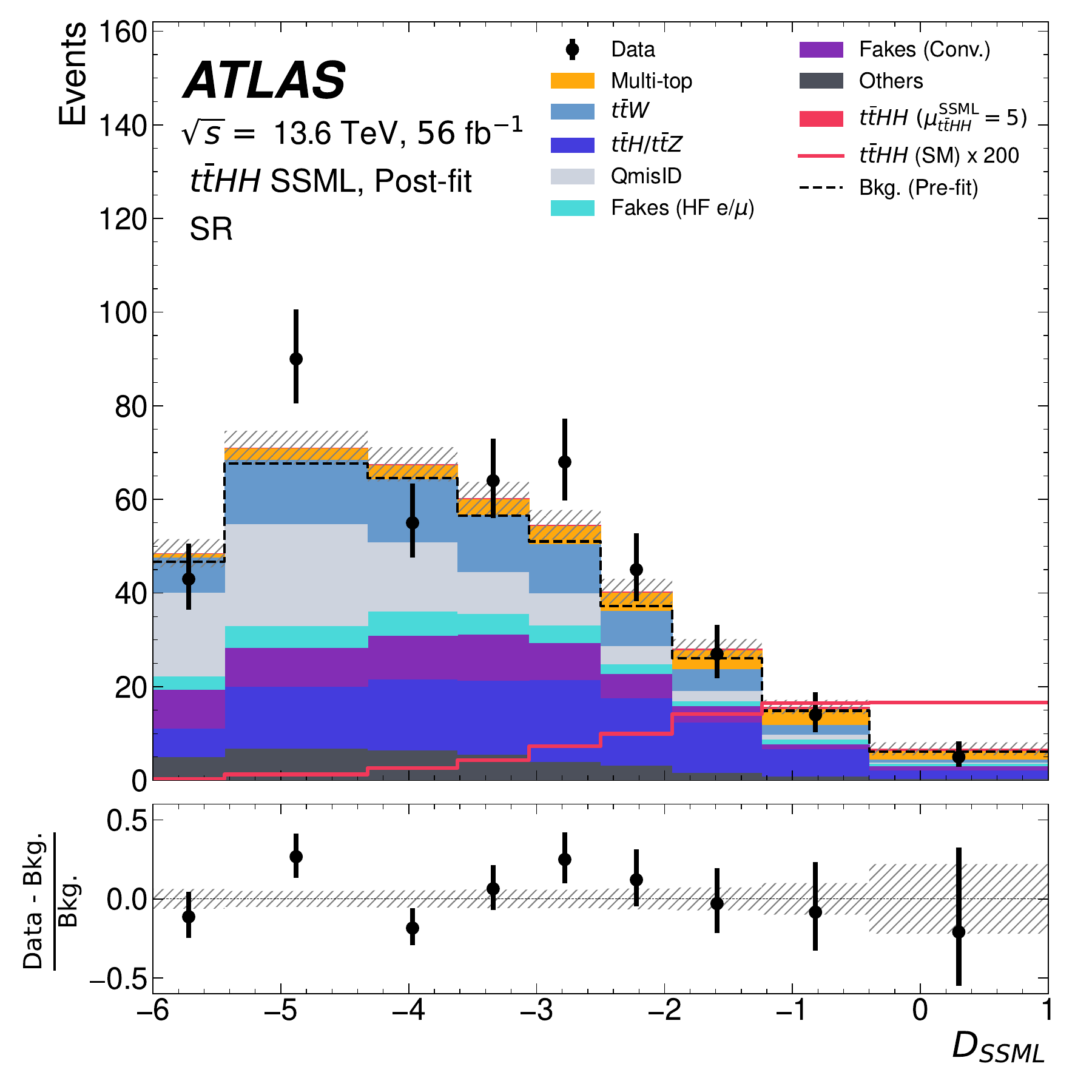}}
\caption{Distributions of the $D_{\SSML}$ discriminant in the \SSML\ signal regions for the (a) Run~2 and (b) Run~3 data samples, comparing the data with the post-fit stacked signal-plus-background predictions.
The observed \ttHH\ signal in this channel, corresponding to a measured signal strength of $\mu^{\SSML}_{\ttHH}=5^{+17}_{-21}$, is displayed stacked on top of the backgrounds.
The first and last bins include contributions from underflow and overflow, respectively.
The lower panels display the relative differences between the data and the background predictions.
The hatched bands indicate the combined statistical and systematic uncertainties in the background predictions.
For reference, the pre-fit background predictions are overlaid as dashed lines, and the SM \ttHH\ signal distribution -- scaled up by a factor of 200 -- are shown as solid lines.
}
\label{fig:SSML-discriminant}
\end{figure}

\begin{figure}[!htbp]
\centering
\subfloat[]{\includegraphics[width=0.33\textwidth]{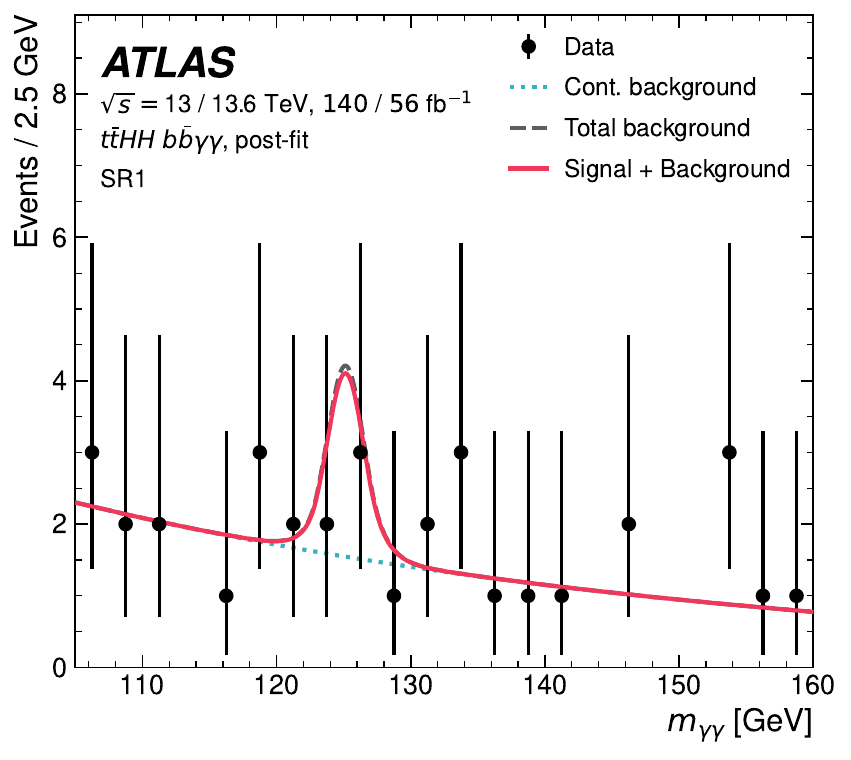}}
\subfloat[]{\includegraphics[width=0.33\textwidth]{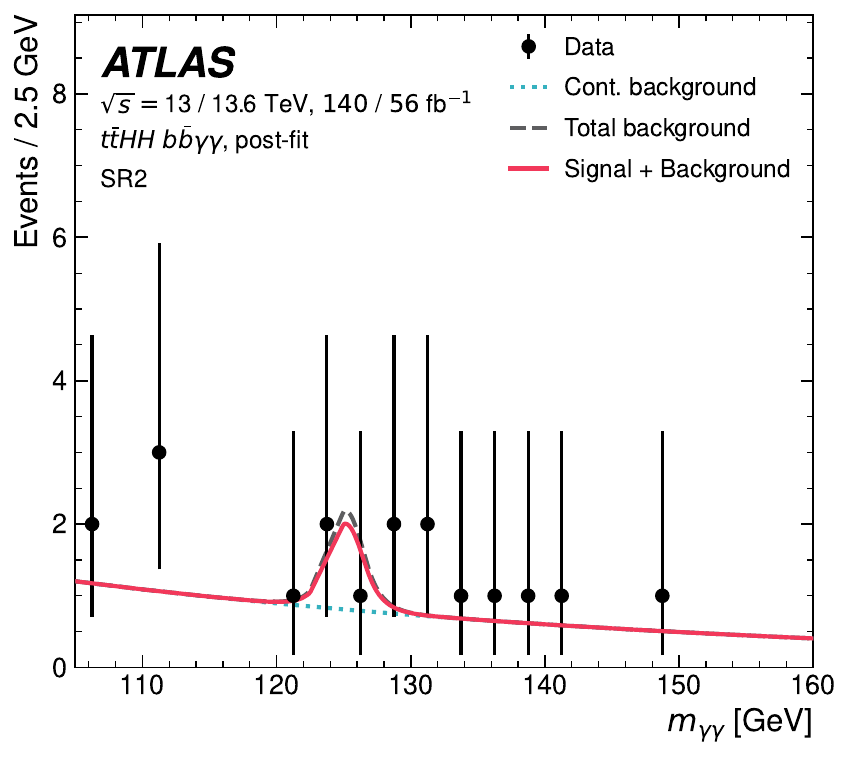}}
\subfloat[]{\includegraphics[width=0.33\textwidth]{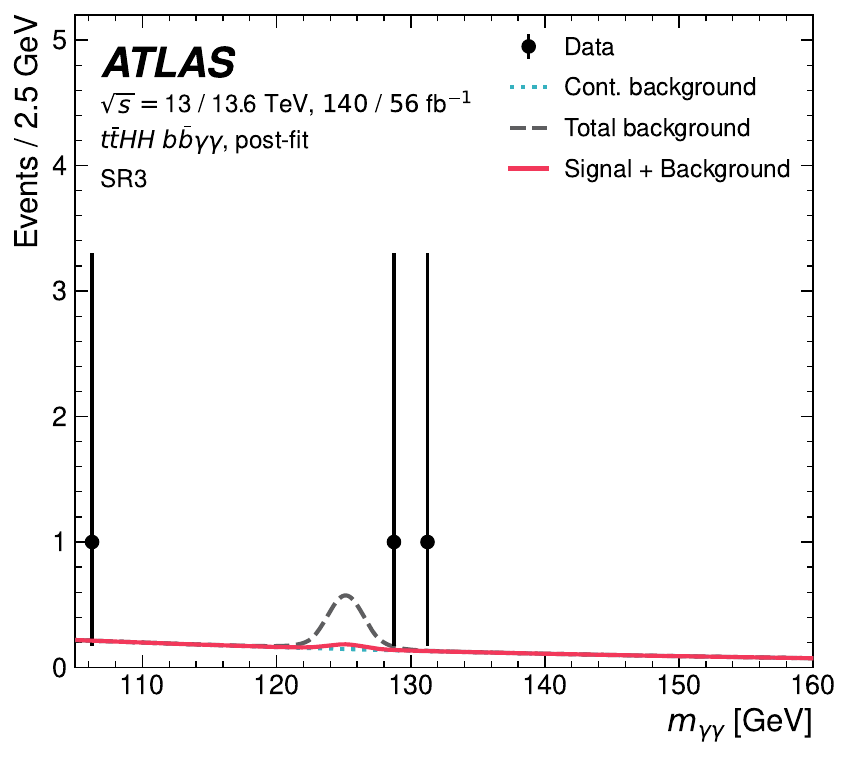}}
\caption{Comparison of the $m_{\gamma\gamma}$ distributions between the observed data and the signal-plus-background fitted estimate (solid) for the \bbyy\ channel in the three signal regions: (a) SR1, (b) SR2 and (c) SR3, where SR1 is the least sensitive and SR3 the most sensitive. The continuous background (dotted) and the total background (dashed) are also shown. The observed \ttHH\ signal strength in this channel is $\mu^{b\bar{b}\gamma\gamma}_{\ttHH}=-13^{+35}_{-6}$.
}
\label{fig:bbyy-discriminant}
\end{figure}

\begin{figure}[!htbp]
\centering
\subfloat[]{\includegraphics[width=0.4\textwidth]{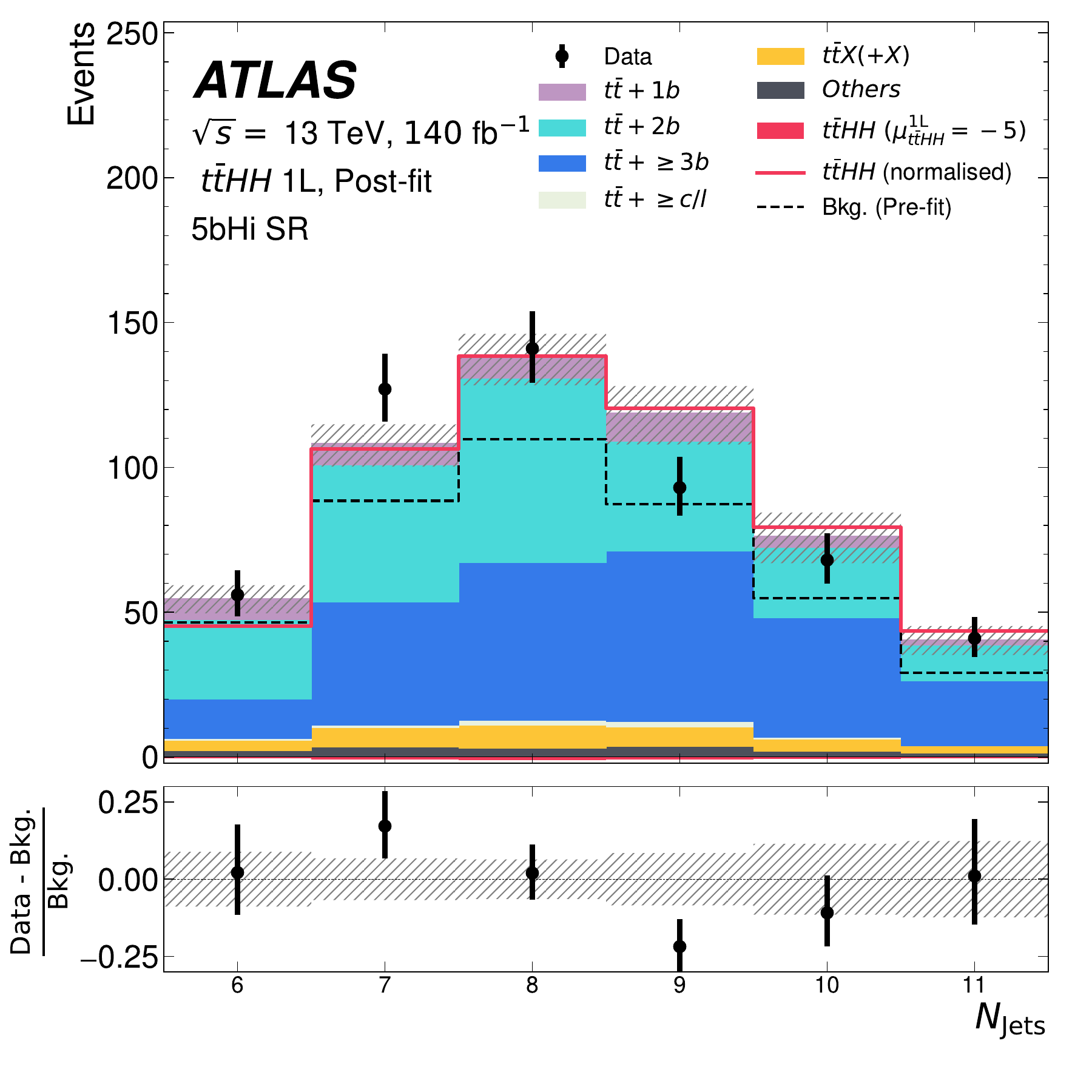}}\ \ \
\subfloat[]{\includegraphics[width=0.4\textwidth]{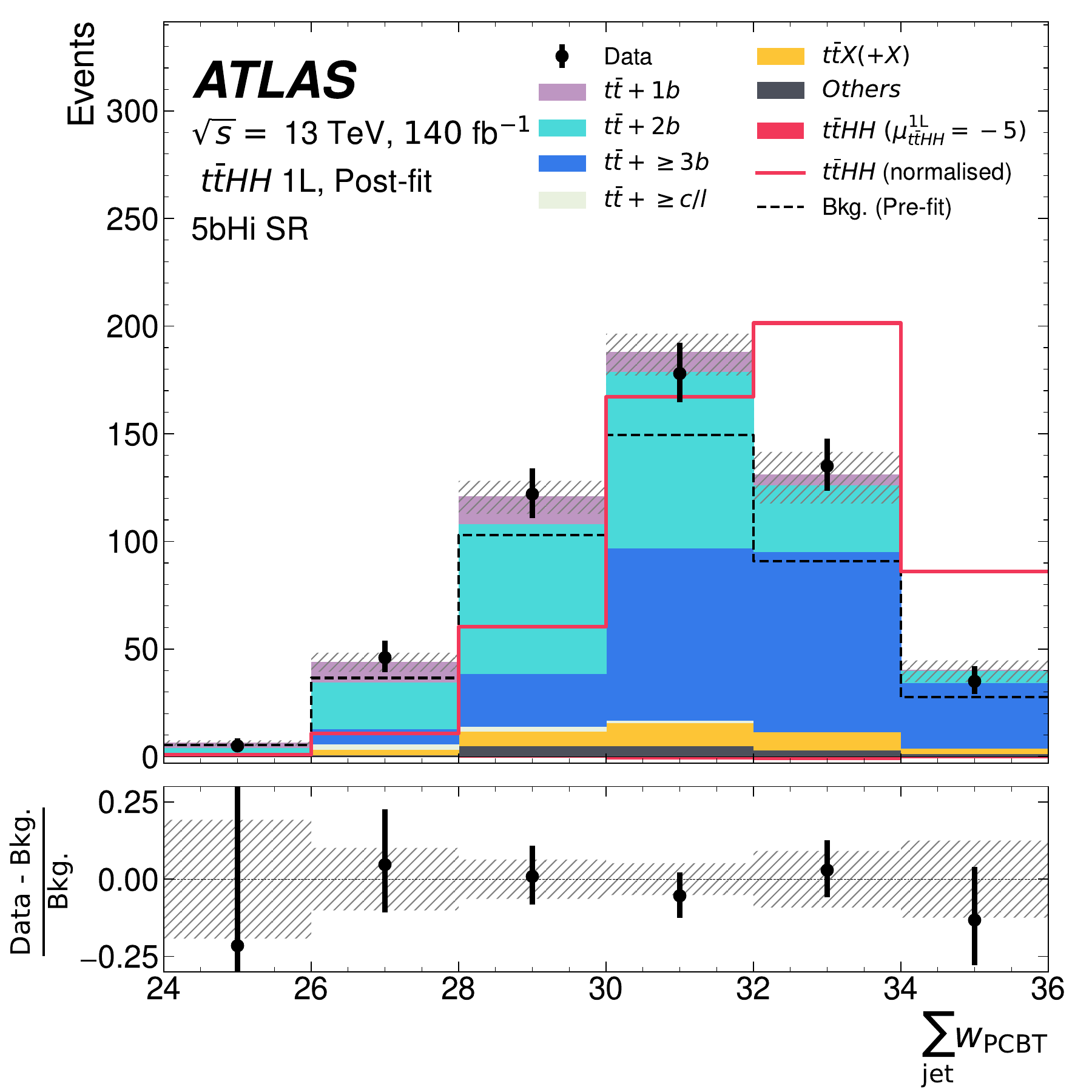}} \\
\subfloat[]{\includegraphics[width=0.4\textwidth]{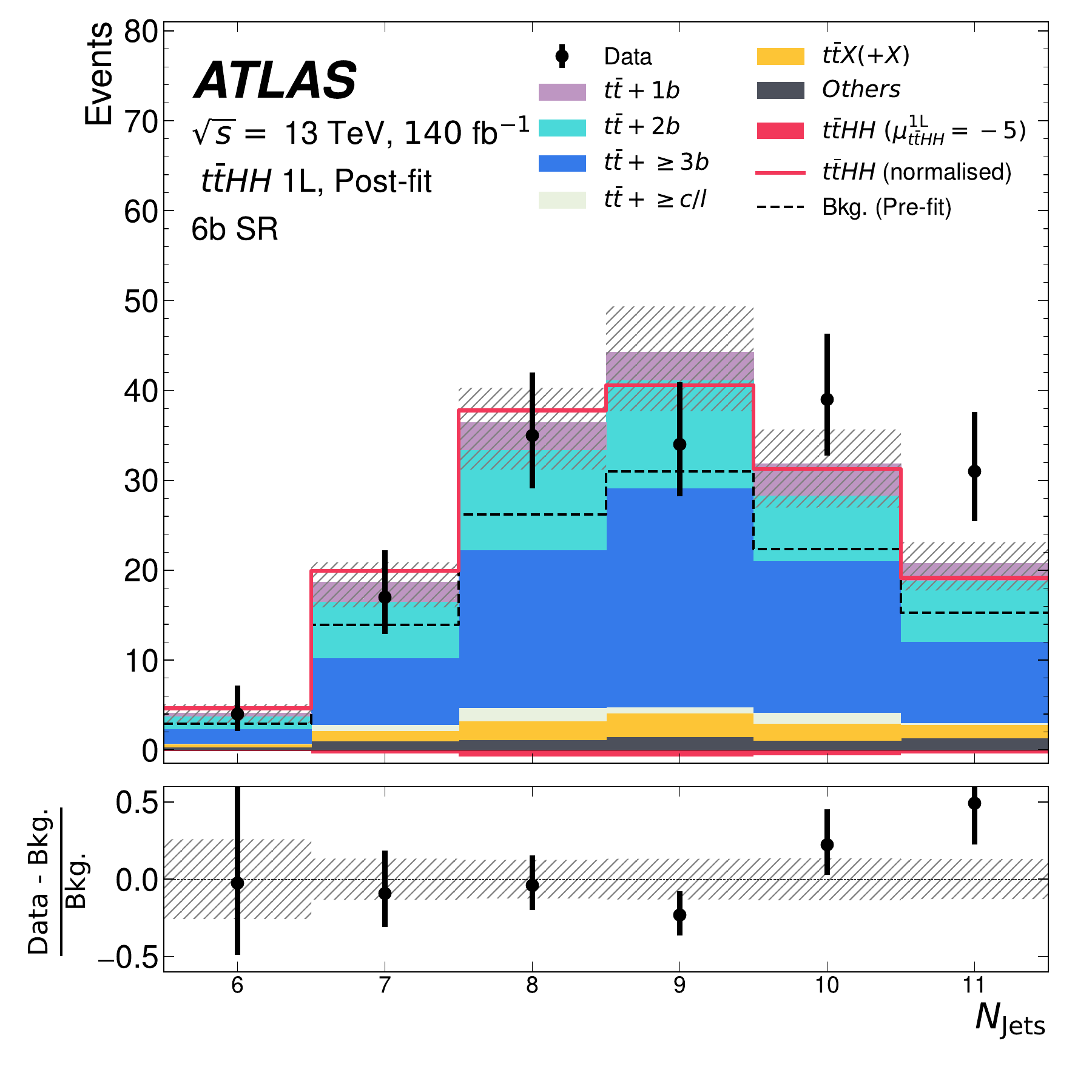}}\ \ \
\subfloat[]{\includegraphics[width=0.4\textwidth]{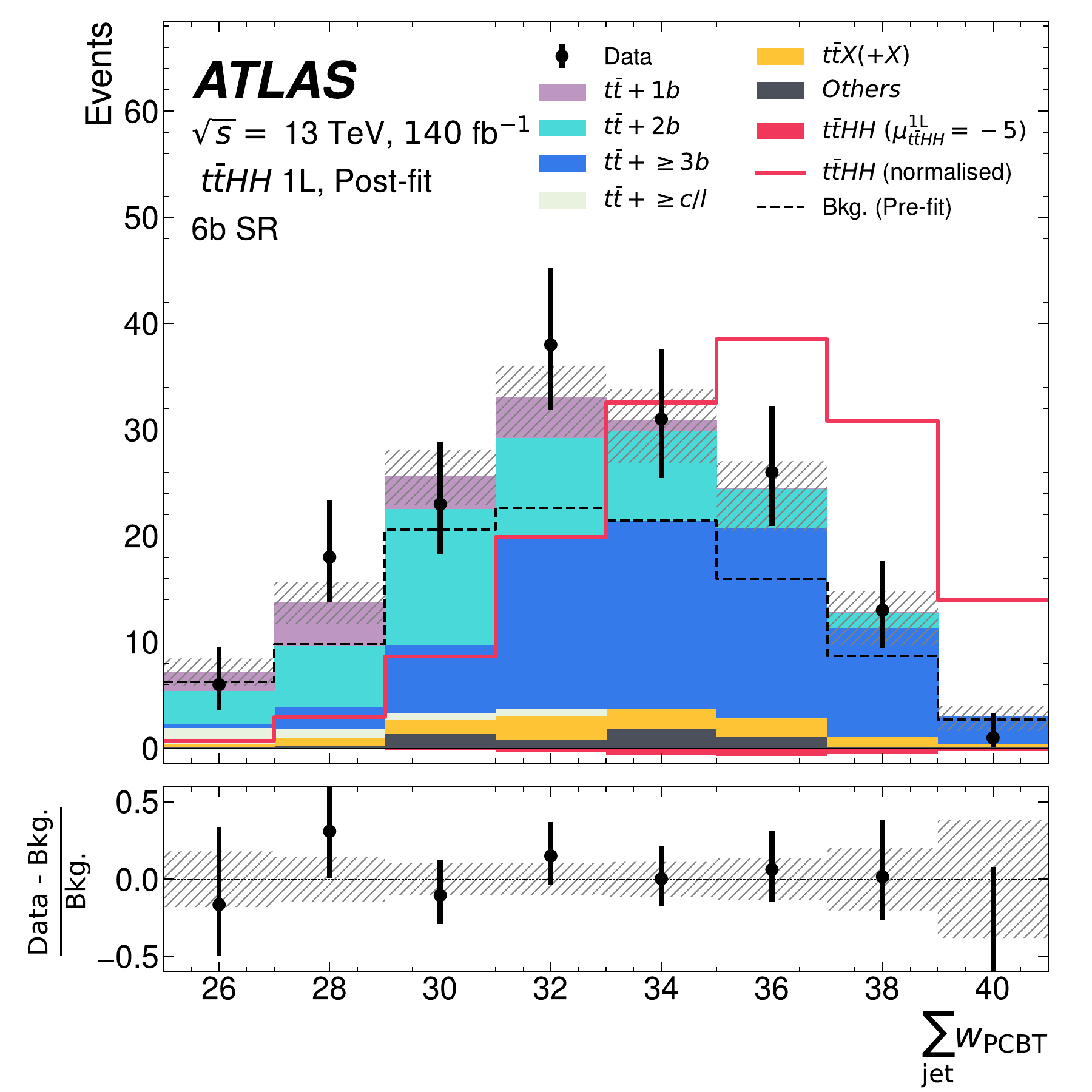}}
\caption{Distributions of the (a, c) jet multiplicity and  (b, d) sum of pseudo-continuous $b$-tagging scores in the (a, b)  $5b_\text{Hi}$ and (c, d) $6b$ signal regions of the \OneL\ channel for the Run~2 data sample, comparing the data with the post-fit stacked background predictions.
The fitted \ttHH\ signal in this channel, $\mu^{\OneL}_{\ttHH}=-5^{+15}_{-15}$, is overlaid as a negative contribution.
The lower panels display the relative differences between the data and the background predictions.
The hatched bands indicate the combined statistical and systematic uncertainties in the background predictions.
For reference, the pre-fit backgrounds are overlaid as dashed lines, and the SM \ttHH\ signal distributions -- normalised to the total background yield -- are shown as solid lines.
}
\label{fig:1L-SR-distribution}
\end{figure}

\begin{figure}[!htbp]
\centering
\subfloat[]{\includegraphics[width=0.45\textwidth]{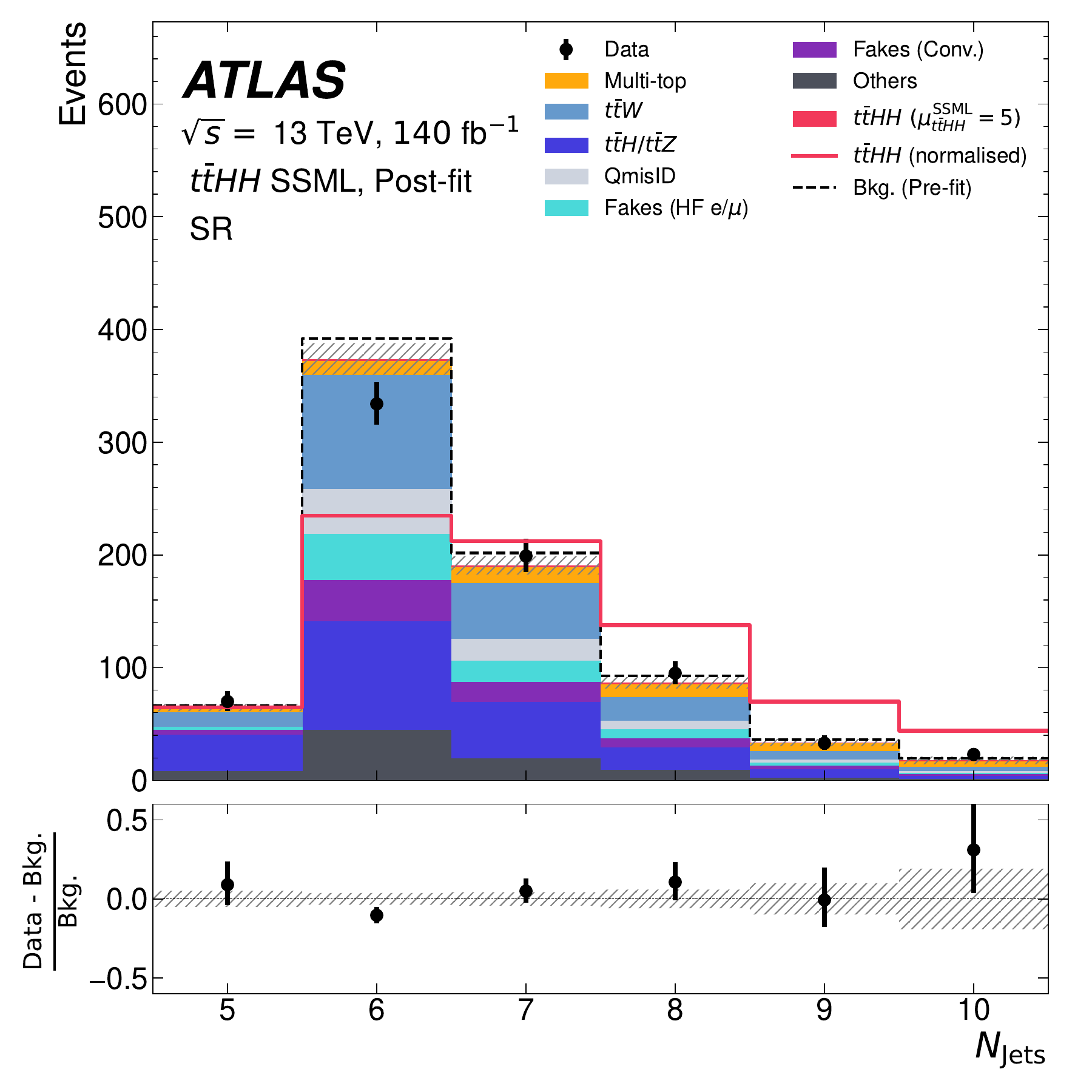}}\hspace*{5mm}
\subfloat[]{\includegraphics[width=0.45\textwidth]{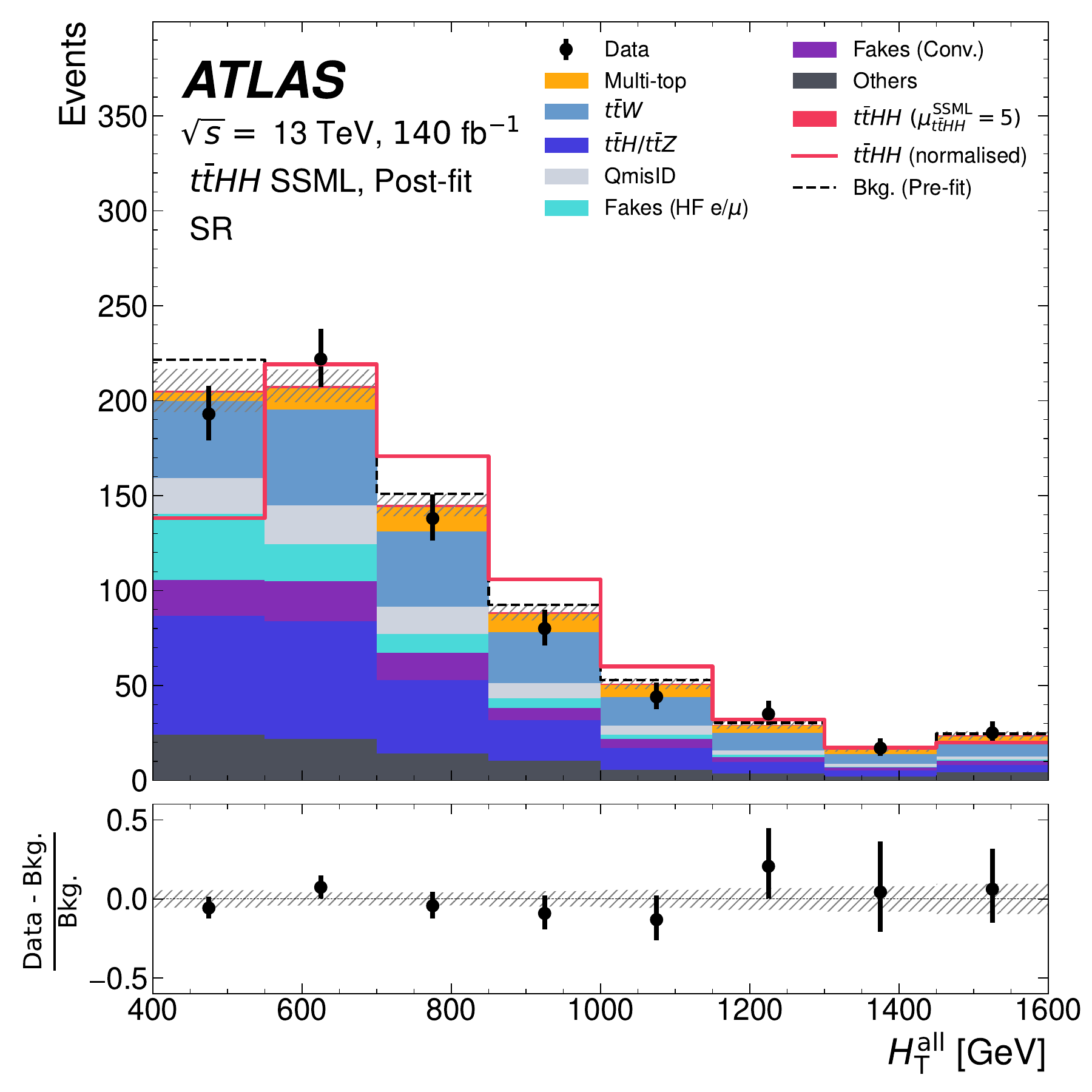}}
\caption{Distributions of the (a) jet multiplicity and (b) $H_\text{T}^\text{all}$ in the \SSML\ signal region for the Run~2 data sample, comparing the data with the post-fit stacked signal-plus-background predictions.
The fitted \ttHH\ signal in this channel, $\mu^{\SSML}_{\ttHH}=5^{+17}_{-21}$, is shown stacked on top of the backgrounds.
The first and last bins include contributions from underflow and overflow, respectively. The lower panels display the relative differences between the data and the background predictions. %
The hatched bands indicate the combined statistical and systematic uncertainties in the background predictions.
For reference, the pre-fit backgrounds are overlaid as dashed lines, and the SM \ttHH\ signal distributions -- normalised to the total background yield -- are shown as solid lines. }
\label{fig:SSML-SR-distribution}
\end{figure}

Upper limits on the signal strength at the 95\% CL are derived using the modified frequentist method~\cite{Read:2002hq}, commonly referred to as the CL$_\text{s}$ method, and employing the $\tilde{q}_{\mu}$ test statistic in the asymptotic approximation~\cite{Cowan:2010js}.
The observed (expected) 95\% CL upper limits on the signal strength are 26 (29), 40 (38), and 75 (79) for the \OneL, \SSML, and \bbyy\ channels, respectively. In all cases, the expected limits are computed under the background-only hypothesis ($\mu_{\ttHH}=0$). Combining the three channels results in an observed upper limit of 20 at 95\% CL, compared with an expected limit of 21. Figure~\ref{fig:combined_limit} shows the observed and expected limits  for the individual \OneL, \SSML, and \bbyy\ channels, as well as their combination. For the \OneL\ and \SSML\ channels, the results from the Run~2 and Run~3 data samples are compatible within uncertainties, with the Run~2 data sample providing more stringent limits due to its larger integrated luminosity.

\begin{figure}[!htbp]
\centering
\includegraphics[width=12.0cm]{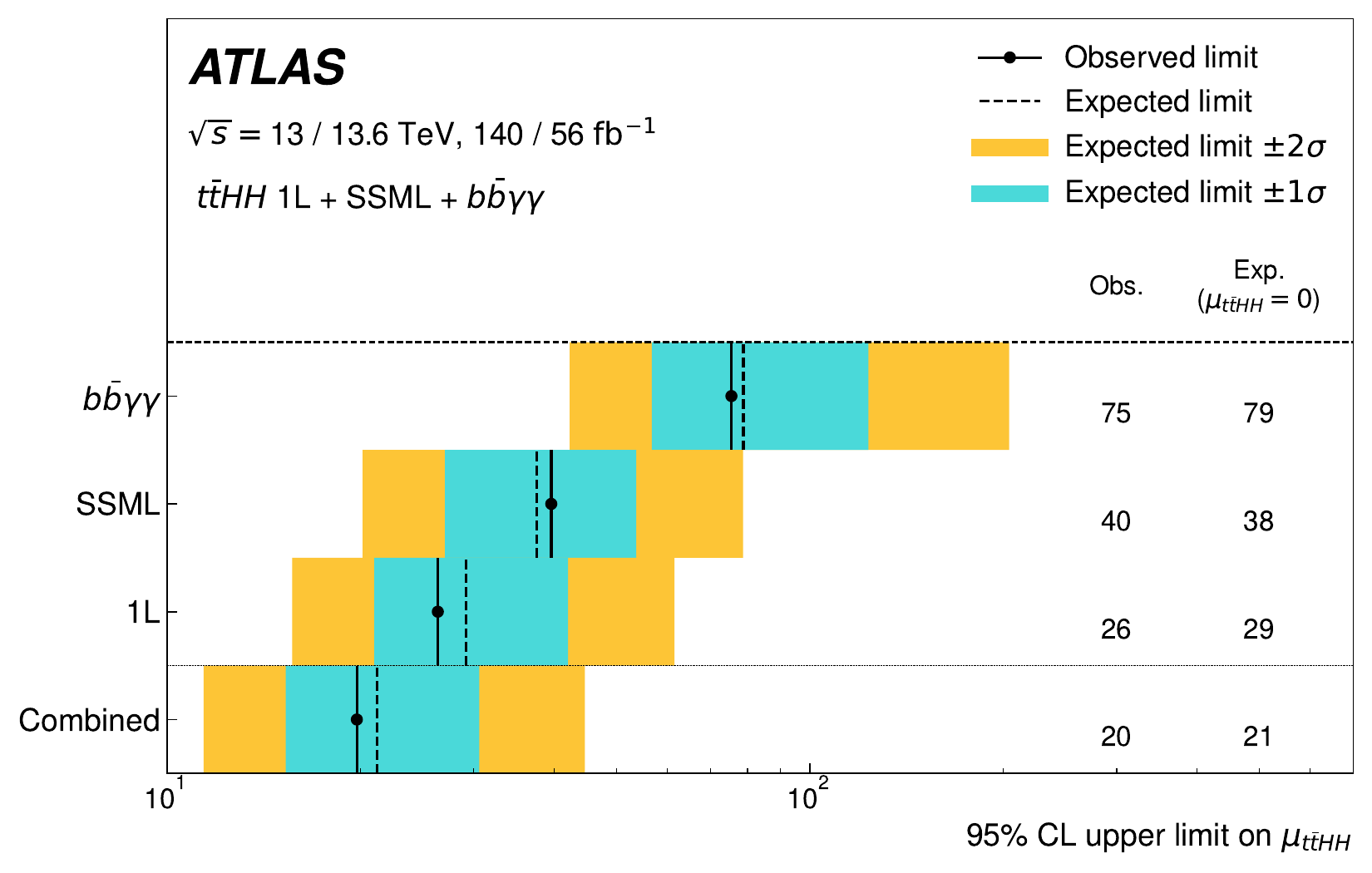}
\caption{Observed and expected 95\% CL upper limits on the signal strength from the individual \OneL, \SSML, and \bbyy\ channels, and from their combination. The expected limits are derived under the background-only hypothesis, i.e. assuming $\mu_{\ttHH}=0$. }
\label{fig:combined_limit}
\end{figure}

Table~\ref{tab:systematics} summarises the breakdown of contributions to the observed and expected uncertainties, $\Delta\mu_{\ttHH}$, in the signal strength of the combined \OneL, \SSML, and \bbyy\ analysis from different groups of uncertainty sources. These breakdowns are derived by performing conditional profile-likelihood fits in which all NPs associated with a given uncertainty group are fixed to their best-fit values, while all remaining parameters are profiled.
The uncertainty attributed to each group is then defined as the difference in quadrature between the total post-fit uncertainty and the uncertainty from the corresponding conditional fit. The expected breakdowns are evaluated assuming $\mu_{\ttHH}=1$, and correlations among different groups are neglected. The systematic uncertainties are categorised into signal modelling, background modelling, number of MC events and detector systematic uncertainties, with the background-modelling  contribution further decomposed by background source. The uncertainties from the floating NFs, which are limited by the available number of data events, are included in the data statistical uncertainty category.
Data statistical and systematic uncertainties contribute comparably to the total uncertainty, while background modelling dominates the systematic uncertainty.

\begin{table}[!htb]
\centering
\caption{Breakdowns of the contributions to the observed and expected uncertainties, $\Delta\mu_{\ttHH}$, in the signal strength of the combined \OneL, \SSML, and \bbyy\ analysis for different groups of uncertainty sources, neglecting correlations among groups.   }
\begin{tabular}{clccc} \toprule
&    Uncertainty Source                &  Observed  & Expected \\ \midrule
& Signal modelling                     &  $(+0.9, -1.8)$     & $(+1.3, -1.0)$  & \\[1mm]
& Background modelling                 &  $(+6.8, -8.2)$     & $(+5.7, -6.2)$  \\
& \hspace{1em} $t\bar{t}+\text{jets}$  &  $(+5.3, -6.5)$  & $(+4.7, -5.0)$  \\
& \hspace{1em} $t\bar{t}t\bar{t}$  &  $(+4.6, -6.4)$  & $(+2.9, -4.0)$  \\
& \hspace{1em} $t\bar{t}H$         &  $(+2.1, -1.6)$  & $(+2.0, -1.9)$  \\
& \hspace{1em} others              &  $(+0.9, -0.7)$  & $(+0.9, -0.8)$  \\[1mm]

& MC statistical              &  $(+2.8, -3.7)$   & $(+2.5, -2.7)$  \\
& Detector systematic       &  $(+2.2, -3.4)$   & $(+1.9, -1.5)$  \\[1mm] \midrule
& Total systematic          &  $(+8.3, -9.4)$   & $(+7.3, -7.4)$  \\
& Data statistical            &  $(+7.2, -6.8)$   & $(+7.2, -6.7)$  \\[1mm] \midrule
& Total                      &  $(+11.0, -11.6)$   & $(+10.3, -10.0)$ \\ \bottomrule
\end{tabular}

\label{tab:systematics}
\end{table}

\FloatBarrier


\section{Higgs effective field theory interpretation}
\label{sec:results_eft}

Beyond-the-SM (BSM) self-interactions of the Higgs boson, or its non-standard interactions with other gauge fields and fermions, could lead to modifications to the \ttHH\ event rate as well as affect the kinematic properties of the signal events. The results obtained with the two most sensitive channels of this analysis, i.e. the \OneL\ and \SSML\ channels, are thus interpreted in the context of the HEFT formalism, with a focus on anomalous interactions between Higgs boson pairs and top quark pairs. %

Similarly to the SM effective field theory (SMEFT)~\cite{Buchmuller:1985jz, Grzadkowski:2010es}, HEFT assumes the SM field content. However, unlike in SMEFT, where the Higgs boson is part of an SU(2) doublet alongside unphysical Goldstone bosons, in HEFT the Higgs boson is a gauge singlet. This allows the Higgs boson dynamics to be decoupled from the symmetry-breaking pattern~\cite{PhysRev.166.1568}. Consequently, HEFT enables more general deviations in Higgs boson couplings, which must be independently constrained through precision measurements, and features a more intricate power counting compared to SMEFT~\cite{Gavela:2016bzc, Buchalla:2013eza, Brivio:2025yrr}. The HEFT approach employed in this study is implemented in a model specifically developed for this purpose and is interfaced through \mg~\cite{Alwall:2014hca} using the \texttt{FeynRules} \texttt{UFO} model format~\cite{Alloul:2013bka, Degrande:2011ua}.

State-of-the-art constraints on \cttHH\ have been derived from the \ggF~$HH$ channel by both the ATLAS~\cite{HDBS-2021-18} and CMS~\cite{CMS-HIG-20-011} collaborations, ($-0.19<\cttHH<0.7$ and $-0.28<\cttHH<0.59$, respectively), which remains the most sensitive channel to date for \cttHH. However, assuming a weakly interacting ultraviolet completion~\cite{Arzt:1994gp}, \ggF\ is parameterised by two additional Higgs--gluon couplings, \cggH\ and \cggHH, in comparison with \ttHH\, where these enter at higher loop order. As a result, obtaining robust bounds from \ggF\ is more challenging unless further assumptions are imposed. The quoted limits are obtained by setting all other couplings of the Higgs boson to their SM values. Relaxing this assumption requires a multidimensional fit, weakening the obtained constraints. The work presented here aims to provide a complementary constraint on \cttHH\, independent of the \ggF~$HH$ channel. The underlying assumption of the analysis, that only \cttHH\ differs from its SM value, is consistent with those used to derive the values for the \ggF~$HH$ channel.
In this context, only cross-section effects are considered, while acceptance and kinematic shape variations are not explicitly parameterised. Acceptance effects are evaluated for values of \cttHH\ close to the exclusion limits and are covered by an additional 30\% uncertainty in the signal yield included in the measurement. An estimate of the kinematic-shape effects is found to impact the 95\% CL interval on \cttHH\ by at most 10\%, which is not included in the quoted limits.

\begin{figure}[!htb]
\centering
\subfloat[]{\includegraphics[width=0.44\textwidth]{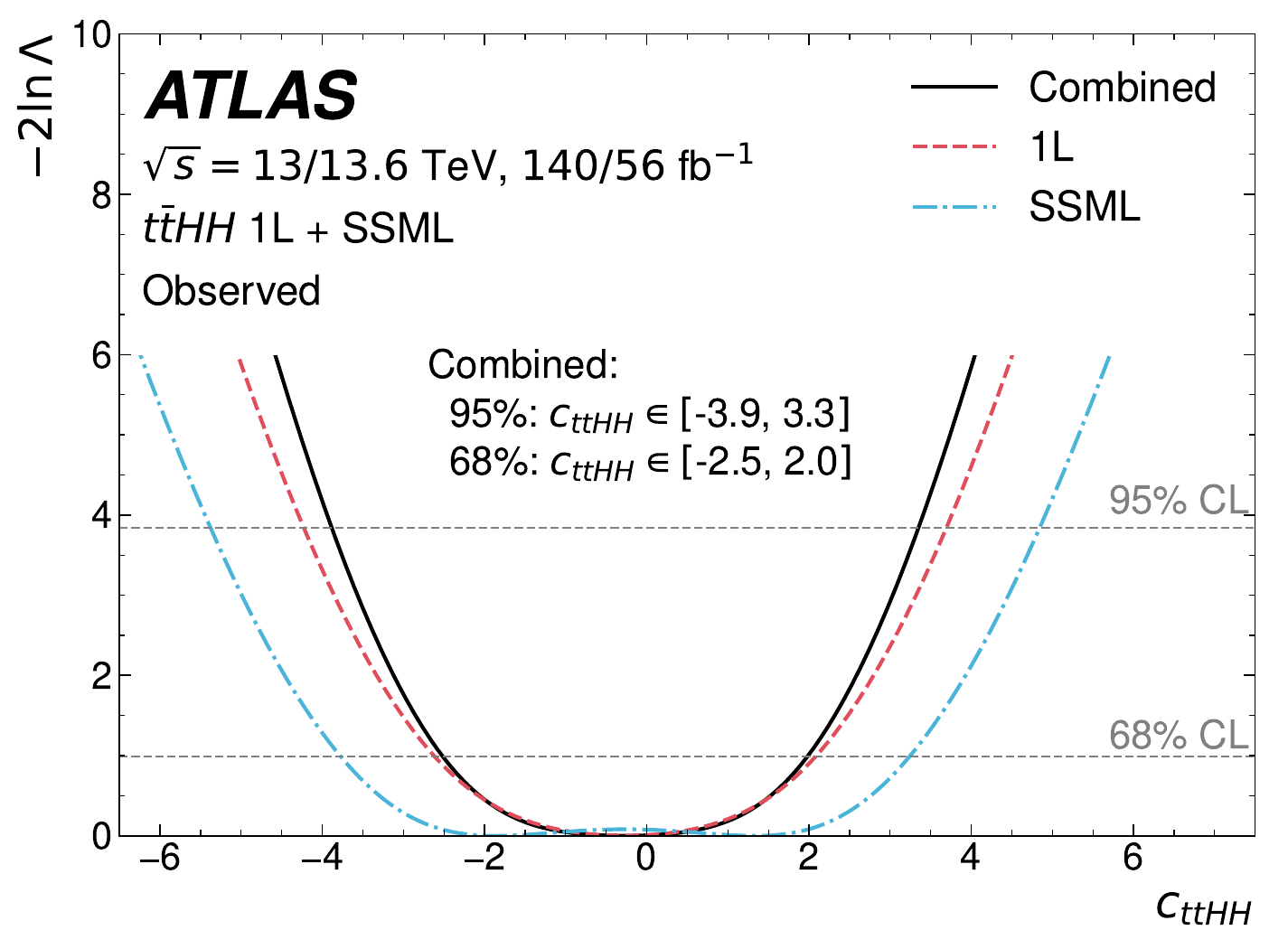}} \hspace*{0.5cm}
\subfloat[]{\includegraphics[width=0.44\textwidth]{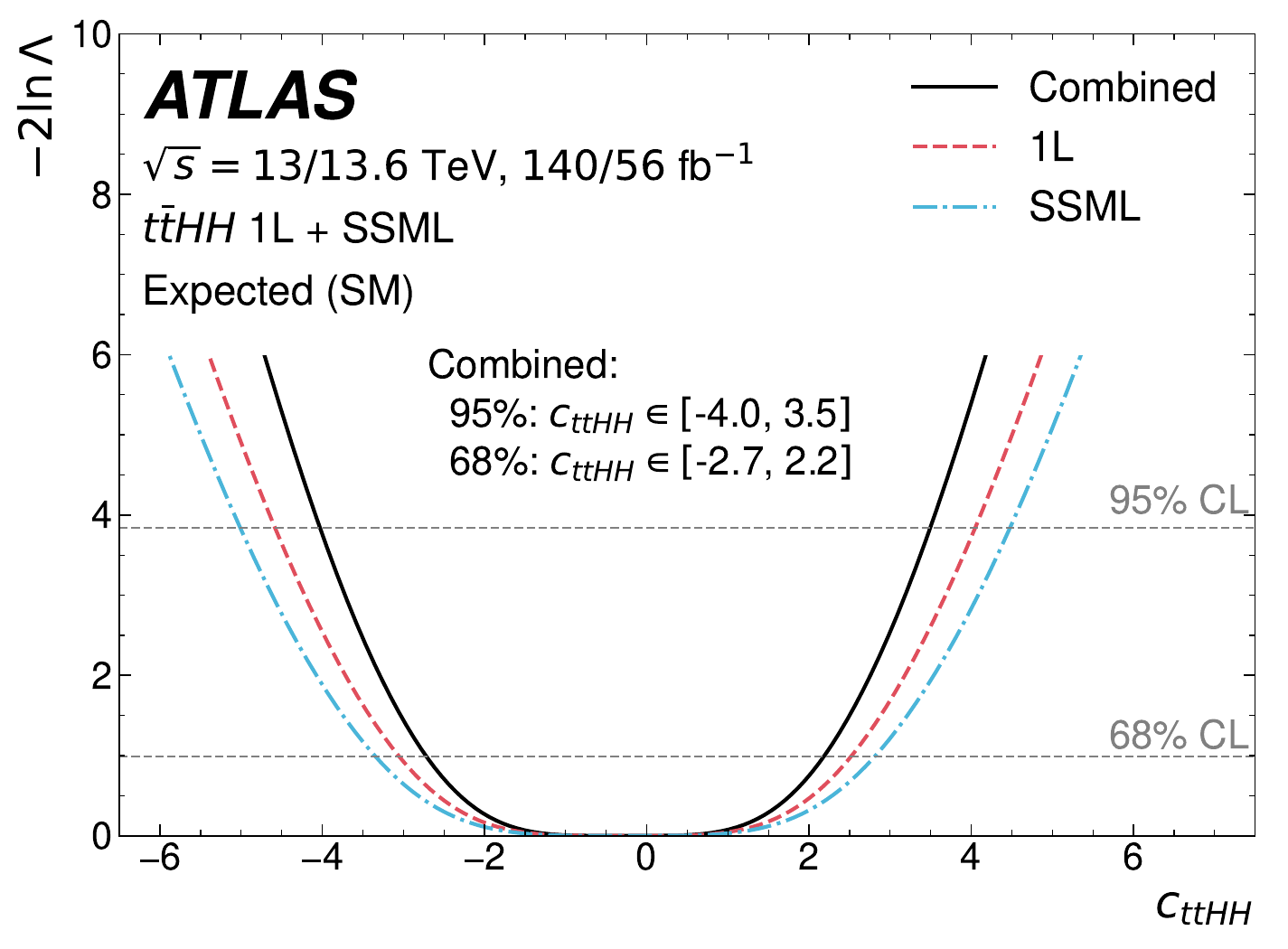}}
\caption{(a) Observed and (b) expected values of the profile likelihood as a function  of \cttHH\ for the \OneL\ channel (dashed curve), the \SSML\ channel (dot-dashed curve), and their combination (solid curve).}
\label{fig:cttHH}
\end{figure}

The constraints on \cttHH\ are extracted from the same phase space as the rest of the analysis using the same multivariate discriminant described in Section~\ref{sec:mva} and in the same analysis regions. The results are presented in Figure~\ref{fig:cttHH}. The \cttHH\ is constrained, at the 95\% CL, in the range of $-3.9<\cttHH<3.3$ for the observed results, while the expected range is $-4.0<\cttHH<3.5$. The expected profile exhibits a double-minimum structure due to the quadratic dependence of the signal cross-section on \cttHH. Given the observed results, the two minima are nearly degenerate, resulting in an almost flat region in the likelihood scan.

Benefiting from the quadratic dependence of the cross-section on \cttHH, the bounds derived in this analysis are only about one order of magnitude weaker than those obtained in the \ggF~$HH$ channel in Ref.~\cite{HDBS-2021-18}, even though the predicted cross-section for \ttHH\ production is smaller by a factor of about 40.

%


\section{Conclusion}
\label{sec:conclusion}
The non-resonant production of Higgs boson pairs in association with top-quark pairs, \ttHH, is predicted to be a rare process in the SM\@. This paper describes the first search for this process using the full Run~2 data sample, corresponding to 140 \ifb\ of $pp$ collisions at $\sqrt{s}=13$~\TeV collected between 2015 and 2018, and a Run~3 data sample of 56 \ifb\ at $\sqrt{s}=13.6$~\TeV collected between 2022 and 2023, recorded with the ATLAS detector at the LHC.

The presented \ttHH\ search targets the \OneL, \SSML, and \bbyy\ final states. The background contributions are estimated through a combination of simulation and data-driven techniques. Machine-learning-based discriminants are employed to improve separation between signal and background.

No significant excess above the background expectation is observed in any channel. A simultaneous fit across all the channels and analysis regions yields a measured signal-strength value of $\mu_{\ttHH}=-3^{+11}_{-12}$, consistent with the SM prediction, and results in an observed upper limit of $\mu_{\ttHH}<20$ at 95\% CL.
The results are further interpreted within the Higgs effective field theory framework, leading to a 95\% CL interval on the \ttHH\ quartic interaction parameter of $-3.9<\cttHH<3.3$.


%

%
%
%
%
%
\section*{Acknowledgements}
%

%

%
%

%
%

We thank CERN for the very successful operation of the LHC and its injectors, as well as the support staff at
CERN and at our institutions worldwide without whom ATLAS could not be operated efficiently.

The crucial computing support from all WLCG partners is acknowledged gratefully, in particular from CERN, the ATLAS Tier-1 facilities at TRIUMF/SFU (Canada), NDGF (Denmark, Norway, Sweden), CC-IN2P3 (France), KIT/GridKA (Germany), INFN-CNAF (Italy), NL-T1 (Netherlands), PIC (Spain), RAL (UK) and BNL (USA), the Tier-2 facilities worldwide and large non-WLCG resource providers. Major contributors of computing resources are listed in Ref.~\cite{ATL-SOFT-PUB-2026-001}.

We gratefully acknowledge the support of ANPCyT, Argentina; YerPhI, Armenia; ARC, Australia; BMWFW and FWF, Austria; ANAS, Azerbaijan; CNPq and FAPESP, Brazil; NSERC, NRC and CFI, Canada; CERN; ANID, Chile; CAS, MOST and NSFC, China; Minciencias, Colombia; MEYS CR, Czech Republic; DNRF and DNSRC, Denmark; IN2P3-CNRS and CEA-DRF/IRFU, France; SRNSFG, Georgia; BMFTR, HGF and MPG, Germany; GSRI, Greece; RGC and Hong Kong SAR, China; ICHEP and Academy of Sciences and Humanities, Israel; INFN, Italy; MEXT and JSPS, Japan; CNRST, Morocco; NWO, Netherlands; RCN, Norway; MNiSW, Poland; FCT, Portugal; MNE/IFA, Romania; MSTDI, Serbia; MSSR, Slovakia; ARIS and MVZI, Slovenia; DSI/NRF, South Africa; MICIU/AEI, Spain; SRC and Wallenberg Foundation, Sweden; SERI, SNSF and Cantons of Bern and Geneva, Switzerland; NSTC, Taipei; TENMAK, T\"urkiye; STFC/UKRI, United Kingdom; DOE and NSF, United States of America.

Individual groups and members have received support from BCKDF, CANARIE, CRC and DRAC, Canada; CERN-CZ, FORTE and PRIMUS, Czech Republic; COST, ERC, ERDF, Horizon 2020 and Marie Sk{\l}odowska-Curie Actions, European Union; Investissements d'Avenir Labex, Investissements d'Avenir Idex and ANR, France; DFG and AvH Foundation, Germany; Herakleitos, Thales and Aristeia programmes co-financed by EU-ESF and the Greek NSRF, Greece; BSF-NSF and MINERVA, Israel; NCN and NAWA, Poland; La Caixa Banking Foundation, CERCA and AGAUR programs from Generalitat de Catalunya and PROMETEO and GenT Programmes Generalitat Valenciana, Spain; G\"{o}ran Gustafssons Stiftelse, Sweden; The Royal Society and Leverhulme Trust, United Kingdom; Eric and Wendy Schmidt Fund for Strategic Innovation, United States of America.

In addition, individual members wish to acknowledge support from Chile: Agencia Nacional de Investigaci\'on y Desarrollo (ANID FONDECYT reg. 1230987, FONDECYT 1230812, FONDECYT 1240864, Fondecyt 3240661, Fondecyt Regular 1240721); China: Chinese Ministry of Science and Technology (MOST-2023YFA1605700, MOST-2023YFA1609300), National Natural Science Foundation of China (NSFC 12275265, NSFC-W2543005); Czech Republic: Czech Science Foundation (GACR - 24-11373S), Ministry of Education Youth and Sports (ERC-CZ-LL2327, FORTE CZ.02.01.01/00/22\_008/0004632), PRIMUS Research Programme (PRIMUS/21/SCI/017); EU: H2020 European Research Council (ERC - 101002463); European Union: European Research Council (BARD No. 101116429, ERC - 101219398, ERC - 948254, ERC 101089007), European Regional Development Fund (HE COFUND GA No.101081355, ERDF), Marie Sklodowska-Curie Actions (GAP-101168829); France: Agence Nationale de la Recherche (ANR-21-CE31-0013, ANR-22-EDIR-0002, ANR-24-CE31-0504-01); Germany: Deutsche Forschungsgemeinschaft (DFG - 469666862); China: Research Grants Council (GRF); Italy: Ministero dell'Università e della Ricerca (NextGenEU 153D23001490006 M4C2.1.1, NextGenEU I53D23000820006 M4C2.1.1, NextGenEU I53D23001490006 M4C2.1.1, SOE2024\_0000023); Japan: Japan Society for the Promotion of Science (JSPS KAKENHI  JP25H0063, JSPS KAKENHI JP22H01227, JSPS KAKENHI JP22H04944, JSPS KAKENHI JP22KK0227, JSPS KAKENHI JP24K23939, JSPS KAKENHI JP24KK0251, JSPS KAKENHI JP25H00650, JSPS KAKENHI JP25H01291, JSPS KAKENHI JP25K01011, JSPS KAKENHI JP25K01023); Norway: Research Council of Norway (RCN-314472); Poland: Polish National Science Centre (NCN 2021/42/E/ST2/00350, NCN OPUS 2023/51/B/ST2/02507, NCN OPUS nr 2022/47/B/ST2/03059, NCN UMO-2019/34/E/ST2/00393, UMO-2022/47/O/ST2/00148, UMO-2023/49/B/ST2/04085, UMO-2023/51/B/ST2/00920, UMO-2024/53/N/ST2/00869); Spain: Agència de Gestió d'Ajuts Universitaris i de Recerca. (AGAUR - 2023 BP 00141), Ministry of Science and Innovation (RYC2019-028510-I, RYC2020-030254-I, RYC2021-031273-I, RYC2022-038164-I), Ministerio de Ciencia, Innovación y Universidades/Agencia Estatal de Investigaci\'on (EU NextGenerationEU (PRTR-C17.I1), PID2022-142604OB-C22); Sweden: Carl Trygger Foundation (Carl Trygger Foundation CTS 22:2312), Swedish Research Council (Swedish Research Council 2023-04654, VR 2021-03651, VR 2022-03845, VR 2022-04683, VR 2023-03403, VR 2024-05451, VR 2025-05940), Knut and Alice Wallenberg Foundation (KAW 2023.0366); Switzerland: Swiss National Science Foundation (SNSF - PCEFP2\_194658); United Kingdom: The Binks Trust, Royal Society (NIF-R1-231091); United States of America: U.S. Department of Energy (ECA DE-AC02-76SF00515), John Templeton Foundation (John Templeton Foundation 63206), Neubauer Family Foundation.

%
%


%
\clearpage
\printbibliography

@article{Englert:1964et,
    author = "Englert, F. and Brout, R.",
    editor = "Taylor, J. C.",
    title = "{Broken Symmetry and the Mass of Gauge Vector Mesons}",
    doi = "10.1103/PhysRevLett.13.321",
    journal = "Phys. Rev. Lett.",
    volume = "13",
    pages = "321--323",
    year = "1964"
}

@article{Higgs:1964pj,
    author = "Higgs, Peter W.",
    editor = "Taylor, J. C.",
    title = "{Broken Symmetries and the Masses of Gauge Bosons}",
    doi = "10.1103/PhysRevLett.13.508",
    journal = "Phys. Rev. Lett.",
    volume = "13",
    pages = "508--509",
    year = "1964"
}

@article{Guralnik:1964eu,
    author = "Guralnik, G. S. and Hagen, C. R. and Kibble, T. W. B.",
    editor = "Taylor, J. C.",
    title = "{Global Conservation Laws and Massless Particles}",
    doi = "10.1103/PhysRevLett.13.585",
    journal = "Phys. Rev. Lett.",
    volume = "13",
    pages = "585--587",
    year = "1964"
}

@article{Arzt:1994gp,
    author = "Arzt, C. and Einhorn, M. B. and Wudka, J.",
    title = "{Patterns of deviation from the standard model}",
    eprint = "hep-ph/9405214",
    archivePrefix = "arXiv",
    reportNumber = "UM-TH-94-15, UCRHEP-125, CALT-68-1932",
    doi = "10.1016/0550-3213(94)00336-D",
    journal = "Nucl. Phys. B",
    volume = "433",
    pages = "41--66",
    year = "1995"
}

@article{PhysRev.166.1568,
  title = {Nonlinear Realizations of Chiral Symmetry},
  author = {Weinberg, Steven},
  journal = {Phys. Rev.},
  volume = {166},
  issue = {5},
  pages = {1568--1577},
  numpages = {0},
  year = {1968},
  month = feb,
  publisher = {American Physical Society},
  doi = {10.1103/PhysRev.166.1568},
}

@article{Feruglio:1992wf,
    author = "Feruglio, F.",
    title = "{The Chiral approach to the electroweak interactions}",
    eprint = "hep-ph/9301281",
    archivePrefix = "arXiv",
    reportNumber = "DFPD-92-TH-50",
    doi = "10.1142/S0217751X93001946",
    journal = "Int. J. Mod. Phys. A",
    volume = "8",
    pages = "4937--4972",
    year = "1993"
}

@article{Buchalla:2012qq,
    author = "Buchalla, Gerhard and Cat\`a, Oscar",
    title = "{Effective Theory of a Dynamically Broken Electroweak Standard Model at NLO}",
    eprint = "1203.6510",
    archivePrefix = "arXiv",
    primaryClass = "hep-ph",
    reportNumber = "LMU-ASC-19-12, FLAVOUR-267104-ERC-9, LMU-ASC\textasciitilde{}19-12",
    doi = "10.1007/JHEP07(2012)101",
    journal = "JHEP",
    volume = "07",
    pages = "101",
    year = "2012"
}

@article{ALONSO2013926,
    author = "Alonso, R. and Gavela, M. B. and Merlo, L. and Rigolin, S. and Yepes, J.",
    title = "{The Effective Chiral Lagrangian for a Light Dynamical ''Higgs Particle''}",
    eprint = "1212.3305",
    archivePrefix = "arXiv",
    primaryClass = "hep-ph",
    reportNumber = "FTUAM-12-115, IFT-UAM-CSIC-12-113, CERN-PH-TH-2012-335, DFPD-2012-TH-23",
    doi = "10.1016/j.physletb.2013.04.037",
    journal = "Phys. Lett. B",
    volume = "722",
    pages = "330--335",
    year = "2013",
    note = "[Erratum: Phys.Lett.B 726, 926 (2013)]"
}

@article{Brivio:2013pma,
    author = "Brivio, I. and Corbett, T. and \'Eboli, O. J. P. and Gavela, M. B. and Gonzalez-Fraile, J. and Gonzalez-Garcia, M. C. and Merlo, L. and Rigolin, S.",
    title = "{Disentangling a dynamical Higgs}",
    eprint = "1311.1823",
    archivePrefix = "arXiv",
    primaryClass = "hep-ph",
    reportNumber = "FTUAM-13-32, IFT-UAM-CSIC-13-119, YITP-SB-13-33, DFPD-2013-TH-20",
    doi = "10.1007/JHEP03(2014)024",
    journal = "JHEP",
    volume = "03",
    pages = "024",
    year = "2014"
}

@article{Buchalla_2014,
    author = "Buchalla, Gerhard and Cat{\`a}, Oscar and Krause, Claudius",
    title = "{Complete Electroweak Chiral Lagrangian with a Light Higgs at NLO}",
    eprint = "1307.5017",
    archivePrefix = "arXiv",
    primaryClass = "hep-ph",
    reportNumber = "LMU-ASC-42-13, LMU-ASC{\textasciitilde}42-13",
    doi = "10.1016/j.nuclphysb.2014.01.018",
    journal = "Nucl. Phys. B",
    volume = "880",
    pages = "552--573",
    year = "2014",
    note = "[Erratum: Nucl.Phys.B 913, 475--478 (2016)]"
}

@article{Brivio_2019,
    author = "Brivio, Ilaria and Trott, Michael",
    title = "{The Standard Model as an Effective Field Theory}",
    eprint = "1706.08945",
    archivePrefix = "arXiv",
    primaryClass = "hep-ph",
    doi = "10.1016/j.physrep.2018.11.002",
    journal = "Phys. Rept.",
    volume = "793",
    pages = "1--98",
    year = "2019"
}

@article{Buchalla:2013eza,
    author = "Buchalla, Gerhard and Cata, Oscar and Krause, Claudius",
    title = "{On the Power Counting in Effective Field Theories}",
    eprint = "1312.5624",
    archivePrefix = "arXiv",
    primaryClass = "hep-ph",
    reportNumber = "LMU-ASC-81-13",
    doi = "10.1016/j.physletb.2014.02.015",
    journal = "Phys. Lett. B",
    volume = "731",
    pages = "80--86",
    year = "2013"
}

@article{Gavela:2016bzc,
    author = "Gavela, B. and Jenkins, E. E. and Manohar, A. V. and Merlo, L.",
    title = "{Analysis of General Power Counting Rules in Effective Field Theory}",
    eprint = "1601.07551",
    archivePrefix = "arXiv",
    primaryClass = "hep-ph",
    reportNumber = "CERN-TH-2016-015, FTUAM-16-2, IFT-UAM-CSIC-16-006",
    doi = "10.1140/epjc/s10052-016-4332-1",
    journal = "Eur. Phys. J. C",
    volume = "76",
    number = "9",
    pages = "485",
    year = "2016"
}

@article{Buchmuller:1985jz,
    author = "Buchmüller, W. and Wyler, D.",
    title = "{Effective lagrangian analysis of new interactions and flavour conservation}",
    reportNumber = "CERN-TH-4254/85",
    doi = "10.1016/0550-3213(86)90262-2",
    journal = "Nucl. Phys. B",
    volume = "268",
    pages = "621--653",
    year = "1986"
}

@article{Grzadkowski:2010es,
  title        = "{Dimension-Six Terms in the Standard Model Lagrangian}",
  author       = "Grzadkowski, B. and Iskrzyński, M. and Misiak, M. and Rosiek, J.",
  year         = 2010,
  journal      = "JHEP",
  volume       = 10,
  pages        = "085",
  doi          = "10.1007/JHEP10(2010)085",
  eprint       = "1008.4884",
  archiveprefix = "arXiv",
  primaryclass = "hep-ph",
  reportnumber = "IFT-9-2010, TTP10-35"
}

@article{CMS:2026nuu,
    author = "{ATLAS and CMS collaborations}",
    collaboration = "CMS, ATLAS",
    title = "{Combination of ATLAS and CMS searches for Higgs boson pair production at $\sqrt{s} = 13$ TeV}",
    eprint = "2602.23991",
    archivePrefix = "arXiv",
    primaryClass = "hep-ex",
    reportNumber = "CERN-EP-2026-011",
    month = "2",
    year = "2026"
}

@article{Alloul:2013bka,
    author = "Alloul, Adam and Christensen, Neil D. and Degrande, C{\'e}line and Duhr, Claude and Fuks, Benjamin",
    title = "{FeynRules  2.0 - A complete toolbox for tree-level phenomenology}",
    eprint = "1310.1921",
    archivePrefix = "arXiv",
    primaryClass = "hep-ph",
    reportNumber = "CERN-PH-TH-2013-239, MCNET-13-14, IPPP-13-71, DCPT-13-142, PITT-PACC-1308",
    doi = "10.1016/j.cpc.2014.04.012",
    journal = "Comput. Phys. Commun.",
    volume = "185",
    pages = "2250--2300",
    year = "2014"
}

@article{Degrande:2011ua,
  title        = "{UFO - The Universal FeynRules Output}",
  author       = "Degrande, Celine and Duhr, Claude and Fuks, Benjamin and Grellscheid, David and Mattelaer, Olivier and Reiter, Thomas",
  year         = 2012,
  journal      = "Comput. Phys. Commun.",
  volume       = 183,
  pages        = "1201--1214",
  doi          = "10.1016/j.cpc.2012.01.022",
  eprint       = "1108.2040",
  archiveprefix = "arXiv",
  primaryclass = "hep-ph",
  reportnumber = "CP3-11-25, IPHC-PHENO-11-04, IPPP-11-39, DCPT-11-78, MPP-2011-68"
}

@article{Vaswani:2017lxt,
    author = "Vaswani, Ashish and Shazeer, Noam and Parmar, Niki and Uszkoreit, Jakob and Jones, Llion and Gomez, Aidan N. and Kaiser, Lukasz and Polosukhin, Illia",
    title = "{Attention Is All You Need}",
    booktitle = "{31st International Conference on Neural Information Processing Systems}",
    eprint = "1706.03762",
    archivePrefix = "arXiv",
    primaryClass = "cs.CL",
    month = "6",
    year = "2023"
}

@article{Frederix:2014hta,
    author = "Frederix, R. and Frixione, S. and Hirschi, V. and Maltoni, F. and Mattelaer, O. and Torrielli, P. and Vryonidou, E. and Zaro, M.",
    title = "{Higgs pair production at the LHC with NLO and parton-shower effects}",
    eprint = "1401.7340",
    archivePrefix = "arXiv",
    primaryClass = "hep-ph",
    reportNumber = "CERN-PH-TH-2014-014, CP3-14-07, ZU-TH03-14, LPN14-055",
    doi = "10.1016/j.physletb.2014.03.026",
    journal = "Phys. Lett. B",
    volume = "732",
    pages = "142--149",
    year = "2014"
}

@article{Chen:2016btl,
    author = "Chen, Tianqi and Guestrin, Carlos",
    title = "{XGBoost: A Scalable Tree Boosting System}",
    eprint = "1603.02754",
    archivePrefix = "arXiv",
    primaryClass = "cs.LG",
    doi = "10.1145/2939672.2939785",
    month = "3",
    year = "2016"
}

@article{Akiba:2019lwq,
    author = "Akiba, Takuya and Sano, Shotaro and Yanase, Toshihiko and Ohta, Takeru and Koyama, Masanori",
    title = "{Optuna: A Next-generation Hyperparameter Optimization Framework}",
    eprint = "1907.10902",
    archivePrefix = "arXiv",
    primaryClass = "cs.LG",
    month = "7",
    year = "2019"
}

@article{Buonocore:2023ljm,
    author = "Buonocore, Luca and Devoto, Simone and Grazzini, Massimiliano and Kallweit, Stefan and Mazzitelli, Javier and Rottoli, Luca and Savoini, Chiara",
    title = "{Precise Predictions for the Associated Production of a W Boson with a Top-Antitop Quark Pair at the LHC}",
    eprint = "2306.16311",
    archivePrefix = "arXiv",
    primaryClass = "hep-ph",
    reportNumber = "ZU-TH 27/23, TIF-UNIMI-2023-12, PSI-PR-23-19",
    doi = "10.1103/PhysRevLett.131.231901",
    journal = "Phys. Rev. Lett.",
    volume = "131",
    number = "23",
    pages = "231901",
    year = "2023"
}

@article{Gutschow:2018tuk,
    author = {G{\"u}tschow, Christian and Lindert, Jonas M. and Sch{\"o}nherr, Marek},
    title = "{Multi-jet merged top-pair production including electroweak corrections}",
    eprint = "1803.00950",
    archivePrefix = "arXiv",
    primaryClass = "hep-ph",
    reportNumber = "CERN-TH-18-039, CERN-TH-2018-039, IPPP/18/13, MCnet-18-05, IPPP-18-13, MCNET-18-05",
    doi = "10.1140/epjc/s10052-018-5804-2",
    journal = "Eur. Phys. J. C",
    volume = "78",
    number = "4",
    pages = "317",
    year = "2018"
}

@software{atlas_collaboration_2025_15927813,
  author       = "{ATLAS Collaboration}",
  title        = {Easyjet analysis framework},
  month        = jul,
  year         = 2025,
  publisher    = {Zenodo},
  version      = {1.0.0},
  doi          = {10.5281/zenodo.15927813},
  url          = {https://doi.org/10.5281/zenodo.15927813},
}

@software{aly_2025_16684099,
  author       = {Aly, Mohamed and
                  Dado, Tomas and
                  Gagnon, Louis-Guillaume and
                  Held, Alexander and
                  Pinamonti, Michele and
                  Valery, Loic},
  title        = {TRExFitter},
  month        = aug,
  year         = 2025,
  publisher    = {Zenodo},
  version      = {1.4.0},
  doi          = {10.5281/zenodo.16684099},
  url          = {https://doi.org/10.5281/zenodo.16684099},
  swhid        = {swh:1:dir:5d3bdfc58b99346f7867229a4e0e07b3299f1c3c
                   ;origin=https://doi.org/10.5281/zenodo.14845712;vi
                   sit=swh:1:snp:5348b16dafbb6dceff0d8f18c69a63487d24
                   7d38;anchor=swh:1:rel:e696a1cb42dbab33d289fca9b41d
                   f54274104385;path=/
                  },
}

@inproceedings{Porteboeuf:2010um,
    author = "Porteboeuf, S. and Pierog, T. and Werner, K.",
    title = "{Producing Hard Processes Regarding the Complete Event: The EPOS Event Generator}",
    booktitle = "{45th Rencontres de Moriond on QCD and High Energy Interactions}",
    eprint = "1006.2967",
    archivePrefix = "arXiv",
    primaryClass = "hep-ph",
    publisher = "Gioi Publishers",
    pages = "135--140",
    year = "2010"
}

@article{Beenakker:2002nc,
    author = "Beenakker, W. and Dittmaier, S. and Kramer, M. and Plumper, B. and Spira, M. and Zerwas, P. M.",
    title = "{NLO QCD corrections to $t\bar{t}H$ production in hadron collisions}",
    eprint = "hep-ph/0211352",
    archivePrefix = "arXiv",
    reportNumber = "DESY-02-177, EDINBURGH-2002-18, MPI-PHT-2002-70, PSI-PR-02-22",
    doi = "10.1016/S0550-3213(03)00044-0",
    journal = "Nucl. Phys. B",
    volume = "653",
    pages = "151--203",
    year = "2003"
}

@article{Dawson:2003zu,
    author = "Dawson, S. and Jackson, C. and Orr, L. H. and Reina, L. and Wackeroth, D.",
    title = "{Associated Higgs boson production with top quarks at the CERN Large Hadron Collider: NLO QCD corrections}",
    eprint = "hep-ph/0305087",
    archivePrefix = "arXiv",
    reportNumber = "BNL-HET-03-9, FSU-HEP-2003-0503, UB-HET-03-02",
    doi = "10.1103/PhysRevD.68.034022",
    journal = "Phys. Rev. D",
    volume = "68",
    pages = "034022",
    year = "2003"
}

@article{Zhang:2014gcy,
    author = "Zhang, Yu and Ma, Wen-Gan and Zhang, Ren-You and Chen, Chong and Guo, Lei",
    title = "{QCD NLO and EW NLO corrections to $t\bar{t}H$ production with top quark decays at hadron collider}",
    eprint = "1407.1110",
    archivePrefix = "arXiv",
    primaryClass = "hep-ph",
    doi = "10.1016/j.physletb.2014.09.022",
    journal = "Phys. Lett. B",
    volume = "738",
    pages = "1--5",
    year = "2014"
}

@article{Frixione:2014qaa,
    author = "Frixione, S. and Hirschi, V. and Pagani, D. and Shao, H.-S. and Zaro, M.",
    title = "{Weak corrections to Higgs hadroproduction in association with a top-quark pair}",
    eprint = "1407.0823",
    archivePrefix = "arXiv",
    primaryClass = "hep-ph",
    reportNumber = "CERN-PH-TH-2014-123, CP3-14-49",
    doi = "10.1007/JHEP09(2014)065",
    journal = "JHEP",
    volume = "09",
    pages = "065",
    year = "2014"
}

@article{Frixione:2015zaa,
    author = "Frixione, S. and Hirschi, V. and Pagani, D. and Shao, H. -S. and Zaro, M.",
    title = "{Electroweak and QCD corrections to top-pair hadroproduction in association with heavy bosons}",
    eprint = "1504.03446",
    archivePrefix = "arXiv",
    primaryClass = "hep-ph",
    reportNumber = "CERN-PH-TH-2015-083, CP3-15-10, LPN15-022, SLAC-PUB-16253",
    doi = "10.1007/JHEP06(2015)184",
    journal = "JHEP",
    volume = "06",
    pages = "184",
    year = "2015"
}

@article{Brivio:2025yrr,
    author = {Brivio, Ilaria and Gr{\"o}ber, Ramona and Schmid, Konstantin},
    title = "{The Art of Counting: a reappraisal of the HEFT expansion}",
    eprint = "2511.23410",
    archivePrefix = "arXiv",
    primaryClass = "hep-ph",
    reportNumber = "COMETA-2025-51",
    doi = "10.1007/JHEP04(2026)202",
    journal = "JHEP",
    volume = "04",
    pages = "202",
    year = "2026"
}

@Article{Cacciari:2008gp,
     author    = "Cacciari, Matteo and Salam, Gavin P. and Soyez, Gregory",
     title     = "{The anti-\(k_{t}\) jet clustering algorithm}",
     journal   = "JHEP",
     volume    = "04",
     year      = "2008",
     pages     = "063",
     eprint    = "0802.1189",
     archivePrefix = "arXiv",
     primaryClass  =  "hep-ph",
     doi       = "10.1088/1126-6708/2008/04/063",
     SLACcitation  = "%%CITATION = 0802.1189;%%"
}

@Article{Butterworth:2015oua,
      author         = "Butterworth, Jon and others",
      title          = "{PDF4LHC recommendations for LHC Run II}",
      journal        = "J. Phys. G",
      volume         = "43",
      year           = "2016",
      pages          = "023001",
      doi            = "10.1088/0954-3899/43/2/023001",
      eprint         = "1510.03865",
      archivePrefix  = "arXiv",
      primaryClass   = "hep-ph",
      reportNumber   = "OUTP-15-17P, SMU-HEP-15-12, TIF-UNIMI-2015-14,
                        LCTS-2015-27, CERN-PH-TH-2015-249",
      SLACcitation   = "%%CITATION = ARXIV:1510.03865;%%"
}

@article{Harland-Lang:2014zoa,
      author         = "Harland-Lang, L. A. and Martin, A. D. and Motylinski, P.
                        and Thorne, R. S.",
      title          = "{Parton distributions in the LHC era: MMHT 2014 PDFs}",
      journal        = "Eur. Phys. J. C",
      number         = "5",
      volume         = "75",
      pages          = "204",
      doi            = "10.1140/epjc/s10052-015-3397-6",
      year           = "2015",
      eprint         = "1412.3989",
      archivePrefix  = "arXiv",
      primaryClass   = "hep-ph",
      reportNumber   = "LCTS-2014-47, IPPP-14-97, DCPT-14-194",
      SLACcitation   = "%%CITATION = ARXIV:1412.3989;%%",
}

@article{Ball:2012cx,
      author         = "{NNPDF Collaboration} and Ball, Richard D. and others",
      title          = "{Parton distributions with LHC data}",
      journal        = "Nucl. Phys. B",
      volume         = "867",
      year           = "2013",
      pages          = "244",
      doi            = "10.1016/j.nuclphysb.2012.10.003",
      eprint         = "1207.1303",
      archivePrefix  = "arXiv",
      primaryClass   = "hep-ph",
      reportNumber   = "EDINBURGH-2012-08, IFUM-FT-997, FR-PHENO-2012-014,
                        RWTH-TTK-12-25, CERN-PH-TH-2012-037, SFB-CPP-12-47\,
                        --CERN-PH-TH-2012-037",
      SLACcitation   = "%%CITATION = ARXIV:1207.1303;%%"
}

@article{Ball:2014uwa,
      author         = "{NNPDF Collaboration} and Ball, Richard D. and others",
      title          = "{Parton distributions for the LHC run II}",
      journal        = "JHEP",
      volume         = "04",
      year           = "2015",
      pages          = "040",
      doi            = "10.1007/JHEP04(2015)040",
      eprint         = "1410.8849",
      archivePrefix  = "arXiv",
      primaryClass   = "hep-ph",
      reportNumber   = "EDINBURGH-2014-15, IFUM-1034-FT, CERN-PH-TH-2013-253,
                        OUTP-14-11P, CAVENDISH-HEP-14-11",
      SLACcitation   = "%%CITATION = ARXIV:1410.8849;%%"
}

@article{Sjostrand:2014zea,
      author         = "Sj{\"o}strand, Torbj{\"o}rn and Ask, Stefan and Christiansen,
                        Jesper R. and Corke, Richard and Desai, Nishita and Ilten,
                        Philip and Mrenna, Stephen and Prestel, Stefan and
                        Rasmussen, Christine O. and Skands, Peter Z.",
      title          = "{An introduction to PYTHIA 8.2}",
      journal        = "Comput. Phys. Commun.",
      volume         = "191",
      year           = "2015",
      pages          = "159",
      doi            = "10.1016/j.cpc.2015.01.024",
      eprint         = "1410.3012",
      archivePrefix  = "arXiv",
      primaryClass   = "hep-ph",
      reportNumber   = "LU-TP-14-36, MCNET-14-22, CERN-PH-TH-2014-190,
                        FERMILAB-PUB-14-316-CD, DESY-14-178, SLAC-PUB-16122,
                        --FERMILAB-PUB-14-316-CD",
      SLACcitation   = "%%CITATION = ARXIV:1410.3012;%%"
}

@Article{Lange:2001uf,
      author         = "Lange, D. J.",
      title          = "{The EvtGen particle decay simulation package}",
      booktitle      = "{Proceedings, 7th International Conference on B physics
                        at hadron machines (BEAUTY 2000)}",
      journal        = "Nucl. Instrum. Meth. A",
      volume         = "462",
      year           = "2001",
      pages          = "152",
      doi            = "10.1016/S0168-9002(01)00089-4",
      SLACcitation   = "%%CITATION = NUIMA,A462,152;%%"
}

@Article{Alwall:2011uj,
      author         = "Alwall, Johan and Herquet, Michel and Maltoni, Fabio and
                        Mattelaer, Olivier and Stelzer, Tim",
      title          = "{MadGraph 5: going beyond}",
      journal        = "JHEP",
      volume         = "06",
      pages          = "128",
      doi            = "10.1007/JHEP06(2011)128",
      year           = "2011",
      eprint         = "1106.0522",
      archivePrefix  = "arXiv",
      primaryClass   = "hep-ph"
}

@Article{Alwall:2014hca,
      author         = "Alwall, J. and Frederix, R. and Frixione, S. and Hirschi,
                        V. and Maltoni, F. and Mattelaer, O. and Shao, H. -S. and
                        Stelzer, T. and Torrielli, P. and Zaro, M.",
      title          = "{The automated computation of tree-level and
                        next-to-leading order differential cross sections, and
                        their matching to parton shower simulations}",
      journal        = "JHEP",
      volume         = "07",
      year           = "2014",
      pages          = "079",
      doi            = "10.1007/JHEP07(2014)079",
      eprint         = "1405.0301",
      archivePrefix  = "arXiv",
      primaryClass   = "hep-ph",
      reportNumber   = "CERN-PH-TH-2014-064, CP3-14-18, LPN14-066, MCNET-14-09,
                        ZU-TH-14-14",
      SLACcitation   = "%%CITATION = ARXIV:1405.0301;%%"
}

@article{Artoisenet:2012st,
      author         = "Artoisenet, Pierre and Frederix, Rikkert and Mattelaer,
                        Olivier and Rietkerk, Robbert",
      title          = "{Automatic spin-entangled decays of heavy resonances in
                        Monte Carlo simulations}",
      journal        = "JHEP",
      volume         = "03",
      pages          = "015",
      doi            = "10.1007/JHEP03(2013)015",
      year           = "2013",
      eprint         = "1212.3460",
      archivePrefix  = "arXiv",
      primaryClass   = "hep-ph",
      reportNumber   = "NIKHEF-2012-021, CERN-PH-TH-2012-329",
      SLACcitation   = "%%CITATION = ARXIV:1212.3460;%%",
}

@article{Frixione:2007zp,
      author         = "Frixione, Stefano and Laenen, Eric and Motylinski,
                        Patrick and Webber, Bryan R.",
      title          = "{Angular correlations of lepton pairs from vector boson
                        and top quark decays in Monte Carlo simulations}",
      journal        = "JHEP",
      volume         = "04",
      year           = "2007",
      pages          = "081",
      doi            = "10.1088/1126-6708/2007/04/081",
      eprint         = "hep-ph/0702198",
      archivePrefix  = "arXiv",
      reportNumber   = "CAVENDISH-HEP-07-01, GEF-TH-09-2007, ITP-UU-07-10,
                        NIKHEF-2007-004",
      SLACcitation   = "%%CITATION = HEP-PH/0702198;%%"
}

@Article{Bahr:2008pv,
      author         = "B{\"a}hr, M. and others",
      title          = "{Herwig++ physics and manual}",
      journal        = "Eur. Phys. J. C",
      volume         = "58",
      year           = "2008",
      pages          = "639",
      doi            = "10.1140/epjc/s10052-008-0798-9",
      eprint         = "0803.0883",
      archivePrefix  = "arXiv",
      primaryClass   = "hep-ph",
      reportNumber   = "CERN-PH-TH-2008-038, CAVENDISH-HEP-08-03, KA-TP-05-2008,
                        DCPT-08-22, IPPP-08-11, CP3-08-05",
      SLACcitation   = "%%CITATION = ARXIV:0803.0883;%%"
}

@article{Bellm:2017jjp,
      author         = "Bellm, Johannes and others",
      title          = "{Herwig 7.1 Release Note}",
      eprint         = "1705.06919",
      year           = "2017",
      archivePrefix  = "arXiv",
      primaryClass   = "hep-ph",
      reportNumber   = "CERN-PH-TH-2017-109, MAN/HEP/2017/08, UWTHPH-2017-10, 
                        IFJPAN-IV-2017-7, NIKHEF 2017-026, HERWIG-2017-02, 
                        KA-TP-19-2017, MCnet-17-08, IPPP/17/40",
      SLACcitation   = "%%CITATION = ARXIV:1705.06919;%%"
}

@Article{Bellm:2019zci,
    author = "Bellm, Johannes and others",
    title = "{Herwig 7.2 release note}",
    eprint = "1912.06509",
    archivePrefix = "arXiv",
    primaryClass = "hep-ph",
    reportNumber = "MAN/HEP/2019/011, CERN-TH-2019-213, IFJPAN-IV-2019-18, HERWIG-2019-02, UWTHPH-19-36, KA-TP-24-2019, LU-TP 19-57, MCnet-19-28, IPPP/19/91",
    doi = "10.1140/epjc/s10052-020-8011-x",
    journal = "Eur. Phys. J. C",
    volume = "80",
    pages = "452",
    year = "2020"
}

@Article{Nason:2004rx,
      author         = "Nason, Paolo",
      title          = "{A new method for combining NLO QCD with shower Monte Carlo algorithms}",
      journal        = "JHEP",
      volume         = "11",
      pages          = "040",
      doi            = "10.1088/1126-6708/2004/11/040",
      year           = "2004",
      eprint         = "hep-ph/0409146",
      archivePrefix  = "arXiv",
}

@Article{Frixione:2007vw,
      author         = "Frixione, Stefano and Nason, Paolo and Oleari, Carlo",
      title          = "{Matching NLO QCD computations with parton shower
                        simulations: the POWHEG method}",
      journal        = "JHEP",
      volume         = "11",
      pages          = "070",
      doi            = "10.1088/1126-6708/2007/11/070",
      year           = "2007",
      eprint         = "0709.2092",
      archivePrefix  = "arXiv",
      primaryClass   = "hep-ph",
}

@Article{Alioli:2010xd,
      author         = "Alioli, Simone and Nason, Paolo and Oleari, Carlo and Re,
                        Emanuele",
      title          = "{A general framework for implementing NLO calculations in
                        shower Monte Carlo programs: the POWHEG BOX}",
      journal        = "JHEP",
      volume         = "06",
      pages          = "043",
      doi            = "10.1007/JHEP06(2010)043",
      year           = "2010",
      eprint         = "1002.2581",
      archivePrefix  = "arXiv",
      primaryClass   = "hep-ph",
}

@article{Alioli:2009je,
      author         = "Alioli, Simone and Nason, Paolo and Oleari, Carlo and Re,
                        Emanuele",
      title          = "{NLO single-top production matched with shower in POWHEG:
                        \(s\)- and \(t\)-channel contributions}",
      journal        = "JHEP",
      volume         = "09",
      year           = "2009",
      pages          = "111",
      doi            = "10.1088/1126-6708/2009/09/111",
      eprint         = "0907.4076",
      archivePrefix  = "arXiv",
      primaryClass   = "hep-ph",
      SLACcitation   = "%%CITATION = ARXIV:0907.4076;%%",
      related        = "Alioli:2009je-err",
      relatedstring  = "Erratum:",
}

@article{Re:2010bp,
      author         = "Re, Emanuele",
      title          = "{Single-top \(Wt\)-channel production matched with parton
                        showers using the POWHEG method}",
      journal        = "Eur. Phys. J. C",
      volume         = "71",
      year           = "2011",
      pages          = "1547",
      doi            = "10.1140/epjc/s10052-011-1547-z",
      eprint         = "1009.2450",
      archivePrefix  = "arXiv",
      primaryClass   = "hep-ph",
      reportNumber   = "IPPP-10-74, DCPT-10-148",
      SLACcitation   = "%%CITATION = ARXIV:1009.2450;%%"
}

@article{Frixione:2007nw,
      author         = "Frixione, Stefano and Ridolfi, Giovanni and Nason, Paolo",
      title          = "{A positive-weight next-to-leading-order Monte Carlo for
                        heavy flavour hadroproduction}",
      journal        = "JHEP",
      volume         = "09",
      pages          = "126",
      doi            = "10.1088/1126-6708/2007/09/126",
      year           = "2007",
      eprint         = "0707.3088",
      archivePrefix  = "arXiv",
      primaryClass   = "hep-ph",
      SLACcitation   = "%%CITATION = ARXIV:0707.3088;%%",
}

@article{Frederix:2017wme,
      author         = "Frederix, Rikkert and Pagani, Davide and Zaro, Marco",
      title          = "{Large NLO corrections in \(t\bar{t}W^{\pm}\) and
                        \(t\bar{t}t\bar{t}\) hadroproduction from supposedly
                        subleading EW contributions}",
      journal        = "JHEP",
      volume         = "02",
      year           = "2018",
      pages          = "031",
      doi            = "10.1007/JHEP02(2018)031",
      eprint         = "1711.02116",
      archivePrefix  = "arXiv",
      primaryClass   = "hep-ph",
      SLACcitation   = "%%CITATION = ARXIV:1711.02116;%%"
}

@article{deFlorian:2016spz,
      author         = "de Florian, D. and others",
      title          = "{Handbook of LHC Higgs Cross Sections: 4. Deciphering the Nature of the Higgs Sector}",
      collaboration  = "LHC Higgs Cross Section Working Group",
      doi            = "10.23731/CYRM-2017-002",
      year           = "2017",
      eprint         = "1610.07922",
      archivePrefix  = "arXiv",
      primaryClass   = "hep-ph",
      reportNumber   = "FERMILAB-FN-1025-T, CERN-2017-002-M",
      SLACcitation   = "%%CITATION = ARXIV:1610.07922;%%"
}

@article{Hoeche:2012yf,
      author         = "H{\"o}che, Stefan and Krauss, Frank and Sch{\"o}nherr, Marek and
                        Siegert, Frank",
      title          = "{QCD matrix elements + parton showers. The NLO case}",
      journal        = "JHEP",
      volume         = "04",
      year           = "2013",
      pages          = "027",
      doi            = "10.1007/JHEP04(2013)027",
      eprint         = "1207.5030",
      archivePrefix  = "arXiv",
      primaryClass   = "hep-ph",
      reportNumber   = "SLAC-PUB-15191, IPPP-12-52, DCPT-12-104, LPN12-081,
                        FR-PHENO-2012-017, MCNET-12-09, --FR-PHENO-2012-017",
      SLACcitation   = "%%CITATION = ARXIV:1207.5030;%%"
}

@article{Bothmann:2019yzt,
      author = "Bothmann, Enrico and others",
      title = "{Event generation with Sherpa 2.2}",
      journal = "SciPost Phys.",
      volume = "7",
      year = "2019",
      number = "3",
      pages = "034",
      doi = "10.21468/SciPostPhys.7.3.034",
      reportNumber = "FERMILAB-PUB-19-218-T, SLAC-PUB-17433, IPPP/19/42, MCNET-19-11",
      eprint = "1905.09127",
      archivePrefix = "arXiv",
      primaryClass = "hep-ph",
}

@article{Cascioli:2011va,
      author         = "Cascioli, Fabio and Maierh{\"o}fer, Philipp and Pozzorini, Stefano",
      title          = "{Scattering Amplitudes with Open Loops}",
      journal        = "Phys. Rev. Lett.",
      volume         = "108",
      year           = "2012",
      pages          = "111601",
      doi            = "10.1103/PhysRevLett.108.111601",
      eprint         = "1111.5206",
      archivePrefix  = "arXiv",
      primaryClass   = "hep-ph",
      reportNumber   = "ZU-TH-23-11, LPN11-66",
      SLACcitation   = "%%CITATION = ARXIV:1111.5206;%%"
}

@article{Buccioni:2019sur,
    author = "Buccioni, Federico and Lang, Jean-Nicolas and Lindert, Jonas M. and Maierh{\"o}fer, Philipp and Pozzorini, Stefano and Zhang, Hantian and Zoller, Max F.",
    title = "{OpenLoops 2}",
    eprint = "1907.13071",
    archivePrefix = "arXiv",
    primaryClass = "hep-ph",
    reportNumber = "IPPP/19/62, FR-PHENO-2019-12, PSI-PR-19-15, ZU-TH 37/19",
    doi = "10.1140/epjc/s10052-019-7306-2",
    journal = "Eur. Phys. J. C",
    volume = "79",
    number = "10",
    pages = "866",
    year = "2019"
}

@article{Denner:2016kdg,
      author         = "Denner, Ansgar and Dittmaier, Stefan and Hofer, Lars",
      title          = "{\textsc{Collier}: A fortran-based complex one-loop library in
                        extended regularizations}",
      journal        = "Comput. Phys. Commun.",
      volume         = "212",
      year           = "2017",
      pages          = "220-238",
      doi            = "10.1016/j.cpc.2016.10.013",
      eprint         = "1604.06792",
      archivePrefix  = "arXiv",
      primaryClass   = "hep-ph",
      reportNumber   = "FR-PHENO-2016-003, ICCUB-16-016",
      SLACcitation   = "%%CITATION = ARXIV:1604.06792;%%"
}

@article{Kallweit:2015dum,
    author = "Kallweit, Stefan and Lindert, Jonas M. and Maierh{\"o}fer, Philipp and Pozzorini, Stefano and Sch{\"o}nherr, Marek",
    title = "{NLO QCD+EW predictions for V + jets including off-shell vector-boson decays and multijet merging}",
    eprint = "1511.08692",
    archivePrefix = "arXiv",
    primaryClass = "hep-ph",
    reportNumber = "DCPT-15-140, FR-PHENO-2015-014, IPPP-15-70, MCNET-15-23, ZU-TH-41-15, MITP-15-108",
    doi = "10.1007/JHEP04(2016)021",
    journal = "JHEP",
    volume = "04",
    pages = "021",
    year = "2016"
}

@article{Czakon:2011xx,
  author         = "Czakon, Michal and Mitov, Alexander",
  title          = "{Top++: A program for the calculation of the top-pair cross-section at hadron colliders}",
  journal        = "Comput. Phys. Commun.",
  volume         = "185",
  year           = "2014",
  pages          = "2930",
  doi            = "10.1016/j.cpc.2014.06.021",
  eprint         = "1112.5675",
  archivePrefix  = "arXiv",
  primaryClass   = "hep-ph",
  reportNumber   = "CERN-PH-TH-2011-303, TTK-11-58",
}

@article{Hamilton:2013fea,
      author         = "Hamilton, Keith and Nason, Paolo and Re, Emanuele and
                        Zanderighi, Giulia",
      title          = "{NNLOPS simulation of Higgs boson production}",
      journal        = "JHEP",
      volume         = "10",
      year           = "2013",
      pages          = "222",
      doi            = "10.1007/JHEP10(2013)222",
      eprint         = "1309.0017",
      archivePrefix  = "arXiv",
      primaryClass   = "hep-ph",
      reportNumber   = "MCNET-13-11, CERN-PH-TH-2013-205, OUTP-13-18P",
      SLACcitation   = "%%CITATION = ARXIV:1309.0017;%%"
}

@article{Hamilton:2015nsa,
      author         = "Hamilton, Keith and Nason, Paolo and Zanderighi, Giulia",
      title          = "{Finite quark-mass effects in the NNLOPS POWHEG+MiNLO
                        Higgs generator}",
      journal        = "JHEP",
      volume         = "05",
      year           = "2015",
      pages          = "140",
      doi            = "10.1007/JHEP05(2015)140",
      eprint         = "1501.04637",
      archivePrefix  = "arXiv",
      primaryClass   = "hep-ph",
      reportNumber   = "CERN-PH-TH-2015-006",
      SLACcitation   = "%%CITATION = ARXIV:1501.04637;%%"
}

@article{Agostinelli:2002hh,
      author         = "Agostinelli, S. and others",
      title          = "{\textsc{Geant4} -- a simulation toolkit}",
      journal        = "Nucl. Instrum. Meth. A",
      volume         = "506",
      year           = "2003",
      pages          = "250",
      doi            = "10.1016/S0168-9002(03)01368-8",
      reportNumber   = "SLAC-PUB-9350, FERMILAB-PUB-03-339",
      SLACcitation   = "%%CITATION = NUIMA,A506,250;%%"
}

@software{TRExFitter,
  author       = {{ATLAS Collaboration}},
  title        = {TRExFitter},
  publisher    = {Zenodo},
  doi          = {10.5281/zenodo.14845712},
  url          = {https://doi.org/10.5281/zenodo.14845712}
}

@Article{Read:2002hq,
    Author = {Read, Alexander L.},
    Journal = {J. Phys. G},
    Pages = {2693},
    doi = "10.1088/0954-3899/28/10/313",
    Title = "{Presentation of search results: the \(CL_s\) technique}",
    Volume = {28},
    Year = {2002}
}

@Article{Cowan:2010js,
   Author = {Cowan, Glen and Cranmer, Kyle and Gross, Eilam and Vitells, Ofer},
   Title = "{Asymptotic formulae for likelihood-based tests of new physics}",
   Journal = {Eur. Phys. J. C},
   Volume = {71},
   Year = {2011},
   Pages = {1554},
   doi = "10.1140/epjc/s10052-011-1554-0",
   Eprint = "1007.1727",
   Archiveprefix = {arXiv},
   Primaryclass = {physics.data-an},
   related = "Cowan:2010js-err",
   relatedstring  = "Erratum:",
}

@Article{PERF-2007-01,
    author         = "{ATLAS Collaboration}",
    title          = "{The ATLAS Experiment at the CERN Large Hadron Collider}",
    journal        = "JINST",
    volume         = "3",
    year           = "2008",
    pages          = "S08003",
    doi            = "10.1088/1748-0221/3/08/S08003",
    primaryClass   = "hep-ex",
}

@Article{SOFT-2010-01,
    author         = "{ATLAS Collaboration}",
    title          = "{The ATLAS Simulation Infrastructure}",
    journal        = "Eur. Phys. J. C",
    volume         = "70",
    year           = "2010",
    pages          = "823",
    doi            = "10.1140/epjc/s10052-010-1429-9",
    eprint         = "1005.4568",
    archivePrefix  = "arXiv",
    primaryClass   = "physics.ins-det",
}

@Article{HIGG-2012-27,
    author         = "{ATLAS Collaboration}",
    title          = "{Observation of a new particle in the search for the Standard Model Higgs boson with the ATLAS detector at the LHC}",
    journal        = "Phys. Lett. B",
    volume         = "716",
    year           = "2012",
    pages          = "1",
    doi            = "10.1016/j.physletb.2012.08.020",
    reportNumber   = "CERN-PH-EP-2012-218",
    eprint         = "1207.7214",
    archivePrefix  = "arXiv",
    primaryClass   = "hep-ex",
}

@Article{HIGG-2014-04,
    author         = "{ATLAS Collaboration}",
    title          = "{Search for Scalar Diphoton Resonances in the Mass Range 65--600\,GeV with the ATLAS Detector in \(pp\) Collision Data at \(\sqrt{s} = 8\,\text{TeV}\)}",
    journal        = "Phys. Rev. Lett.",
    volume         = "113",
    year           = "2014",
    pages          = "171801",
    doi            = "10.1103/PhysRevLett.113.171801",
    reportNumber   = "CERN-PH-EP-2014-142",
    eprint         = "1407.6583",
    archivePrefix  = "arXiv",
    primaryClass   = "hep-ex",
}

@Article{PERF-2014-03,
    author         = "{ATLAS Collaboration}",
    title          = "{Performance of pile-up mitigation techniques for jets in \(pp\) collisions at \(\sqrt{s} = 8\,\text{TeV}\) using the ATLAS detector}",
    journal        = "Eur. Phys. J. C",
    volume         = "76",
    year           = "2016",
    pages          = "581",
    doi            = "10.1140/epjc/s10052-016-4395-z",
    reportNumber   = "CERN-PH-EP-2015-206",
    eprint         = "1510.03823",
    archivePrefix  = "arXiv",
    primaryClass   = "hep-ex",
}

@Article{PERF-2015-09,
    author         = "{ATLAS Collaboration}",
    title          = "{Jet reconstruction and performance using particle flow with the ATLAS Detector}",
    journal        = "Eur. Phys. J. C",
    volume         = "77",
    year           = "2017",
    pages          = "466",
    doi            = "10.1140/epjc/s10052-017-5031-2",
    reportNumber   = "CERN-EP-2017-024",
    eprint         = "1703.10485",
    archivePrefix  = "arXiv",
    primaryClass   = "hep-ex",
}

@Article{PERF-2015-10,
    author         = "{ATLAS Collaboration}",
    title          = "{Muon reconstruction performance of the ATLAS detector in proton--proton collision data at \(\sqrt{s} = 13\,\text{TeV}\)}",
    journal        = "Eur. Phys. J. C",
    volume         = "76",
    year           = "2016",
    pages          = "292",
    doi            = "10.1140/epjc/s10052-016-4120-y",
    reportNumber   = "CERN-EP-2016-033",
    eprint         = "1603.05598",
    archivePrefix  = "arXiv",
    primaryClass   = "hep-ex",
}

@Article{PERF-2016-07,
    author         = "{ATLAS Collaboration}",
    title          = "{Performance of missing transverse momentum reconstruction with the ATLAS detector using proton--proton collisions at \(\sqrt{s} = 13\,\text{TeV}\)}",
    journal        = "Eur. Phys. J. C",
    volume         = "78",
    year           = "2018",
    pages          = "903",
    doi            = "10.1140/epjc/s10052-018-6288-9",
    reportNumber   = "CERN-EP-2017-274",
    eprint         = "1802.08168",
    archivePrefix  = "arXiv",
    primaryClass   = "hep-ex",
}

@Article{TRIG-2016-01,
    author         = "{ATLAS Collaboration}",
    title          = "{Performance of the ATLAS trigger system in 2015}",
    journal        = "Eur. Phys. J. C",
    volume         = "77",
    year           = "2017",
    pages          = "317",
    doi            = "10.1140/epjc/s10052-017-4852-3",
    reportNumber   = "CERN-EP-2016-241",
    eprint         = "1611.09661",
    archivePrefix  = "arXiv",
    primaryClass   = "hep-ex",
}

@Article{TOPQ-2017-12,
    author         = "{ATLAS Collaboration}",
    title          = "{Measurements of inclusive and differential fiducial cross-sections of \(t\bar{t}\) production with additional heavy-flavour jets in proton--proton collisions at \(\sqrt{s} = 13\,\text{TeV}\) with the ATLAS detector}",
    journal        = "JHEP",
    volume         = "04",
    year           = "2019",
    pages          = "046",
    doi            = "10.1007/JHEP04(2019)046",
    reportNumber   = "CERN-EP-2018-276",
    eprint         = "1811.12113",
    archivePrefix  = "arXiv",
    primaryClass   = "hep-ex",
}

@Article{DAPR-2018-01,
    author         = "{ATLAS Collaboration}",
    title          = "{ATLAS data quality operations and performance for 2015--2018 data-taking}",
    journal        = "JINST",
    volume         = "15",
    year           = "2020",
    pages          = "P04003",
    doi            = "10.1088/1748-0221/15/04/P04003",
    reportNumber   = "CERN-EP-2019-207",
    eprint         = "1911.04632",
    archivePrefix  = "arXiv",
    primaryClass   = "physics.ins-det",
}

@Article{EGAM-2018-01,
    author         = "{ATLAS Collaboration}",
    title          = "{Electron and photon performance measurements with the ATLAS detector using the 2015--2017 LHC proton--proton collision data}",
    journal        = "JINST",
    volume         = "14",
    year           = "2019",
    pages          = "P12006",
    doi            = "10.1088/1748-0221/14/12/P12006",
    reportNumber   = "CERN-EP-2019-145",
    eprint         = "1908.00005",
    archivePrefix  = "arXiv",
    primaryClass   = "hep-ex",
}

@Article{JETM-2018-05,
    author         = "{ATLAS Collaboration}",
    title          = "{Jet energy scale and resolution measured in proton--proton collisions at \(\sqrt{s} = 13\,\text{TeV}\) with the ATLAS detector}",
    journal        = "Eur. Phys. J. C",
    volume         = "81",
    year           = "2021",
    pages          = "689",
    doi            = "10.1140/epjc/s10052-021-09402-3",
    reportNumber   = "CERN-EP-2020-083",
    eprint         = "2007.02645",
    archivePrefix  = "arXiv",
    primaryClass   = "hep-ex",
}

@Article{MUON-2018-03,
    author         = "{ATLAS Collaboration}",
    title          = "{Muon reconstruction and identification efficiency in ATLAS using the full Run~2 \(pp\) collision data set at \(\sqrt{s} = 13\,\text{TeV}\)}",
    journal        = "Eur. Phys. J. C",
    volume         = "81",
    year           = "2021",
    pages          = "578",
    doi            = "10.1140/epjc/s10052-021-09233-2",
    reportNumber   = "CERN-EP-2020-199",
    eprint         = "2012.00578",
    archivePrefix  = "arXiv",
    primaryClass   = "hep-ex",
}

@Article{TRIG-2018-01,
    author         = "{ATLAS Collaboration}",
    title          = "{Performance of the ATLAS muon triggers in Run~2}",
    journal        = "JINST",
    volume         = "15",
    year           = "2020",
    pages          = "P09015",
    doi            = "10.1088/1748-0221/15/09/p09015",
    reportNumber   = "CERN-EP-2020-031",
    eprint         = "2004.13447",
    archivePrefix  = "arXiv",
    primaryClass   = "physics.ins-det",
}

@Article{TRIG-2018-05,
    author         = "{ATLAS Collaboration}",
    title          = "{Performance of electron and photon triggers in ATLAS during LHC Run~2}",
    journal        = "Eur. Phys. J. C",
    volume         = "80",
    year           = "2020",
    pages          = "47",
    doi            = "10.1140/epjc/s10052-019-7500-2",
    reportNumber   = "CERN-EP-2019-169",
    eprint         = "1909.00761",
    archivePrefix  = "arXiv",
    primaryClass   = "hep-ex",
}

@Article{EXOT-2019-26,
    author         = "{ATLAS Collaboration}",
    title          = "{Search for \(t\bar{t}H/A \rightarrow t\bar{t} t\bar{t}\) production in the multilepton final state in proton--proton collisions at \(\sqrt{s} = 13\,\text{TeV}\) with the ATLAS detector}",
    journal        = "JHEP",
    volume         = "07",
    year           = "2023",
    pages          = "203",
    doi            = "10.1007/JHEP07(2023)203",
    reportNumber   = "CERN-EP-2022-170",
    eprint         = "2211.01136",
    archivePrefix  = "arXiv",
    primaryClass   = "hep-ex",
    related        = "EXOT-2019-26-err",
    relatedstring  = "Erratum:",
}

@Article{FTAG-2019-02,
    author         = "{ATLAS Collaboration}",
    title          = "{Calibration of the light-flavour jet mistagging efficiency of the \(b\)-tagging algorithms with \(Z\)+jets events using \(139\,\text{fb}^{-1}\) of ATLAS proton--proton collision data at \(\sqrt{s} = 13\,\text{TeV}\)}",
    journal        = "Eur. Phys. J. C",
    volume         = "83",
    year           = "2023",
    pages          = "728",
    doi            = "10.1140/epjc/s10052-023-11736-z",
    reportNumber   = "CERN-EP-2022-211",
    eprint         = "2301.06319",
    archivePrefix  = "arXiv",
    primaryClass   = "hep-ex",
}

@Article{FTAG-2019-07,
    author         = "{ATLAS Collaboration}",
    title          = "{ATLAS flavour-tagging algorithms for the LHC Run~2 \(pp\) collision dataset}",
    journal        = "Eur. Phys. J. C",
    volume         = "83",
    year           = "2023",
    pages          = "681",
    doi            = "10.1140/epjc/s10052-023-11699-1",
    reportNumber   = "CERN-EP-2022-226",
    eprint         = "2211.16345",
    archivePrefix  = "arXiv",
    primaryClass   = "physics.data-an",
}

@Article{GENR-2019-02,
    author         = "{ATLAS Collaboration}",
    title          = "{The ATLAS experiment at the CERN Large Hadron Collider: a description of the detector configuration for Run~3}",
    journal        = "JINST",
    volume         = "19",
    year           = "2024",
    pages          = "P05063",
    doi            = "10.1088/1748-0221/19/05/P05063",
    reportNumber   = "CERN-EP-2022-259",
    eprint         = "2305.16623",
    archivePrefix  = "arXiv",
    primaryClass   = "physics.ins-det",
}

@Article{HDBS-2019-31,
    author         = "{ATLAS Collaboration}",
    title          = "{Search for Higgs boson pair production in association with a vector boson in \(pp\) collisions at \(\sqrt{s} = 13\,\text{TeV}\) with the ATLAS detector}",
    journal        = "Eur. Phys. J. C",
    volume         = "83",
    year           = "2023",
    pages          = "519",
    doi            = "10.1140/epjc/s10052-023-11559-y",
    reportNumber   = "CERN-EP-2022-159",
    eprint         = "2210.05415",
    archivePrefix  = "arXiv",
    primaryClass   = "hep-ex",
}

@Article{TOPQ-2019-03,
    author         = "{ATLAS Collaboration}",
    title          = "{Measurement of \(t\bar{t}\) production in association with additional \(b\)-jets in the \(e\mu\) final state in proton--proton collisions at \(\sqrt{s} = 13\,\text{TeV}\) with the ATLAS detector}",
    journal        = "JHEP",
    volume         = "01",
    year           = "2025",
    pages          = "068",
    doi            = "10.1007/JHEP01(2025)068",
    reportNumber   = "CERN-EP-2024-191",
    eprint         = "2407.13473",
    archivePrefix  = "arXiv",
    primaryClass   = "hep-ex",
}

@Article{TOPQ-2019-30,
    author         = "{ATLAS Collaboration}",
    title          = "{Measurement of the total and differential cross-sections of \(t\bar{t}W\) production in \(pp\) collisions at \(\sqrt{s}=13\,\text{TeV}\) with the ATLAS detector}",
    journal        = "JHEP",
    volume         = "05",
    year           = "2024",
    pages          = "131",
    doi            = "10.1007/JHEP05(2024)131",
    reportNumber   = "CERN-EP-2023-260",
    eprint         = "2401.05299",
    archivePrefix  = "arXiv",
    primaryClass   = "hep-ex",
    related        = "TOPQ-2019-30-err",
    relatedstring  = "Erratum:",
}

@Article{FTAG-2020-08,
    author         = "{ATLAS Collaboration}",
    title          = "{Measurement of the \(c\)-jet mistagging efficiency in \(t\bar{t}\) events using \(pp\) collision data at \(\sqrt{s} = 13\,\text{TeV}\) collected with the ATLAS detector}",
    journal        = "Eur. Phys. J. C",
    volume         = "82",
    year           = "2022",
    pages          = "95",
    doi            = "10.1140/epjc/s10052-021-09843-w",
    reportNumber   = "CERN-EP-2021-166",
    eprint         = "2109.10627",
    archivePrefix  = "arXiv",
    primaryClass   = "hep-ex",
}

@Article{HIGG-2020-24,
    author         = "{ATLAS Collaboration}",
    title          = "{Measurement of the associated production of a top-antitop-quark pair and a Higgs boson decaying into a \(b\bar{b}\) pair in \(pp\) collisions at \(\sqrt{s} = 13\,\text{TeV}\) using the ATLAS detector at the LHC}",
    journal        = "Eur. Phys. J. C",
    volume         = "85",
    year           = "2025",
    pages          = "210",
    doi            = "10.1140/epjc/s10052-025-13740-x",
    reportNumber   = "CERN-EP-2024-194",
    eprint         = "2407.10904",
    archivePrefix  = "arXiv",
    primaryClass   = "hep-ex",
}

@Article{JETM-2020-03,
    author         = "{ATLAS Collaboration}",
    title          = "{The performance of missing transverse momentum reconstruction and its significance with the ATLAS detector using \(140\,\text{fb}^{-1}\) of \(\sqrt{s} = 13\,\text{TeV}\) \(pp\) collisions}",
    journal        = "Eur. Phys. J. C",
    volume         = "85",
    year           = "2025",
    pages          = "606",
    doi            = "10.1140/epjc/s10052-025-14062-8",
    reportNumber   = "CERN-EP-2024-023",
    eprint         = "2402.05858",
    archivePrefix  = "arXiv",
    primaryClass   = "hep-ex",
}

@Article{SIMU-2020-01,
    author         = "{ATLAS Collaboration}",
    title          = "{Emulating the impact of additional proton--proton interactions in the ATLAS simulation by presampling sets of inelastic Monte Carlo events}",
    journal        = "Comput. Softw. Big Sci.",
    volume         = "6",
    year           = "2022",
    pages          = "3",
    doi            = "10.1007/s41781-021-00062-2",
    reportNumber   = "CERN-EP-2021-007",
    eprint         = "2102.09495",
    archivePrefix  = "arXiv",
    primaryClass   = "hep-ex",
}

@Article{DAPR-2021-01,
    author         = "{ATLAS Collaboration}",
    title          = "{Luminosity determination in \(pp\) collisions at \(\sqrt{s} = 13\,\text{TeV}\) using the ATLAS detector at the LHC}",
    journal        = "Eur. Phys. J. C",
    volume         = "83",
    year           = "2023",
    pages          = "982",
    doi            = "10.1140/epjc/s10052-023-11747-w",
    reportNumber   = "CERN-EP-2022-281",
    eprint         = "2212.09379",
    archivePrefix  = "arXiv",
    primaryClass   = "hep-ex",
}

@Article{EGAM-2021-01,
    author         = "{ATLAS Collaboration}",
    title          = "{Electron and photon efficiencies in LHC Run~2 with the ATLAS experiment}",
    journal        = "JHEP",
    volume         = "05",
    year           = "2024",
    pages          = "162",
    doi            = "10.1007/JHEP05(2024)162",
    reportNumber   = "CERN-EP-2023-182",
    eprint         = "2308.13362",
    archivePrefix  = "arXiv",
    primaryClass   = "hep-ex",
}

@Article{EGAM-2021-02,
    author         = "{ATLAS Collaboration}",
    title          = "{Electron and photon energy calibration with the ATLAS detector using LHC Run~2 data}",
    journal        = "JINST",
    volume         = "19",
    year           = "2024",
    pages          = "P02009",
    doi            = "10.1088/1748-0221/19/02/P02009",
    reportNumber   = "CERN-EP-2023-128",
    eprint         = "2309.05471",
    archivePrefix  = "arXiv",
    primaryClass   = "hep-ex",
}

@Article{HDBS-2021-18,
    author         = "{ATLAS Collaboration}",
    title          = "{Combination of Searches for Higgs Boson Pair Production in \(pp\) Collisions at \(\sqrt{s}=13\,\text{TeV}\) with the ATLAS detector}",
    journal        = "Phys. Rev. Lett.",
    volume         = "133",
    year           = "2024",
    pages          = "101801",
    doi            = "10.1103/PhysRevLett.133.101801",
    reportNumber   = "CERN-EP-2024-160",
    eprint         = "2406.09971",
    archivePrefix  = "arXiv",
    primaryClass   = "hep-ex",
}

@Article{HIGG-2021-23,
    author         = "{ATLAS Collaboration}",
    title          = "{A detailed map of Higgs boson interactions by the ATLAS experiment ten years after the discovery}",
    journal        = "Nature",
    volume         = "607",
    year           = "2022",
    pages          = "52--59",
    doi            = "10.1038/s41586-022-04893-w",
    reportNumber   = "CERN-EP-2022-057",
    eprint         = "2207.00092",
    archivePrefix  = "arXiv",
    primaryClass   = "hep-ex",
    related        = "HIGG-2021-23-err",
    relatedstring  = "Erratum:",
}

@Article{TOPQ-2021-08,
    author         = "{ATLAS Collaboration}",
    title          = "{Observation of four-top-quark production in the multilepton final state with the ATLAS detector}",
    journal        = "Eur. Phys. J. C",
    volume         = "83",
    year           = "2023",
    pages          = "496",
    doi            = "10.1140/epjc/s10052-023-11573-0",
    reportNumber   = "CERN-EP-2023-055",
    eprint         = "2303.15061",
    archivePrefix  = "arXiv",
    primaryClass   = "hep-ex",
    related        = "TOPQ-2021-08-err",
    relatedstring  = "Erratum:",
}

@Article{TOPQ-2021-26,
    author         = "{ATLAS Collaboration}",
    title          = "{Measurement of top-quark pair production in association with charm quarks in proton--proton collisions at \(\sqrt{s}=13\,\text{TeV}\) with the ATLAS detector}",
    journal        = "Phys. Lett. B",
    volume         = "860",
    year           = "2025",
    pages          = "139177",
    doi            = "10.1016/j.physletb.2024.139177",
    reportNumber   = "CERN-EP-2024-242",
    eprint         = "2409.11305",
    archivePrefix  = "arXiv",
    primaryClass   = "hep-ex",
}

@Article{JETM-2022-01,
    author         = "{ATLAS Collaboration}",
    title          = "{New techniques for jet calibration with the ATLAS detector}",
    journal        = "Eur. Phys. J. C",
    volume         = "83",
    year           = "2023",
    pages          = "761",
    doi            = "10.1140/epjc/s10052-023-11837-9",
    reportNumber   = "CERN-EP-2023-028",
    eprint         = "2303.17312",
    archivePrefix  = "arXiv",
    primaryClass   = "hep-ex",
}

@Article{MUON-2022-01,
    author         = "{ATLAS Collaboration}",
    title          = "{Studies of the muon momentum calibration and performance of the ATLAS detector with \(pp\) collisions at \(\sqrt{s} = 13\,\text{TeV}\)}",
    journal        = "Eur. Phys. J. C",
    volume         = "83",
    year           = "2023",
    pages          = "686",
    doi            = "10.1140/epjc/s10052-023-11584-x",
    reportNumber   = "CERN-EP-2022-274",
    eprint         = "2212.07338",
    archivePrefix  = "arXiv",
    primaryClass   = "hep-ex",
}

@Article{SOFT-2022-02,
    author         = "{ATLAS Collaboration}",
    title          = "{Software and computing for Run~3 of the ATLAS experiment at the LHC}",
    journal        = "Eur. Phys. J. C",
    volume         = "85",
    year           = "2025",
    pages          = "234",
    doi            = "10.1140/epjc/s10052-024-13701-w",
    reportNumber   = "CERN-EP-2024-100",
    eprint         = "2404.06335",
    archivePrefix  = "arXiv",
    primaryClass   = "hep-ex",
    related        = "SOFT-2022-02-err",
    relatedstring  = "Erratum:",
}

@Article{TRIG-2022-01,
    author         = "{ATLAS Collaboration}",
    title          = "{The ATLAS trigger system for LHC Run~3 and trigger performance in 2022}",
    journal        = "JINST",
    volume         = "19",
    year           = "2024",
    pages          = "P06029",
    doi            = "10.1088/1748-0221/19/06/P06029",
    reportNumber   = "CERN-EP-2023-299",
    eprint         = "2401.06630",
    archivePrefix  = "arXiv",
    primaryClass   = "hep-ex",
}

@Article{FTAG-2023-05,
    author         = "{ATLAS Collaboration}",
    title          = "{Transforming jet flavour tagging at ATLAS}",
    journal        = "Nature Commun.",
    volume         = "17",
    year           = "2026",
    pages          = "541",
    doi            = "10.1038/s41467-025-65059-6",
    reportNumber   = "CERN-EP-2025-103",
    eprint         = "2505.19689",
    archivePrefix  = "arXiv",
    primaryClass   = "hep-ex",
}

@Article{HIGP-2024-08,
    author         = "{ATLAS Collaboration}",
    title          = "{Measurement of the Higgs boson production in association with top quarks in multilepton final states in \(pp\) collisions at \(\sqrt{s}=13\,\text{TeV}\) with the ATLAS detector}",
    journal        = "JHEP",
    volume         = "05",
    year           = "2026",
    pages          = "183",
    doi            = "10.1007/JHEP05(2026)183",
    reportNumber   = "CERN-EP-2025-239",
    eprint         = "2510.23755",
    archivePrefix  = "arXiv",
    primaryClass   = "hep-ex",
}

@Article{HIGP-2024-33,
    author         = "{ATLAS Collaboration}",
    title          = "{Search for Higgs boson pair production in association with top-quark pairs using \(196\,\text{fb}^{-1}\) of proton--proton collision data at \(\sqrt{s} = 13\) and \(13.6\,\text{TeV}\) with the ATLAS detector}",
    year           = "2026",
    reportNumber   = "CERN-EP-2026-045",
    eprint         = "2603.13113",
    archivePrefix  = "arXiv",
    primaryClass   = "hep-ex",
}

@Article{HIGP-2025-10,
    author         = "{ATLAS Collaboration}",
    title          = "{Study of Higgs boson pair production in the \(HH \rightarrow b\bar{b} \gamma\gamma\) final state with \(308\,\text{fb}^{-1}\) of data collected at \(\sqrt{s} = 13\,\text{TeV}\) and \(13.6\,\text{TeV}\) by the ATLAS experiment}",
    journal        = "Phys. Lett. B",
    volume         = "876",
    year           = "2026",
    pages          = "140280",
    doi            = "10.1016/j.physletb.2026.140280",
    reportNumber   = "CERN-EP-2025-140",
    eprint         = "2507.03495",
    archivePrefix  = "arXiv",
    primaryClass   = "hep-ex",
}

@Booklet{ATL-SOFT-PUB-2026-001,
    author         = "{ATLAS Collaboration}",
    title          = "{ATLAS Computing Acknowledgements}",
    howpublished   = "{ATL-SOFT-PUB-2026-001}",
    url            = "https://cds.cern.ch/record/2952666",
    year           = "2026",
}

@Article{LUCID2,
    author         = "Avoni, G. and others",
    title          = "{The new LUCID-2 detector for luminosity measurement and monitoring in ATLAS}",
    journal        = "JINST",
    volume         = "13",
    year           = "2018",
    pages          = "P07017",
    doi            = "10.1088/1748-0221/13/07/P07017",
}

@Article{CMS-HIG-12-028,
    author         = "{CMS Collaboration}",
    title          = "{Observation of a new boson at a mass of 125 GeV with the CMS experiment at the LHC}",
    journal        = "Phys. Lett. B",
    volume         = "716",
    year           = "2012",
    pages          = "30",
    doi            = "10.1016/j.physletb.2012.08.021",
    reportNumber   = "CERN-PH-EP-2012-220",
    eprint         = "1207.7235",
    archivePrefix  = "arXiv",
    primaryClass   = "hep-ex",
}

@Article{CMS-HIG-20-011,
    author         = "{CMS Collaboration}",
    title          = "{Combination of searches for nonresonant Higgs boson pair production in proton--proton collisions at \(\sqrt{s} = 13\,\text{TeV}\)}",
    year           = "2025",
    reportNumber   = "CERN-EP-2025-214",
    eprint         = "2510.07527",
    archivePrefix  = "arXiv",
    primaryClass   = "hep-ex",
}

@Article{CMS-TOP-21-011,
    author         = "{CMS Collaboration}",
    title          = "{Measurement of the cross section of top quark--antiquark pair production in association with a \(W\) boson in proton--proton collisions at \(\sqrt{s} = 13\,\text{TeV}\)}",
    journal        = "JHEP",
    volume         = "07",
    year           = "2023",
    pages          = "219",
    doi            = "10.1007/JHEP07(2023)219",
    reportNumber   = "CERN-EP-2022-168",
    eprint         = "2208.06485",
    archivePrefix  = "arXiv",
    primaryClass   = "hep-ex",
}

@Article{CMS-HIG-22-001,
    author         = "{CMS Collaboration}",
    title          = "{A portrait of the Higgs boson by the CMS experiment ten years after the discovery}",
    journal        = "Nature",
    volume         = "607",
    year           = "2022",
    pages          = "60--68",
    doi            = "10.1038/s41586-022-04892-x",
    reportNumber   = "CERN-EP-2022-039",
    eprint         = "2207.00043",
    archivePrefix  = "arXiv",
    primaryClass   = "hep-ex",
    related        = "CMS-HIG-22-001-err",
    relatedstring  = "Erratum:",
}

@Article{CMS-HIG-22-006,
    author         = "{CMS Collaboration}",
    title          = "{Search for Higgs boson pair production with one associated vector boson in proton--proton collisions at \(\sqrt{s} = 13\,\text{TeV}\)}",
    journal        = "JHEP",
    volume         = "10",
    year           = "2024",
    pages          = "061",
    doi            = "10.1007/JHEP10(2024)061",
    reportNumber   = "CERN-EP-2024-064",
    eprint         = "2404.08462",
    archivePrefix  = "arXiv",
    primaryClass   = "hep-ex",
}

@Article{CMS-TOP-22-009,
    author         = "{CMS Collaboration}",
    title          = "{Inclusive and differential cross section measurements of \(t\bar{t}b\bar{b}\) production in the lepton+jets channel at \(\sqrt{s} = 13\,\text{TeV}\)}",
    journal        = "JHEP",
    volume         = "05",
    year           = "2024",
    pages          = "042",
    doi            = "10.1007/JHEP05(2024)042",
    reportNumber   = "CERN-EP-2023-201",
    eprint         = "2309.14442",
    archivePrefix  = "arXiv",
    primaryClass   = "hep-ex",
}

@Booklet{ATL-PHYS-PUB-2010-013,
    author         = "{ATLAS Collaboration}",
    title          = "{The simulation principle and performance of the ATLAS fast calorimeter simulation FastCaloSim}",
    howpublished   = "{ATL-PHYS-PUB-2010-013}",
    url            = "https://cds.cern.ch/record/1300517",
    year           = "2010",
}

@Booklet{ATL-PHYS-PUB-2014-021,
    author         = "{ATLAS Collaboration}",
    title          = "{ATLAS Pythia~8 tunes to \(7\,\text{TeV}\) data}",
    howpublished   = "{ATL-PHYS-PUB-2014-021}",
    url            = "https://cds.cern.ch/record/1966419",
    year           = "2014",
}

@Booklet{ATL-PHYS-PUB-2015-026,
    author         = "{ATLAS Collaboration}",
    title          = "{Vertex Reconstruction Performance of the ATLAS Detector at \(\sqrt{s} = 13\,\text{TeV}\)}",
    howpublished   = "{ATL-PHYS-PUB-2015-026}",
    url            = "https://cds.cern.ch/record/2037717",
    year           = "2015",
}

@Booklet{ATL-PHYS-PUB-2016-020,
    author         = "{ATLAS Collaboration}",
    title          = "{Studies on top-quark Monte Carlo modelling for Top2016}",
    howpublished   = "{ATL-PHYS-PUB-2016-020}",
    url            = "https://cds.cern.ch/record/2216168",
    year           = "2016",
}

@Booklet{ATL-PHYS-PUB-2020-028,
    author         = "{ATLAS Collaboration}",
    title          = "{Recommendations for the Modeling of Smooth Backgrounds}",
    howpublished   = "{ATL-PHYS-PUB-2020-028}",
    url            = "https://cds.cern.ch/record/2743717",
    year           = "2020",
}

@Booklet{ATL-DAPR-PUB-2023-001,
    author         = "{ATLAS Collaboration}",
    title          = "{Preliminary analysis of the luminosity calibration of the ATLAS \(13.6\,\text{TeV}\) data recorded in 2022}",
    howpublished   = "{ATL-DAPR-PUB-2023-001}",
    url            = "https://cds.cern.ch/record/2853525",
    year           = "2023",
}

@Booklet{ATL-DAPR-PUB-2024-001,
    author         = "{ATLAS Collaboration}",
    title          = "{Preliminary analysis of the luminosity calibration for the ATLAS \(13.6\,\text{TeV}\) data recorded in 2023}",
    howpublished   = "{ATL-DAPR-PUB-2024-001}",
    url            = "https://cds.cern.ch/record/2900949",
    year           = "2024",
}
\clearpage
 
\begin{flushleft}
\hypersetup{urlcolor=black}
{\Large The ATLAS Collaboration}

\bigskip

\AtlasOrcid[0000-0002-6665-4934]{G.~Aad}$^\textrm{\scriptsize 102}$,
\AtlasOrcid[0000-0001-7616-1554]{E.~Aakvaag}$^\textrm{\scriptsize 17}$,
\AtlasOrcid[0000-0002-5888-2734]{B.~Abbott}$^\textrm{\scriptsize 121}$,
\AtlasOrcid[0000-0002-0287-5869]{S.~Abdelhameed}$^\textrm{\scriptsize 83b}$,
\AtlasOrcid[0000-0002-1002-1652]{K.~Abeling}$^\textrm{\scriptsize 54}$,
\AtlasOrcid[0000-0001-5763-2760]{N.J.~Abicht}$^\textrm{\scriptsize 48}$,
\AtlasOrcid[0000-0002-8496-9294]{S.H.~Abidi}$^\textrm{\scriptsize 30}$,
\AtlasOrcid[0009-0003-6578-220X]{M.~Aboelela}$^\textrm{\scriptsize 44}$,
\AtlasOrcid[0000-0002-9987-2292]{A.~Aboulhorma}$^\textrm{\scriptsize 36e}$,
\AtlasOrcid[0000-0001-5329-6640]{H.~Abramowicz}$^\textrm{\scriptsize 154}$
% ...
\bigskip
\\

% ...
$^{17}$Department for Physics and Technology, University of Bergen, Bergen; Norway.\\
% ...
$^{30}$Physics Department, Brookhaven National Laboratory, Upton NY; United States of America.\\
% ...
$^{36}$$^{(a)}$Facult\'e des Sciences Ain Chock, Universit\'e Hassan II de Casablanca;$^{(b)}$Facult\'{e} des Sciences, Universit\'{e} Ibn-Tofail, K\'{e}nitra;$^{(c)}$Facult\'e des Sciences Semlalia, Universit\'e Cadi Ayyad, LPHEA-Marrakech;$^{(d)}$LPMR, Facult\'e des Sciences, Universit\'e Mohamed Premier, Oujda;$^{(e)}$Facult\'e des sciences, Universit\'e Mohammed V, Rabat;$^{(f)}$Institute of Applied Physics, Mohammed VI Polytechnic University, Ben Guerir; Morocco.\\
% ...
$^{44}$Physics Department, Southern Methodist University, Dallas TX; United States of America.\\
% ...
$^{48}$Fakult\"{a}t Physik, Technische Universit{\"a}t Dortmund, Dortmund; Germany.\\
% ...
$^{54}$II. Physikalisches Institut, Georg-August-Universit\"{a}t G\"ottingen, G\"ottingen; Germany.\\
% ...
$^{83}$$^{(a)}$Khalifa University of Science and Technology, Abu Dhabi;$^{(b)}$New York University Abu Dhabi, Abu Dhabi;$^{(c)}$United Arab Emirates University, Al Ain;$^{(d)}$University of Sharjah, Sharjah; United Arab Emirates.\\
% ...
$^{102}$CPPM, Aix-Marseille Universit\'e, CNRS/IN2P3, Marseille; France.\\
% ...
$^{121}$Homer L. Dodge Department of Physics and Astronomy, University of Oklahoma, Norman OK; United States of America.\\
% ...
$^{154}$Raymond and Beverly Sackler School of Physics and Astronomy, Tel Aviv University, Tel Aviv; Israel.\\
% ...
$^{a}$ Also at Affiliated with an institute formerly covered by a cooperation agreement with CERN.\\
$^{b}$ Also at An-Najah National University, Nablus; Palestine.\\
$^{c}$ Also at Borough of Manhattan Community College, City University of New York, New York NY; United States of America.\\
$^{d}$ Also at Center for Interdisciplinary Research and Innovation (CIRI-AUTH), Thessaloniki; Greece.\\
$^{e}$ Also at Centre of Physics of the Universities of Minho and Porto (CF-UM-UP); Portugal.\\
$^{f}$ Also at CERN, Geneva; Switzerland.\\
% ...

\end{flushleft}


%

%
%

%

%
%
%
%

\end{document}